\documentclass{aa}
\usepackage[varg]{txfonts}
\usepackage{graphicx}
\usepackage{lscape}

\usepackage{natbib,twoopt}
\usepackage{xcolor}
\usepackage[
  breaklinks=true,
  ]{hyperref} %% to avoid \citeads line fills
\bibpunct{(}{)}{;}{a}{}{,}             %% natbib format for A&A and ApJ
\makeatletter
  \newcommandtwoopt{\citeads}[3][][]{\href{http://adsabs.harvard.edu/abs/#3}%
    {\def\hyper@linkstart##1##2{}%
     \let\hyper@linkend\@empty\citealp[#1][#2]{#3}}}
  \newcommandtwoopt{\citepads}[3][][]{\href{http://adsabs.harvard.edu/abs/#3}%
    {\def\hyper@linkstart##1##2{}%
     \let\hyper@linkend\@empty\citep[#1][#2]{#3}}}
  \newcommandtwoopt{\citetads}[3][][]{\href{http://adsabs.harvard.edu/abs/#3}%
    {\def\hyper@linkstart##1##2{}%
     \let\hyper@linkend\@empty\citet[#1][#2]{#3}}}
  \newcommandtwoopt{\citeyearads}[3][][]%
    {\href{http://adsabs.harvard.edu/abs/#3}
    {\def\hyper@linkstart##1##2{}%
     \let\hyper@linkend\@empty\citeyear[#1][#2]{#3}}}
\makeatother

\newcommand{\rphk}{\ensuremath{R'_\mathrm{HK}}\xspace}
\newcommand{\rhk}{\ensuremath{R_\mathrm{HK}}\xspace}
\newcommand{\rphot}{\ensuremath{R_\mathrm{HK,\,phot}}\xspace}

\newcommand{\lha}{\ensuremath{L_\mathrm{H\alpha}/L_\mathrm{bol}}\xspace}
\newcommand{\fha}{\ensuremath{\mathcal{F}_\mathrm{H\alpha}/\mathcal{F}_\mathrm{bol}}\xspace}
\newcommand{\fpha}{\ensuremath{\mathcal{F}'_\mathrm{H\alpha}/\mathcal{F}_\mathrm{bol}}\xspace}

\newcommand{\teff}{\ensuremath{T_\mathrm{eff}}\xspace}
\newcommand{\logg}{\ensuremath{\log(g)}\xspace}

\newcommand{\feh}{\ensuremath{[\mathrm{Fe/H}]}\xspace}
\newcommand{\teffspt}{\ensuremath{T_\mathrm{eff,\,SpT}}\xspace}
\newcommand{\teffmass}{\ensuremath{T_\mathrm{eff,\,\mathcal{M}_\star}}\xspace}
\newcommand{\teffmal}{\ensuremath{T_\mathrm{eff,\,M15}}\xspace}
\begin{document}

\title{Absolute Ca II H \& K and H-alpha flux measurements of low-mass stars: Extending $R'_\mathrm{HK}$ to M dwarfs}

  \author{C. J. Marvin\inst{1}
    \and A. Reiners\inst{1}
    \and G. Anglada-Escud{\'e}\inst{1,2,3}
    \and S. V. Jeffers\inst{5}
    \and S. Boro Saikia\inst{1,4}
    }
    
  \offprints{C.J. Marvin, \email{chris.j.marvin@gmail.com}}
    
  \titlerunning{Absolute Ca II H \& K and H-alpha flux measurements of low-mass stars}
  \authorrunning{Marvin et al.}

  \institute{Universit{\"a}t G{\"o}ttingen, Institut f{\"u}r Astrophysik und Geophysik, Friedrich-Hund-Platz 1, 37077 G{\"o}ttingen, Germany
    \and School of Physics and Astronomy, Queen Mary, University of London, 327 Mile End Rd. London, United Kingdom
    \and Instituto de Astrofísica de Andaluc{\'i}a, Glorieta de la Astronom{\'i}a 1, E-18008 Granada, Spain
    \and University of Vienna, Department of Astrophysics, T{\"u}rkenschanzstrasse 17, 1180 Vienna, Austria
    \and Max-Planck-Institut f{\"u}r Sonnensystemforschung, Justus-von-Liebig-Weg-3, 37077 Goettingen}

%  \date{Received 1 July 2006 /
%    Accepted 2 July 2006}

  \abstract
  % CONTEXT
  {
  With the recent surge of planetary surveys focusing on detecting
  Earth-mass planets around M dwarfs,
  it is becoming more important to understand chromospheric activity
  in M dwarfs.
  Stellar chromospheric calcium emission is typically measured using the 
  $R'_\mathrm{HK}$ calibrations of \citet{Noyes1984}, which are only
  valid for $0.44 \le B-V \le 0.82$.
  Measurements of calcium emission for cooler dwarfs $B-V \ge 0.82$
  are difficult because of their intrinsic dimness in the blue end
  of the visible spectrum.
  }
  % AIMS
  {
  We measure the absolute \ion{Ca}{II} H \& K and H$\alpha$ flux
  of a sample of 110 HARPS M dwarfs
  and also extend the calibration of \rphk{} to the M dwarf regime
  using PHOENIX stellar atmosphere models.
  }
  % METHODS
  {
  We normalized a template spectrum with a high signal-to-noise ratio that was
  obtained by coadding multiple spectra of the same star
  to a PHOENIX stellar atmosphere model
  to measure the chromospheric \ion{Ca}{II} H \& K and H$\alpha$ flux in physical units.
  We used three different \teff calibrations and investigated their effect on
  \ion{Ca}{II} H \& K and H$\alpha$ activity measurements.
  We performed conversions of the Mount Wilson S index to $R'_\mathrm{HK}$
  as a function of effective temperature for the range
  $2300\ \mbox{K}\ \le \teff \le 7200$~K.
  Last, we calculated continuum luminosity $\chi$ values for
  \ion{Ca}{II} H \& K and H$\alpha$ in the same manner
  as \citet{WestHawley2008} for $-1.0 \le \feh \le +1.0$
  in steps of $\Delta \feh = 0.5$.
  }
  % RESULTS
  {
  We compare different \teff calibrations and find 
  $\Delta \teff \sim\,\mathrm{several}\,100$~K for mid- to late-M dwarfs.
  Using these different \teff calibrations,
  we establish a catalog of $\log \rphk$ and $\fpha$ measurements for 110 HARPS M dwarfs.
  The difference between our results
  and the calibrations of \citet{Noyes1984}
  is $\Delta \log \rphk = 0.01$~dex for a Sun-like star.
  Our $\chi$ values agree well with those of \citet{WestHawley2008}.
  We 
  confirm
   that the lower boundary of chromospheric \ion{Ca}{II} H and K activity
  does not increase toward
  later-M dwarfs: it either stays constant or decreases, depending on the
  \teff calibration used.
  We also confirm that
   for H$\alpha$, the lower boundary of chromospheric flux is in absorption for earlier
-M dwarfs and fills into the continuum toward later M dwarfs.
  }
  % CONCLUSIONS
  {
  We 
  confirm
   that we can effectively measure \rphk in M dwarfs
  using template spectra with a high signal-to-noise ratio.
  We also conclude that our calibrations are a reliable extension
  of previous \rphk calibrations,
  and effective temperature calibration is the main source of
  error in our activity measurements.
  }


  \maketitle

%%%%%%%%%%%%%%%%%%%%%%%%%%%%%%%%%%%%%%%%%%%%%%%%%%%%%%%%%%%%%%%%%%%%%%%%%%%%%%
%%%%%%%%%%%%%%%%%%%%%%%%%%%[ SECTIONS ]%%%%%%%%%%%%%%%%%%%%%%%%%%%%%%%%%%%%%%%
%%%%%%%%%%%%%%%%%%%%%%%%%%%%%%%%%%%%%%%%%%%%%%%%%%%%%%%%%%%%%%%%%%%%%%%%%%%%%%

%%%%%%%%%%%%%%%%%%%%%%%%%%%%%%%%%%%%%%%%%%%%%%%%%%%%%%%%%%%%%%%%%%%%%%%%%%%%%%%
\section{Introduction}

\begin{figure}
    \resizebox{\hsize}{!}{\includegraphics{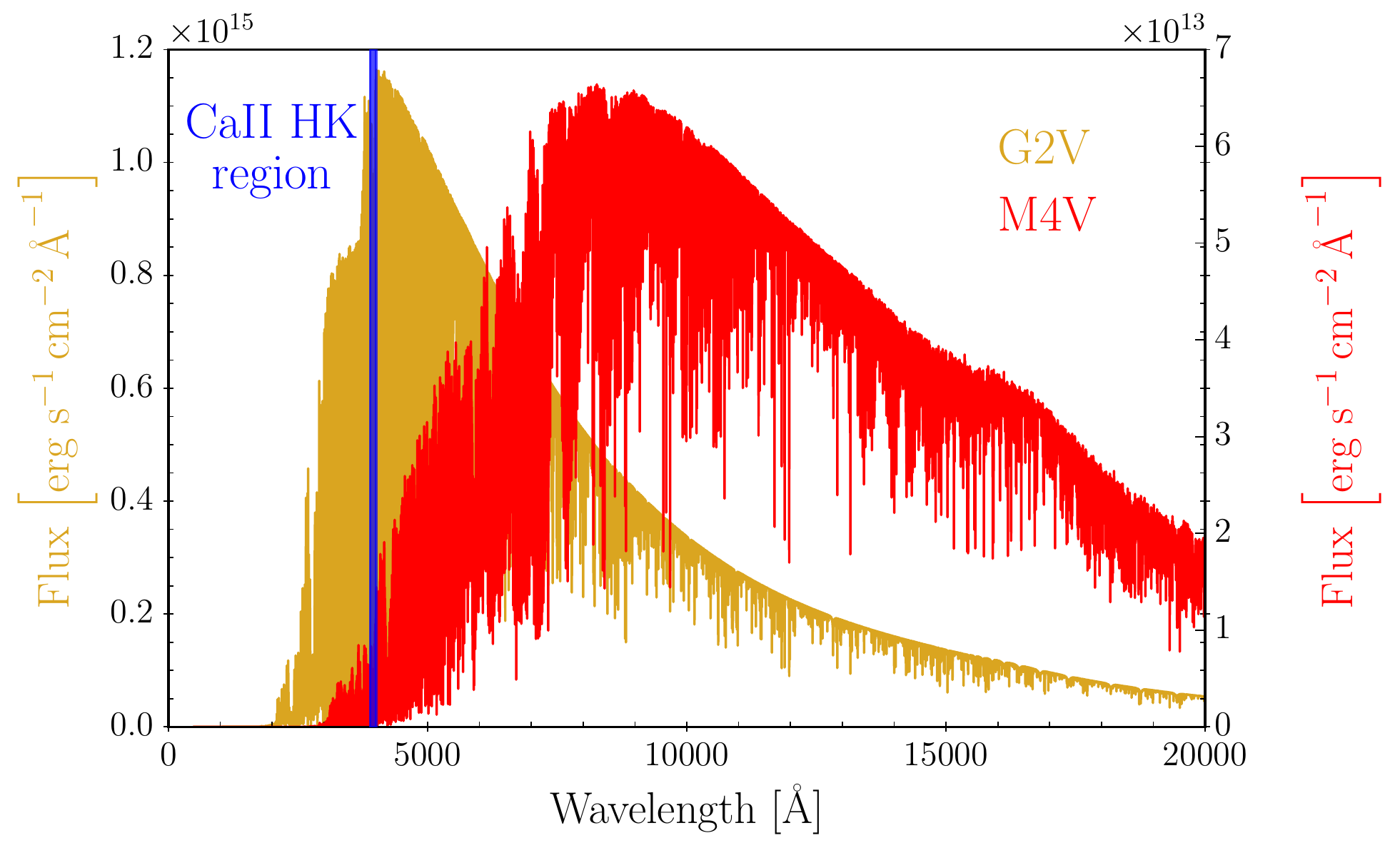}}
    \caption{Model G2V spectra (yellow) and model M4V spectra (red).
    The flux scale of the G2V is displayed on the left, and the flux scale
    of the M4V is given on the right.
    The blue shaded area indicates the region of the \ion{Ca}{II} H and K lines.}
    \label{figure:spt_compare}
  \end{figure}

M dwarfs comprise the majority of the stellar population,
but their fundamental properties present challenges in measuring some
of the most common activity indicators in the optical wavelength region,
in particular, \ion{Ca}{II} H and K,
which lie in the bluer part of the visible spectrum.
The cooler temperatures of M dwarfs
imply that the bulk of their radiation lies toward longer wavelengths
than for their FGK counterparts.
M dwarfs are also intrinsically dimmer,
which either decreases the
signal-to-noise ratio (S/N)
of observations
or requires longer exposure times.
Figure~\ref{figure:spt_compare} demonstrates the differences in brightness
and spectral energy distribution between
a Sun-like G2 dwarf and a typical M4 dwarf.
Additionally,
telescope transmission and detector sensitivity are often higher
in the redder wavelengths, which further exacerbates the problem
of comparing their calcium flux with FGK stars in a consistent manner.

In what is known as the de facto standard of stellar activity surveys,
\citet{Baliunas1995} monitored \ion{Ca}{II} H and K lines of
111 main-sequence FGKM stars for several decades
using the dimensionless measure called the Mount Wilson S index,
or $S_\mathrm{MWO}$.
This S index is a ratio of the \ion{Ca}{II} H and K line core fluxes
normalized to nearby continuum bands.
However, the fluxes of the nearby continuum bands are not constant across spectral types \citep{VaughanPreston1980, Hartmann1984},
which makes the comparison of stellar activity of different spectral types
difficult with $S_\mathrm{MWO}$.
To mitigate the color-dependence, $S_\mathrm{MWO}$ is usually transformed
into a physical quantity known as \rhk,
which is the ratio of \ion{the Ca}{II} H and K surface flux to bolometric flux.
A more desirable measure, known as \rphk,
subtracts the photospheric contribution \rphot,
leaving only the chromospheric flux excess.
Here, the prime denotes that the flux measurement is solely of
chromospheric origin.
The most common method of calculating \rphk is the prescription 
derived by \citet{Noyes1984},
which requires only $S_\mathrm{MWO}$ and $B-V$ to obtain \rphk.
However, this method is only valid for $0.44 \le B-V \le 0.82$,
which means that M dwarfs lie outside of this \rphk calibration range.

Despite these difficulties,
measurements of \ion{Ca}{II} H and K 
and H$\alpha$
flux in M dwarfs have been performed before.
\citet{WalkowiczHawley2009} measured \ion{Ca}{II} H and K equivalent widths
for a sample of M3 dwarfs
with the spectral subtraction method,
using a stellar atmosphere model to correct for the photospheric flux contribution.
The same technique was used by \citet{Montes1995a},
who coined the term "spectral subtraction technique"
and that has been performed as far back as \citet{Barden1985} and \citet{Herbig1985}.
In these studies, a synthetic spectral line profile is used as the
photospheric contribution of a given star and is subtracted from an
observed spectrum, resulting in a measurement of the chromospheric flux excess.
\citet{Montes1995} used the spectral subtraction technique
to measure the chromospheric \ion{Ca}{II} H and K flux excess in 28 FGK stars.
Instead of using a photospheric spectrum,
\citet{Cincunegui2007} measured the surface flux $F_\mathrm{HK}$ of main-sequence stars
from early-F down to M5 spectral types,
and extrapolated the \citet{Noyes1984} photospheric
contribution for the M dwarfs.
To measure the fractional surface flux to bolometric flux of \ion{Ca}{II} H and K,
\citet{WestHawley2008} used the $\chi$ method,
where multiplying an equivalent width by a factor $\chi$, a continuum measurement
nearby the calcium line, results in 
$L_{\ion{Ca}{II}\mathrm{HK}} / L_\mathrm{bol}$.
This study provided $\chi$ factors of \ion{Ca}{II} H and K 
for the spectral range M0 - M8.
However, it did not provide a correction for the photospheric contribution.
To extend the photospheric flux relation down to $B-V = 1.6$ with the spectral subtraction technique,
\citet{MartinezArnaiz2011} 
used a synthetic template photospheric spectrum
obtained by adding together spectra of nonactive stars of similar
spectral type,
and measured excess surface fluxes of 298 main-sequence stars
ranging from F to M.
\citet{Mittag2013} used PHOENIX model atmospheres
to update the relations of \citet{Noyes1984} and measured $R'_\mathrm{HK}$
for 2133 main-sequence stars.
Instead of using stellar models,
\citet{SuarezMascareno2015} used HARPS spectra of main-sequence FGKM dwarfs
to derive their own $R'_\mathrm{HK}$ relations down to $B-V \sim 1.9$,
and measured $R'_\mathrm{HK}$ for 48 late-F to mid-M type stars.
\citet{Scandariato2017} used the spectral subtraction technique with
BT-Settl models as photospheric spectra for 71 early-M dwarfs
and measured \ion{Ca}{II} H,  K, and H$\alpha$.
\citet{Astudillo2017} formulated their own S-index calibration
using HARPS spectra
and used their own conversion from S-index to \rphk for 403 M dwarfs.
\citet{Newton2017} found $\lha$ for 270 nearby M dwarfs
using recomputed $\chi_{H\alpha}$ values of \citet{WestHawley2008}.

The relation between \ion{Ca}{II} H,  K, and H$\alpha$ 
emission (or absorption)
is also of much interest.
Measuring the line profiles of 147 K7-M5 main-sequence stars,
\citet{Rauscher2006} showed that \ion{Ca}{II} H and K lines form at
slightly different heights in the chromosphere, and that
the equivalent width of H$\alpha$ only correlates
with \ion{Ca}{II} H and K high widths. They also reported a possible threshold above
which the lower and upper chromospheres become thermally coupled.
    \citet{Cincunegui2007} found a clear correlation between averaged \ion{Ca}{II} H,  K, and H$\alpha$
with the strongest correlation for stars with the strongest emission.
Conversely, studying stars individually and at different time intervals, \citet{Cincunegui2007} found no clear indication of
how H$\alpha$ varies with \ion{Ca}{II}, with stars showing correlation,
anticorellation, or no correlation.
Also observing individual time measurements for a sample of 30 M dwarfs, \citet{GomesDaSilva2011} found a positive correlation
for the most active stars, and a tendency for a low or negative correlation
in the least active stars.
\citet{WalkowiczHawley2009} found an initial deepening of H$\alpha$ absorption
for the stars that are least active in \ion{Ca}{II} H and K before line filling and going into
emission.
\citet{Scandariato2017} found this same nonlinear relation between \ion{Ca}{II} H, K, and H$\alpha$
in 71 early-M dwarfs.
\citet{Maldonado2017} separated older stars from younger and more active stars
using the distinction of two branches identified by \citet{MartinezArnaiz2011}.
They found that the log-fluxes of \ion{Ca}{II} H, K, and H$\alpha$ relatively follow the
the same linear relation for stars spectral type F to M,
which they identify as being the inactive branch,
and found that stars deviating from this tend to be more active and younger,
and thus lie on the active branch.
More recently, \citet{Reiners2022} reported relations between chromospheric \ion{Ca}{II} H, K, and magnetic flux,
    and also H$\alpha$ emission and magnetic flux. Combining these relations, a relation between
\ion{Ca}{II} H, K, and H$\alpha$ might be derived.

Many calibrations for \rphk{} exist for the main sequence from
early-F to late-M spectral type.
Very few studies have used high S/N coadded spectra
with the spectral subtraction technique (~\citet{BoroSaikia2018};~\citet{Perdelwitz2021}).
In fact, our work in this paper
provided the M18 template-model method and measurements of~\citet{BoroSaikia2018} (see Sec.2 and 3.2.2 of the aforementioned work).
In this work, we measure \ion{Ca}{II} H, K, and H$\alpha$
activity with the spectral subtraction technique in a sample of 110 M dwarfs
using high S/N template spectra that are flux-calibrated
to PHOENIX stellar atmosphere models.
The main difference of this study is that instead of taking the mean value
of \ion{Ca}{II} H and K flux measurements,
we combine all available spectra and coadd them together
before
the flux measurement.
This allows us to not just scale the \ion{Ca}{II} H and K
measurement to an absolute flux unit,
but to fit the 
spectral energy distribution (SED)
of the calcium line to a PHOENIX stellar atmosphere,
and similarly for H$\alpha$.
We compare three different effective temperature calibrations,
and investigate their effect on \ion{Ca}{II} H, K, and H$\alpha$
activity measurements.
We extend the \rphk{} calibrations
to $2300\ \mbox{K} \le \teff \le 7200$~K
using PHOENIX stellar atmosphere models.
We also provide a table of \rphk{} calibrations in this effective temperature
range for different metallicities and surface gravities
of main-sequence stars.
Last, we compute the $\chi$ values of \citet{WestHawley2008} for 
\ion{Ca}{II} H, K, and H$\alpha$ for different metallicities
from the PHOENIX model atmospheres.

This paper is organized as follows:
In Sec.~\ref{SecMethods}
we briefly review the definition of \ion{Ca}{II} H and K and H$\alpha$
activity.
In Sec.~\ref{SecHarps} we discuss the sample of stars, and
we calibrate \teff using three different methods.
We discuss the technique of measuring
\ion{Ca}{II} H, K, and H$\alpha$ in
M~dwarfs with the subtraction method,
using coadded template spectra and model photospheres.
In Sec.~\ref{SecResults} we discuss our
\ion{Ca}{II} H, K, and H$\alpha$ measurements,
provide extended \rphk calibrations,
and compare our calibrations with previous works.
In Sec.~\ref{SecSummary} we summarize our work.
%%%%%%%%%%%%%%%%%%%%%%%%%%%%%%%%%%%%%%%%%%%%%%%%%%%%%%%%%%%%%%%%%%%%%%%%%%%%%%%

%%%%%%%%%%%%%%%%%%%%%%%%%%%%%%%%%%%%%%%%%%%%%%%%%%%%%%%%%%%%%%%%%%%%%%%%%%%%%%%
\section{Overview of the measurement equations}
  \label{SecMethods}

\subsection{Mount Wilson S-index}
\label{SecSindex}

The \mbox{HKP-2} spectrophotometer installed at the Mount Wilson Observatory measures the 
\ion{Ca}{II} H and K line cores with a
triangular $1.09~\mbox{\AA}$ 
full width at half maximum (FWHM)
bandpass while simultaneously measuring two $20~\mbox{\AA}$ wide bands;
$R$, centered on $4001.07~\mbox{\AA}$,
and $V$, centered on $3901.07~\mbox{\AA}$.
To mimic the response of this instrument,
\citet{Duncan1991} prescribed the following S index formula:
\begin{equation}
  \label{EqSHarps}
  S = 8 \alpha \frac{N_\mathrm{H} + N_\mathrm{K}}{N_R + N_V},
\end{equation}
where $N_\mathrm{H}$, $N_\mathrm{K}$, $N_R$, and $N_V$ are the counts in
their respective bands,
and $\alpha$ is a proportionality constant equating measurements made by the
\mbox{HKP-2} spectrophotometer to those made with \mbox{HKP-1};
\citet{Duncan1991} adopted the value $\alpha = 2.4$.
The factor of 8 arises from the 8:1 duty cycle between the line core
and continuum bandpasses.
Since its inception, the S-index has been the most widely used
activity indicator for FGK stars.
% %%%%%%%%%%%%%%%%%%%%%%%%%%%%%%%%%%%%%%%%%%%%%%%%%%%%%%%%%%%%%%%%%%%%%%%%%%%%%

\subsection{Chromospheric Ca II H and K ratio}

To convert the dimensionless $S_\mathrm{MWO}$
into arbitrary surface flux $F_\mathrm{HK}$,
\citet{Middelkoop1982} and \citet{Rutten1984} derived a continuum conversion
factor $C_\mathrm{cf}$.
The arbitrary surface flux is defined as
\begin{equation}
  \label{EqFHKR84}
  F_\mathrm{HK} = S_\mathrm{MWO} C_\mathrm{cf} T_\mathrm{eff}^4 10^{-14},
\end{equation}
and its conversion into absolute units is given by
\begin{equation}
  \label{EqFHKAbs}
  \mathcal{F}_\mathrm{HK} = K F_\mathrm{HK},
\end{equation}
where $\mathcal{F}_\mathrm{HK}$ and $K$ are in units of
erg cm$^{-2}$ s$^{-1}$.

% %%%%%%%%%%%%%%%%%%%%%%%%%%%%%%%%%%%%%%%%%%%%%%%%%%%%%%%%%%%%%%%%%%%%%%%%%%%%%

\begin{equation}
  R'_\mathrm{HK} = \frac{\mathcal{F}'_\mathrm{H} + \mathcal{F}'_\mathrm{K}}{\sigma T_\mathrm{eff}^4}
  = \frac{\mathcal{F}'_\mathrm{HK}}{\mathcal{F}_\mathrm{bol}},
  \label{EqRpHK}
\end{equation}
where $\mathcal{F}'_\mathrm{H} = \mathcal{F}_\mathrm{H} - \mathcal{F}_\mathrm{H,\,phot}$
and $\mathcal{F}'_\mathrm{K} = \mathcal{F}_\mathrm{K} - \mathcal{F}_\mathrm{K,\,phot}$.
Here, $\mathcal{F}'_\mathrm{H}$ and $\mathcal{F}'_\mathrm{K}$ are the
chromospheric fluxes,
$\mathcal{F}_\mathrm{H}$ and $\mathcal{F}_\mathrm{K}$ are the surface fluxes,
and $\mathcal{F}_\mathrm{H,\,phot}$ and $\mathcal{F}_\mathrm{K,\,phot}$
are the photospheric fluxes of the \ion{Ca}{II} H and K lines, respectively.
From a slight rearranging, Eq.~\ref{EqRpHK} can be written as
\begin{equation}
  \label{EqRpHKdiff}
  R'_\mathrm{HK} = \frac{\mathcal{F}_\mathrm{HK} - \mathcal{F}_\mathrm{HK,\,phot}}{\sigma T_\mathrm{eff}^4}
  = R_\mathrm{HK} - R_\mathrm{HK,\,phot},
\end{equation}
with the surface flux ratio given by
\begin{equation}
  \label{EqRHK}
  R_\mathrm{HK} = \frac{\mathcal{F}_\mathrm{HK}}{\sigma T_\mathrm{eff}^4}
\end{equation}
and the photospheric flux ratio given by
\begin{equation}
  \label{EqRphot}
  R_\mathrm{HK,\,phot} = \frac{\mathcal{F}_\mathrm{HK,\,phot}}{\sigma T_\mathrm{eff}^4}.
\end{equation}

\label{SecRpHK}

Typically, \rphk{} is measured through a conversion from the Mount Wilson
S-index.
The method pioneered by \citet{Noyes1984} calculates \rhk{} using the equation
\begin{equation}
  \label{EqRHKN84}
  \rhk = 1.34 \times 10^{-4} C_\mathrm{cf} S_\mathrm{MWO}.
\end{equation}
The quantity $C_\mathrm{cf}$ is a color-dependent conversion factor
to remove the color-dependence of the $S_\mathrm{MWO}$, and \citet{Noyes1984}
used the \citet{Middelkoop1982} relation,
\begin{equation}
\label{EqCcf}
  % \log{C_\mathrm{cf,\,M82}} = 1.13 (B-V)^3 - 3.91 (B-V)^2 + 2.84 (B-V) - 0.47
  \log{C_\mathrm{cf}} = 1.13 (B-V)^3 - 3.91 (B-V)^2 + 2.84 (B-V) - 0.47,
\end{equation}
for $0.45 \le (B-V) \le 1.50$.
To calculate $R_\mathrm{phot}$, \citet{Noyes1984} used the following
relation:
\begin{equation}
  \label{EqRphotN84}
  \log R_\mathrm{phot,\,N84} = -4.898 +1.918(B-V)^2 -2.893(B-V)^3
\end{equation}
for $0.44 \le (B-V) \le 0.82$.
Equation~\ref{EqRHKN84} and Eq.~\ref{EqRphotN84} are then combined to obtain
the chromospheric flux excess,
\begin{equation}
  \label{EqRpHKN84}
  \rphk = \rhk - R_\mathrm{phot,\,N84}.
\end{equation}

\subsection{H-alpha}
\label{SecHAlpha}
% %%%%%%%%%%%%%%%%%%%%%%%%%%%%%%%%%%%%%%%%%%%%%%%%%%%%%%%%%%%%%%%%%%%%%%%%%%%%%
  H$\alpha$ can be measured in a similar way to \rhk in Eq.~\ref{EqRpHK},
  but substituting H$\alpha$ for \ion{Ca}{II} H and K.
  The chromospheric flux ratio of H$\alpha$ might be defined as the
  surface flux subtracted by the photospheric flux,
  divided by the bolometric flux so that
  \begin{equation}
    \label{EqRpHAlpha}
      \frac{ \mathcal{F}'_\mathrm{H\alpha} }{ \mathcal{F}_\mathrm{bol} }  = \frac{ \mathcal{F}_\mathrm{H\alpha} - \mathcal{F}_\mathrm{H\alpha,\,phot} } { \sigma T_\mathrm{eff}^4 }.
  \end{equation}

  In brief, we measure the surface flux of each line, $\mathcal{F}_\mathrm{line}$
  by integrating the flux of the template spectrum, normalized to a stellar atmosphere model, inside an integration window centered on the line core.
  We measure the photospheric flux, $\mathcal{F}_\mathrm{line,\,phot}$, by integrating the flux of the stellar atmosphere model, with the same integration window centered on the line core.
  The bolometric flux $\mathcal{F}_\mathrm{bol}$ is determined from \teff, which is needed to obtain a proper stellar atmosphere model for a given star.
  We provide more detail in Sec.~\ref{SecHarps}.
%%%%%%%%%%%%%%%%%%%%%%%%%%%%%%%%%%%%%%%%%%%%%%%%%%%%%%%%%%%%%%%%%%%%%%%%%%%%%%%

%%%%%%%%%%%%%%%%%%%%%%%%%%%%%%%%%%%%%%%%%%%%%%%%%%%%%%%%%%%%%%%%%%%%%%%%%%%%%%%
\section{Case study: HARPS M dwarf sample}
  \label{SecHarps}

We used high-resolution archival spectra
obtained with the HARPS spectrograph~\citep{Pepe2002}
installed on ESO's 3.6m telescope in La Silla, Chile.
The sample mainly consists of 102 targets from the
HARPS GTO M dwarf sample~\citep{Bonfils2013}\footnote{ESO IDs 082.C-0718 and 183.C-0437}.
We also used data obtained for the Cool Tiny Beats survey~\citep{Anglada2014, Berdinas2015}\footnote{ESO ID 191.C-0505},
which adds the following four stars to the sample:
GJ\,160.2, GJ\,180, GJ\,570B, and GJ\,317.
Last, the following six M dwarfs with published planetary systems,
and that are available in the ESO HARPS public archive, were added:
GJ\,676A, 
GJ\,1214,
HIP\,12961,
GJ\,163,
GJ\,3634,
and GJ\,3740. 
Photometry values, mean radial velocities, proper motions, and parallaxes
were acquired from SIMBAD~\citep{Wenger2000}
(see Table~\ref{TableSimbad}).
  
\subsection{High S/N template spectra}
\label{SecTemplate}
We used the HARPS-TERRA software~\citep{AngladaButler2012}
for all available spectra on all stars in the sample.
HARPS-TERRA is a 
sophisticated
 tool that matches individual spectra
to a high S/N 
"template spectrum"
using a least-squares fit
to compute high-precision radial velocities.
The high S/N template spectrum is obtained by coadding all individual
spectra of a given star using the highest S/N spectrum as a basis spectrum.
Pixel weighting, telluric masking, and outlier filtering are all implemented
in the algorithm to assess and reduce systematic biases (see Sec. 2 in~\citet{AngladaButler2012}
for a much more detailed explanation).
Because of this technique, the template spectrum for a star is essentially
an averaged spectrum with median clipping.
For each star, we obtained a high S/N template spectrum of each spectral order.
We then used the corresponding spectral order that contains the chromospheric
line of interest as the stellar observation spectrum.

After the initial run of HARPS-TERRA on all spectra of each target,
we ran HARPS-TERRA a second time,
excluding spectra that matched any of the following criteria:
  1) program ID 60.A-9036(A),
  where spectra were acquired under nonoptimal conditions
  (an engineering run),
  2) spectra reduced with 
  cross-correlation function (CCF)
  masks earlier than M2
  (the HARPS DRS pipeline uses an M2 mask for all M stars), and
  3) spectra with 
  radial velocity (RV) 
  outliers determined by inspection.
  In general, looking at the RV time-series, spurious observation 
  differences on the order of 1-10
   km/s
with respect to most observations are considered outliers.
All three of the above criteria could negatively influence the resulting
template spectra in a suboptimal way.
After the second run of HARPS-TERRA,
we corrected for the blaze function by running HARPS-TERRA a third time
with the \texttt{-useblaze} option and the blaze file given by the
\texttt{ESO DRS BLAZE FILE} keyword in the template spectrum file header.
Finally, we obtained a blaze-corrected high S/N template spectrum
for each star.

\subsection{Photospheric flux from stellar atmosphere models}
  \label{SecPhotFlux}

%%%%%%%%%%%%%%%%%%%%%%%%%%%%%%%%%%%%%%%%%%%%%%%%%%%%%%%%%%%%%%%%%%%%%%%%%%%%%%%

\begin{table}
  \caption{PHOENIX grid parameter space} % title of Table
  \label{TablePhxGridParams} % is used to refer this table in the text
  \centering % used for centering table
    \begin{tabular}{c c c c} % centered columns (4 columns)
      \hline\hline % inserts double horizontal lines
      & Min & Max & Step size \\ % table heading
      \hline % inserts single horizontal line
      \teff [K]  & 2300 & 7200 & 100 \\ % inserting body of the table
      \feh & -1.0 & +1.0 & 0.5 \\
      \logg & 4.0 & 5.0 & 0.5 \\
      \hline %inserts single line
    \end{tabular}
\end{table}

%%%%%%%%%%%%%%%%%%%%%%%%%%%%%%%%%%%%%%%%%%%%%%%%%%%%%%%%%%%%%%%%%%%%%%%%%%%%%%%

  To calculate photospheric fluxes, we used a grid of
  PHOENIX-ACES stellar atmosphere models from the G\"{o}ttingen Spectral
  Library\footnote{http://phoenix.astro.physik.uni-goettingen.de/}~\citep{Husser2013}.
  The parameters and step sizes of the grid are listed in
  Table~\ref{TablePhxGridParams}.
  We calculated $\mathcal{F}_\mathrm{HK,\,phot}$ by setting a 1.09~\AA\ FWHM
  triangular bandpass centered on the \ion{Ca}{II} H and K lines
  and summing the flux.
  The triangular bandpass mimics the response of the 
  $H$ and $K$ bands of the \mbox{HKP-2} Mount Wilson spectrograph \citep{Duncan1991}
  (see Sec.~\ref{SecSindex}).
  To measure the fractional chromospheric flux \fpha,
  we used a 5.0~\AA\ wide rectangular bandpass
  centered on H$\alpha$.

\subsection{Surface flux from a high S/N template spectum}
  \label{SecSurfaceFlux}

% -----------------------------------------------------------------------------
\begin{figure}
  \resizebox{\hsize}{!}{\includegraphics{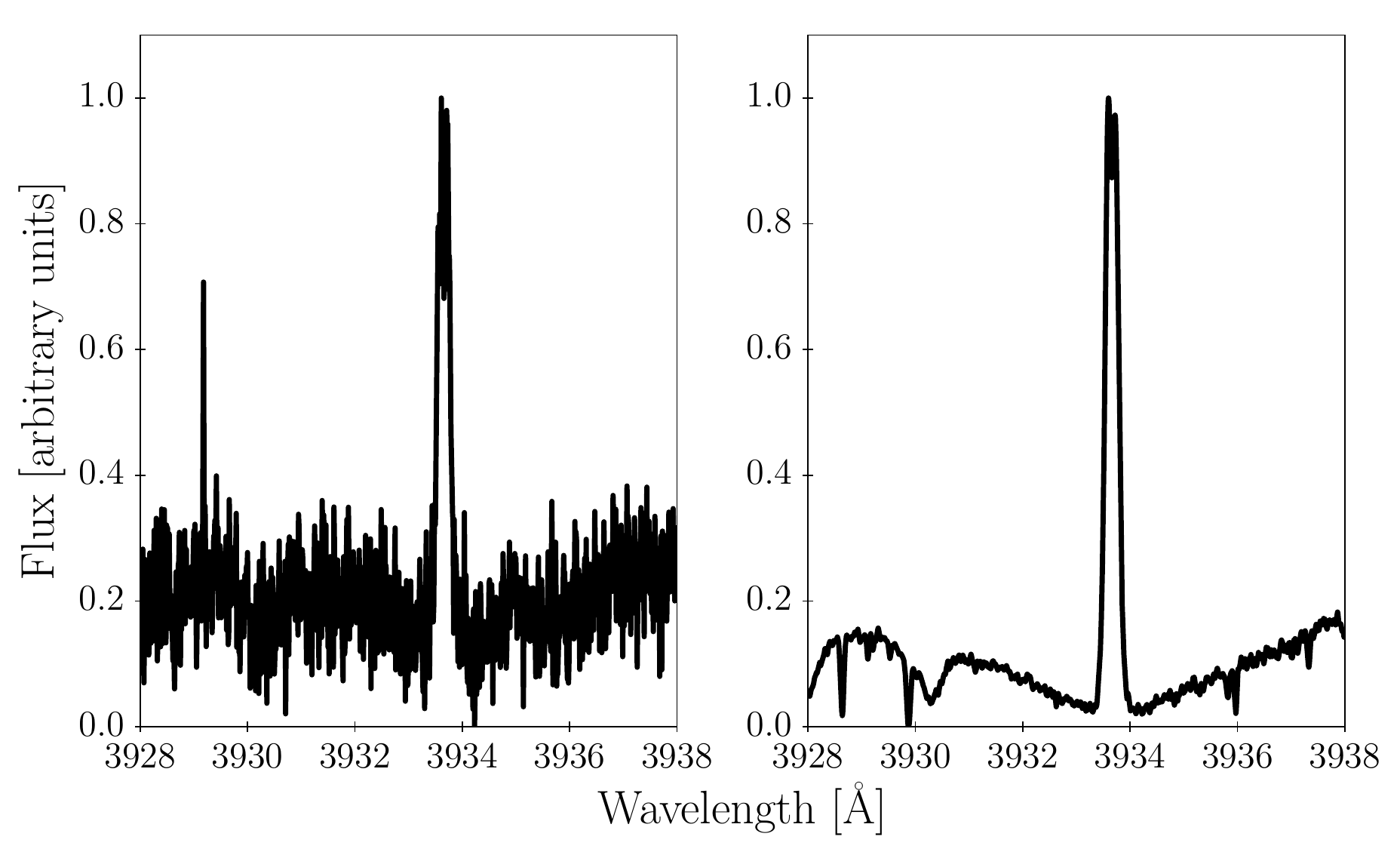}}
  \caption{
 	\ion{Ca}{II} K line of an M1.5 star.
    \emph{Left}: 
    Spectrum of a single HARPS observation.
    \emph{Right}: Template spectrum consisting of 47 coadded spectra of the same star.}
  \label{FigCompareTemplate}
\end{figure}
% -----------------------------------------------------------------------------

For a given star, the surface flux is measured from the high S/N template spectrum
computed in Sec.~\ref{SecTemplate}.
In the left panel of Fig.~\ref{FigCompareTemplate},
we plot a single HARPS observation of the \ion{Ca}{II} K line of an M1.5 star.
  Because HARPS has a high resolution of $R~\sim~110,000$,
the S/N in the region near the calcium line is low, as shown in the figure.
The right panel of Fig.~\ref{FigCompareTemplate} shows the same line
of the same star, but this time, with 47 observations coadded together.
This demonstrates the considerable S/N improvement obtained by coadding spectra.

For a single chromospheric line of a given star,
the entire template order spectrum is normalized to a PHOENIX model
spectrum via a first-degree polynomial least-squares fit,
namely $\mathcal{F}(\lambda) = a f(\lambda) + b$.
The PHOENIX spectra were bilinearly interpolated to
a precision of $\Delta \teff = 1\,\mbox{K}$
and $\Delta \feh = 0.01\,\mbox{dex}$ using the stellar parameters
chosen by the methods outlined in Sec.~\ref{SecParams} and
listed in Table~\ref{TableStellarParams}.
For the \ion{Ca}{II} K line, we used order 6,
for the \ion{Ca}{II} H line, we used order 8,
and for H$\alpha$, we used order 68.
Before normalizing the template spectrum to the PHOENIX spectrum,
we converted counts into energy,
shifted the spectrum to rest wavelengths,
and then transformed wavelengths to vacuum wavelengths.
We masked the active lines, as well as H$\epsilon$ at 3971.2 \AA.
To reduce the influence of low S/N at the spectral order edges
from the blaze function,
we also masked the outer $10 \%$ of each order.
We took the sum of the line flux in the same manner as in Sec.~\ref{SecPhotFlux}.

\begin{figure}
  \resizebox{\hsize}{!}{\includegraphics{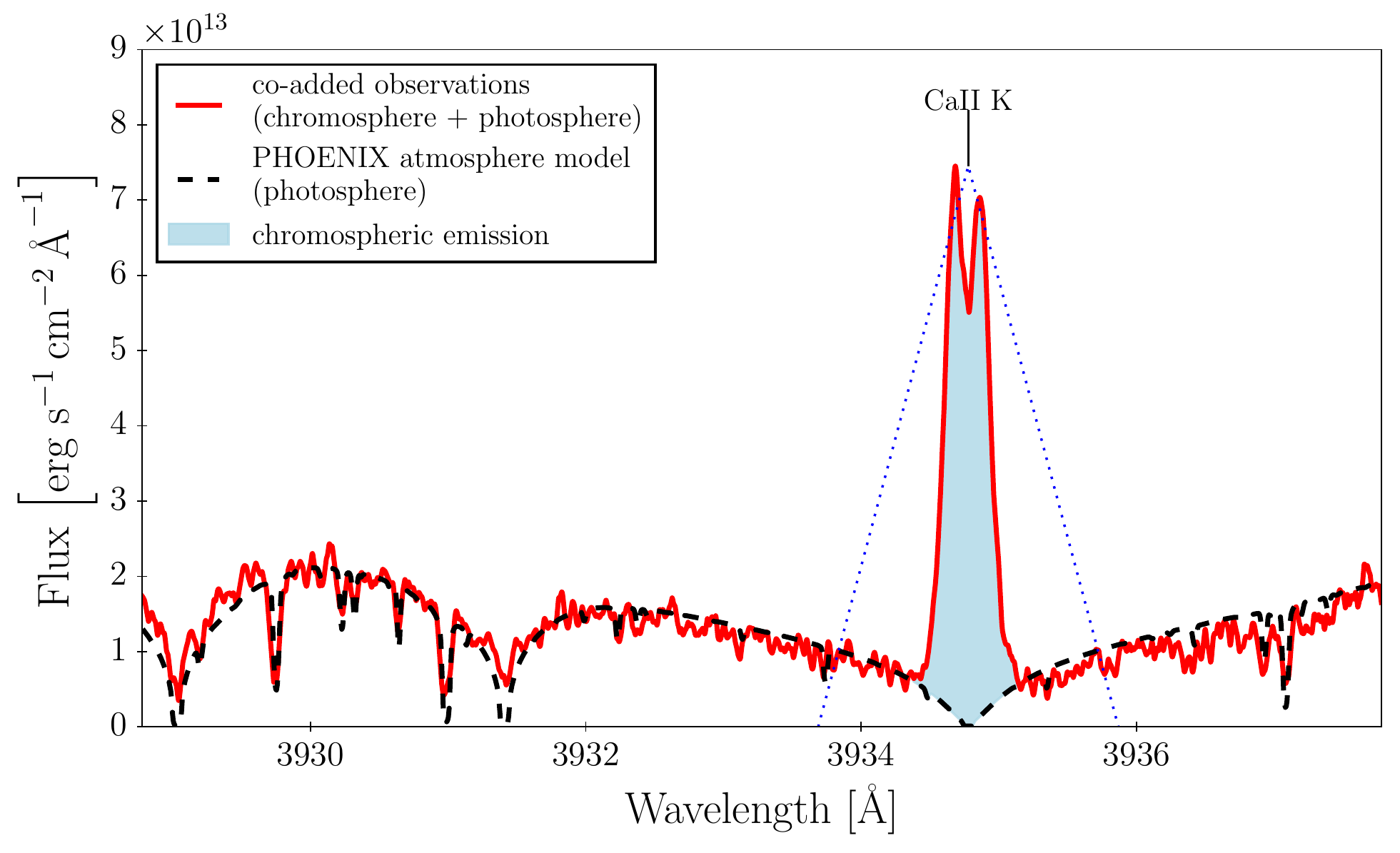}}
  \caption{
    High S/N template spectrum of an early-M dwarf normalized to a PHOENIX
    atmosphere model.
    The solid red line is the high S/N template spectrum,
    while the dashed black line is the PHOENIX model atmosphere.
The blue shaded region shows the chromospheric emission.
    The dotted lines indicate the 1.09 \AA{} FWHM triangular bandpass used to integrate the
    \ion{Ca}{II} H and K lines.
    }
  \label{FigCaKLine}
\end{figure}

\begin{figure}
  \resizebox{\hsize}{!}{\includegraphics{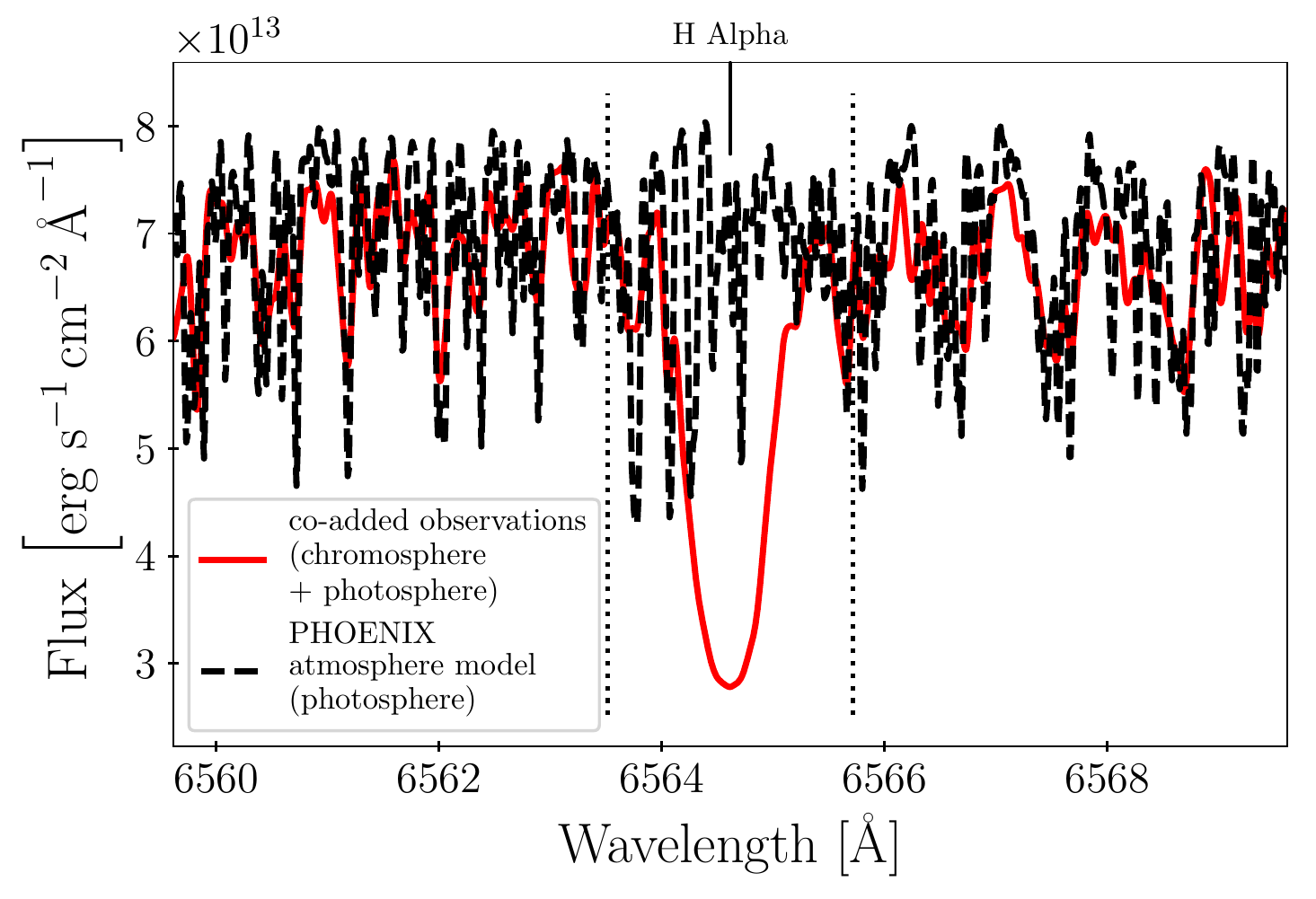}}
  \caption{
    High S/N template spectrum of an M dwarf with H$\alpha$ in absorption
    normalized to a PHOENIX atmosphere model.
    The solid red line is the high S/N template spectrum,
    while the dashed black line is the PHOENIX model atmosphere.
    The vertical dotted lines indicate a 5.0 \AA{} integration region.
    }
  \label{FigHAlphaLineAbs}
\end{figure}

\begin{figure}
  \resizebox{\hsize}{!}{\includegraphics{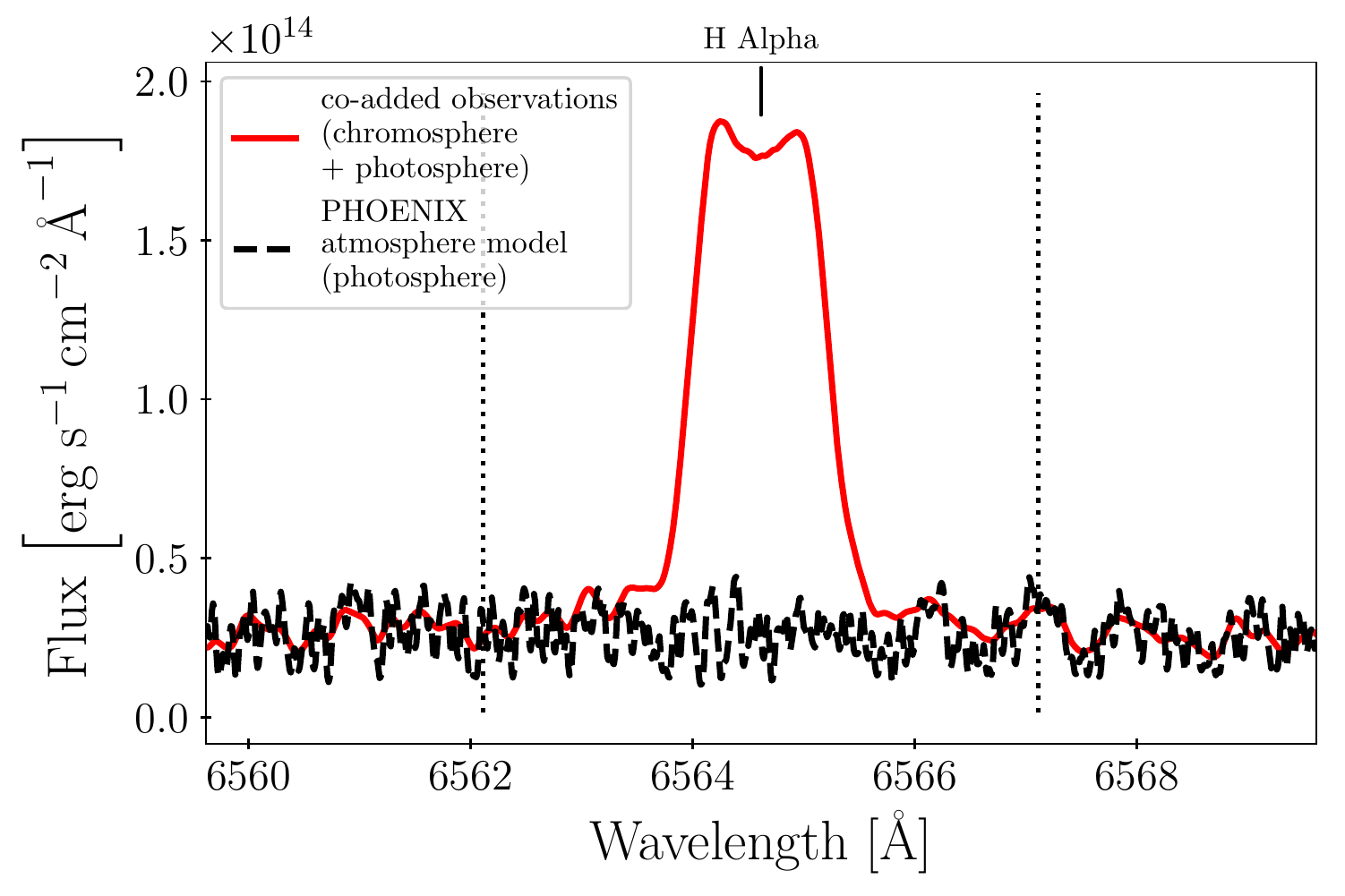}}
  \caption{
    High S/N template spectrum of an active M dwarf with H$\alpha$ in emission
    normalized to a PHOENIX atmosphere model.
    The solid red line is the high S/N template spectrum,
    while the dashed black line is the PHOENIX model atmosphere.
    The vertical dotted lines indicate the 5.0 \AA{} integration region of H$\alpha$.
    }
  \label{FigHAlphaLineEmm}
\end{figure}

Figure~\ref{FigCaKLine} demonstrates this normalization
of a high S/N template spectrum of an early-M dwarf normalized
to a PHOENIX atmosphere model
around 
the \ion{Ca}{II} K line.
The high S/N template spectrum consists of
all available spectra coadded and flux-calibrated to 
the PHOENIX model atmosphere.
Similarly, Fig.~\ref{FigHAlphaLineAbs} shows this normalization
of
an M dwarf with H$\alpha$ in absorption, 
while Fig.~\ref{FigHAlphaLineEmm} shows
another M dwarf with H$\alpha$ in emission.

% -----------------------------------------------------------------------------
% -----------------------------------------------------------------------------
% -----------------------------------------------------------------------------

\subsection{S-index comparison}
\label{SecSindexCalibrationCompare}

\begin{figure}
\resizebox{\hsize}{!}{\includegraphics{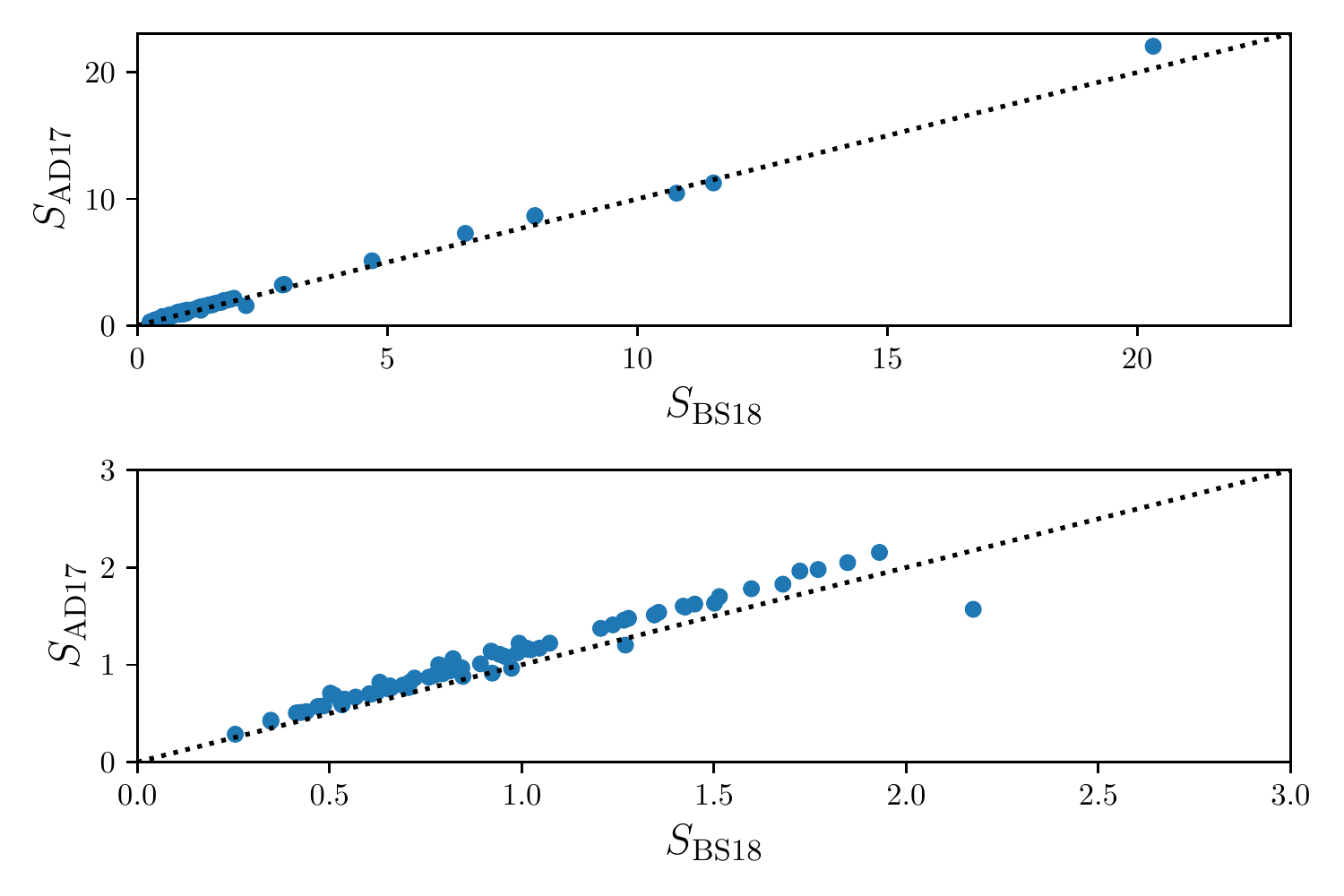}}
\caption{
$S_\mathrm{AD17}$ of \citet{Astudillo2017} vs our work $S_\mathrm{BS18}$~\citep{BoroSaikia2018}.
  \emph{Top}: Entire range of the sample.
  \emph{Bottom}:  Same sample with zoomed in axes.
The dotted line shows the 1:1 relation.
\label{FigSAstVsUs}
}
\end{figure}

As a sanity check, we compared the $S$ -index values of our sample (for their values, we refer to~\citet{BoroSaikia2018} and for the treatment of how they were calculated) with those of~\citet{Astudillo2017}
  in Fig.~\ref{FigSAstVsUs}.
Although it agrees, the linear fit slightly overestimates the values of~\citet{Astudillo2017}
  compared to~\citet{BoroSaikia2018} for $S$ values below 2,
\begin{equation}
  S_\mathrm{AD17} = 1.0487 S_\mathrm{BS18} + 0.008
\end{equation}

\subsection{Stellar parameters}
\label{SecParams}

For an accurate stellar atmosphere model to normalize a template spectrum to,
we must first determine a set of stellar parameters for each star
in a self-consistent manner.
Here we describe the different calibrations we used to determine effective temperature
and how we determined metallicity.
\paragraph{Effective temperature}
For each star in the sample, 
we estimated \teff using three different calibrations.
The first and second calibration used the combined relation of
spectral type, effective temperature, mass, and radius
from \citet{KenyonHartmann1995} and \citet{Golimowski2004}.
This same combination of calibrations was used by
\citet{ReinersBasri2007},
\citet{Shulyak2011},
and \citet{ReinersMohanty2012}.
The first method simply converts spectral type into effective temperature.
We denote effective temperatures obtained from this method with
\teffspt.
The second method uses the $M_{K_S}$--$\mathcal{M}_\star$
calibration of \citet{Delfosse2000}
to first obtain a mass $\mathcal{M}_\star$ from the absolute $M_{K_S}$ magnitude,
and then uses the former relation to convert $\mathcal{M}_\star$ into \teff.
We denote effective temperatures obtained from this method with
\teffmass.
The third method adopts \teff values obtained by \citet{Maldonado2015}
of 53 stars in the HARPS M dwarf GTO sample.
We denote effective temperatures from this study with
\teffmal.

\paragraph{Metallicity}
\citet{Maldonado2015} also calculated metallicities of the same 53 targets
using the pseudo-equivalent width (PEW)
technique of \citet{Neves2014}, who also calculated metallicities
for the entire HARPS GTO M dwarf sample.
\citet{Maldonado2015} reported that their results agreed well overall with those of
\citet{Neves2014}, and we therefore adopted the metallicities of \citet{Neves2014}
for the sample completeness.
Although effective temperatures were also calculated, the authors noted that
their \teff values are systematically underestimated compared with other works;
therefore, we did not adopt the \teff values of \citet{Neves2014}.
For two stars not listed in \citet{Neves2014} (GJ\,570B and GJ\,180),
we used the conversion of 
$M_{K_S}$ and $V-K$ to \feh in \citet{Neves2012}.
We note that the difference in metallicities between
\citet{Neves2012} and \citet{Maldonado2015}
can be up to $\Delta \mbox{[Fe/H]} \pm 0.2$.
However, this difference is much smaller than the resolution
of our grid of models, $\Delta \mbox{[Fe/H]} \pm 0.5$.
For stars with no $M_{K_S}$ or $V-K_S$ measurements, we assumed solar metallicity.
For all stars in the sample, we constrained the surface gravity to $\logg = 5.0$,
which is a typical value for M dwarfs.
The stellar parameters we used are listed in Table~\ref{TableStellarParams}.

\begin{figure}
  \resizebox{\hsize}{!}{\includegraphics{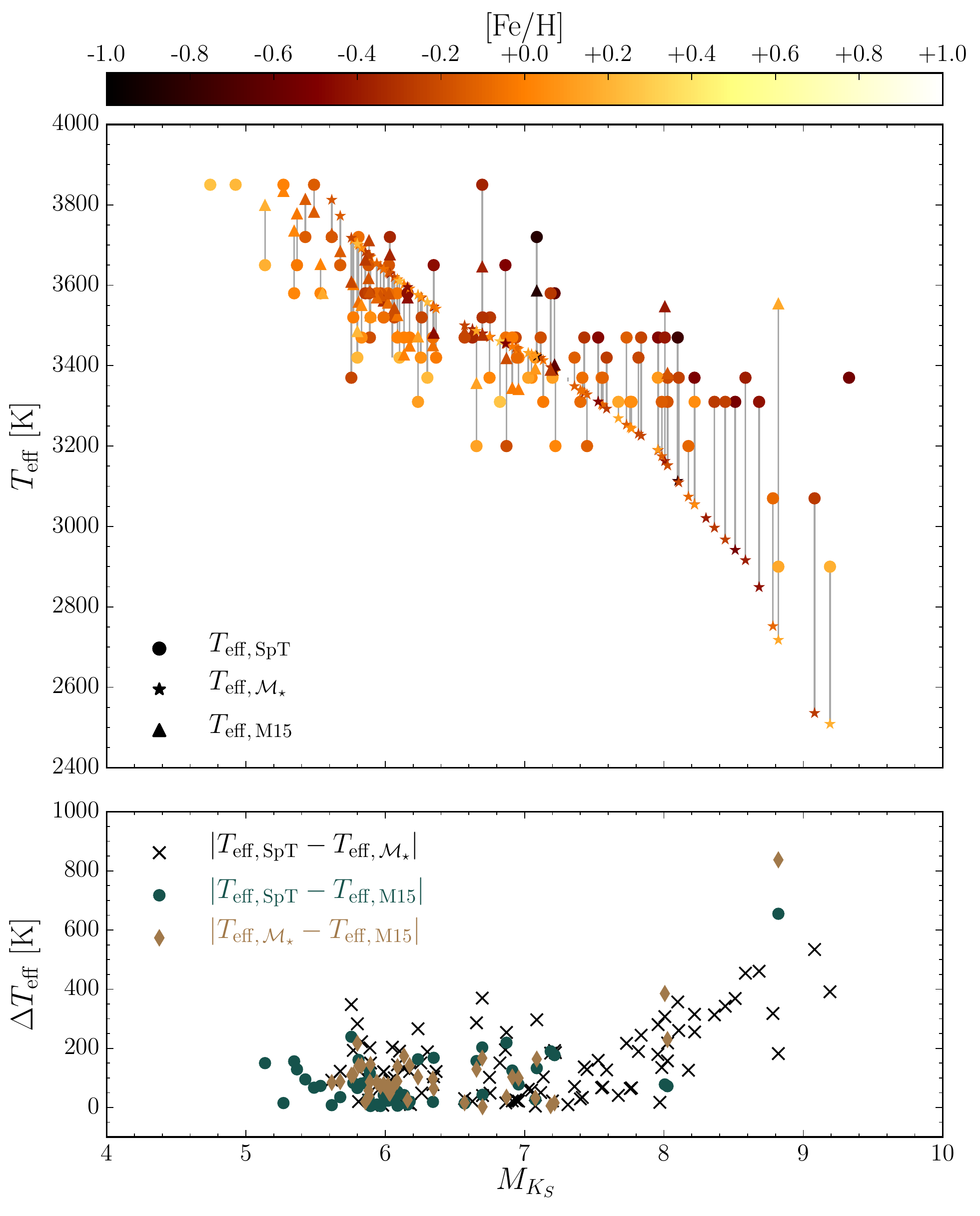}}
  \caption{
  	Effective temperature \teff{} of the stellar sample using different \teff{} determination methods, which are shown by different plot symbols.
    \emph{Top}: Effective temperature \teff{} as a function of
    absolute magnitude $M_{K_S}$ with metallicity $\mbox{[Fe/H]}$
    shown according to the color scale.
Temperatures of the same star are connected with a solid vertical line.
    \emph{Bottom}: Difference of the methods used to obtain \teff in K.
  }
  \label{FigTeffCompare}
\end{figure}

\paragraph{}
The top panel of Fig.~\ref{FigTeffCompare} shows \teff of the three methods we used
as a function of $M_{K_S}$,
with their metallicity represented by color.
For \teffspt, we were able to calibrate 110 stars.
With \teffmass, we were able to calibrate 99 stars,
and for \teffmal, there are 49 stars.
Temperatures of the same star are connected with a solid vertical line.
In the lower panel of Fig.~\ref{FigTeffCompare},
we show the residuals of the different sources of \teff.
As is evident from both panels, the \teff
determined by different methods disagrees to some extent,
and this becomes much more pronounced for stars with $M_{K_S} > 8$.
The mean of all $\Delta \teff$ values is 176 K.
For $\Delta \teff$ values with $M_{K_S} < 8$, the mean difference drops to 122 K.
The mean of $\Delta \teff$ values for $M_{K_S} > 8$
increases to 363~K.
Most of these values with $M_{K_S} > 8$ belong to $\teffmass - \teffspt$.

We note that metallicity has an effect on the \teff determination.
The largest scatter in $\Delta \teff$ is between
\teffspt{} and \teffmal{}.
The determination of the spectral type depends on line ratios that are
sensitive to both \teff and \feh.
Moreover, \citet{Maldonado2015}
determined \teff simultaneously with metallicity,
However, metallicity does not have much impact when determining \teff from $M_{K_S}$-$\mathcal{M}_\star$,
as infrared absolute magnitudes are less sensitive to metallicity
\citep{Delfosse2000}.
For \teffmass, since $\mathcal{M}_\star$ is determined
from a polynomial as a function of $M_{K_S}$, we expect to see a clear
relation between \teff and $M_{K_S}$.
Whether $M_{K_S}$ actually is such a precise indicator of \teff
is beyond the context of this study.
Regardless, accurate effective temperatures of M dwarfs remain elusive;
see \citet{Mann2015} and \citet{Passegger2016} for more thorough analyses
of the current state of M dwarf stellar parameter determination.

%%%%%%%%%%%%%%%%%%%%%%%%%%%%%%%%%%%%%%%%%%%%%%%%%%%%%%%%%%%%%%%%%%%%%%%%%%%%%%%

%%%%%%%%%%%%%%%%%%%%%%%%%%%%%%%%%%%%%%%%%%%%%%%%%%%%%%%%%%%%%%%%%%%%%%%%%%%%%%%

\section{Results}
  \label{SecResults}
 \subsection{Chromospheric Ca II H and K flux}
% ----------------------------------------------------------------------------
\begin{figure*}
\resizebox{\hsize}{!}{\includegraphics{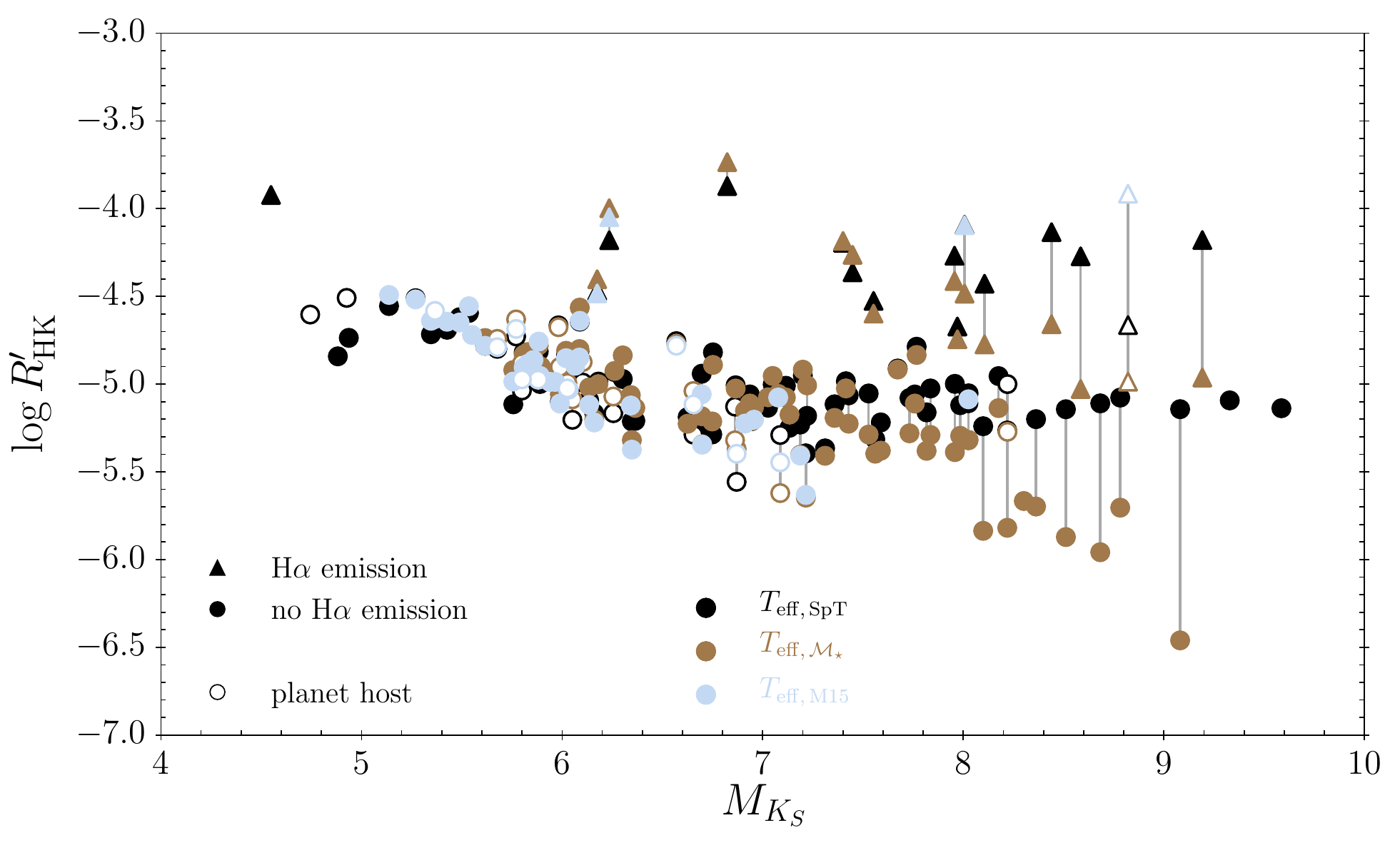}}
\caption{
    Fractional chromospheric \ion{Ca}{II} H and K flux normalized to
    bolometric flux on a logarithmic scale, $\log \rphk$,
    as a function of absolute magnitude $M_{K_S}$.
    For each star, $\log \rphk$ is calculated with different \teff{}
    calibrations, each plotted with a different color.
    Vertical gray lines connect different measurements of the same star.
    Triangles indicate stars exhibiting H$\alpha$ emission.
    Filled symbols indicate stars without known planetary systems,
    and open symbols indicate stars with known planetary systems.
\label{FigRpHKMk}
}
\end{figure*}
% -----------------------------------------------------------------------------
We plot the chromospheric \ion{Ca}{II} H and K flux normalized to
bolometric flux, $\log \rphk$, as a function of
absolute magnitude $M_{K_S}$ in Fig.~\ref{FigRpHKMk}.
$\log \rphk$ was calculated using the 
spectral subtraction technique outlined in Sec.~\ref{SecHarps},
following Eq.~\ref{EqRpHK}.
We estimated \teff using the three different methods
described in Sec.~\ref{SecParams}
and for each \teff , calculated $\log \rphk$.
The plot legend contains the colors of each \teff calibration.
We connect $\log \rphk$ measurements of the same star but different
effective temperatures with a solid vertical line.
Additionally, we plot stars with known planetary systems with open symbols,
and stars without known planetary systems with closed symbols.
Stars without H$\alpha$ in emission are plotted with a circle
and stars exhibiting H$\alpha$ emission with a triangle.

\subsubsection{\teffspt{} calibration}
For 110 stars we have spectral type information and are able
to measure $\log \rphk$ using
the spectral type to \teff{} conversion, \teffspt{}.
Of these stars, 13 exhibit H$\alpha$ emission, and 19 have known
planetary systems.
The earliest-M dwarfs with $M_{K_S} < 5.8$ have higher values
of $\log \rphk$, between -4.8 and -4.5.
There is a drop in the lower boundary of activity levels near $M_{K_S} = 5.7$,
where values range from $-5.3 < \log \rphk < -4.7$.
After this drop, the sequence of lower-activity stars has
no apparent drop, and $\log \rphk$ stays between -5.5 and -4.9.
Stars exhibiting H$\alpha$ emission tend to have much higher
$\log \rphk$ values than those that do not exhibit H$\alpha$ emission.
Their measured activity levels are $\log \rphk > -4.7$
and can be as high as $\log \rphk = -3.8$

\subsubsection{\teffmass{} calibration}
For 99 stars, we have $M_{K_S}$ measurements
and can measure $\log \rphk$ using the $M_{K_S}$
to mass to \teff{} calibration \teffmass{}.
Of these stars, 13 exhibit H$\alpha$ emission, and 16 have known
planetary systems.
Unlike $\log \rphk$ measured with \teffspt,
the lower boundary of activity for \teffmass decreases with $M_{K_S}$.
This relation is expected because in this case, \teffmass is calibrated
using $M_{K_S}$.
This does not have a dramatic effect on $\log \rphk$ values
until $M_{K_S} = 8$.
At higher $M_{K_S}$, the difference in $\log \rphk$ can be as high
as 1.3 dex,
and this arises from the large differences in \teff seen
in Fig.~\ref{FigTeffCompare}.
However, even though the measured values of $\log \rphk$ using
\teffmass are much lower,
stars with H$\alpha$ in emission still have significantly
higher $\log \rphk$ values than their counterparts.

\subsubsection{\teffmal{} calibration}
For 49 stars with adopted values from \citet{Maldonado2015},
we are able to measure $\log \rphk$ using \teffmal{}.
Of these stars, 4 exhibit H$\alpha$ emission, and 9 have known
planetary systems.
Similar with the other \teff calibrations,
the lower level of $\log \rphk$ decreases from -4.5 to -5.4
between $5 < M_{K_S} < 6.4$.
There are only three stars with $M_{K_S} \gtrsim 8$,
and they tend to agree more with \teffspt values than \teffmass.

\paragraph{}

We list the individual $\log \rphk$ measurements
in Table~\ref{TableResults}.
The largest differences of $\log \rphk$ values occur at
the latest spectral types
where it is difficult to determine a consistent \teff
using a color, magnitude, or spectral type relation.
For the entire sequence of stars,
the mean of $\Delta \log \rphk$ is 0.17 dex.
At $M_{K_S}$ < 8, the mean of $\Delta \log \rphk$ is only 0.10 dex,
however for $M_{K_S}$ > 8, the mean is 0.56 dex.
The star with the highest difference in $\log \rphk$
is GJ\,1002, which has $\Delta \log \rphk = 1.31\ \mbox{dex}$.
This is because of its wide range of \teff calibrations,
with $\Delta \teff = 534\ \mbox{K}$.

\begin{figure}
  \resizebox{\hsize}{!}{\includegraphics{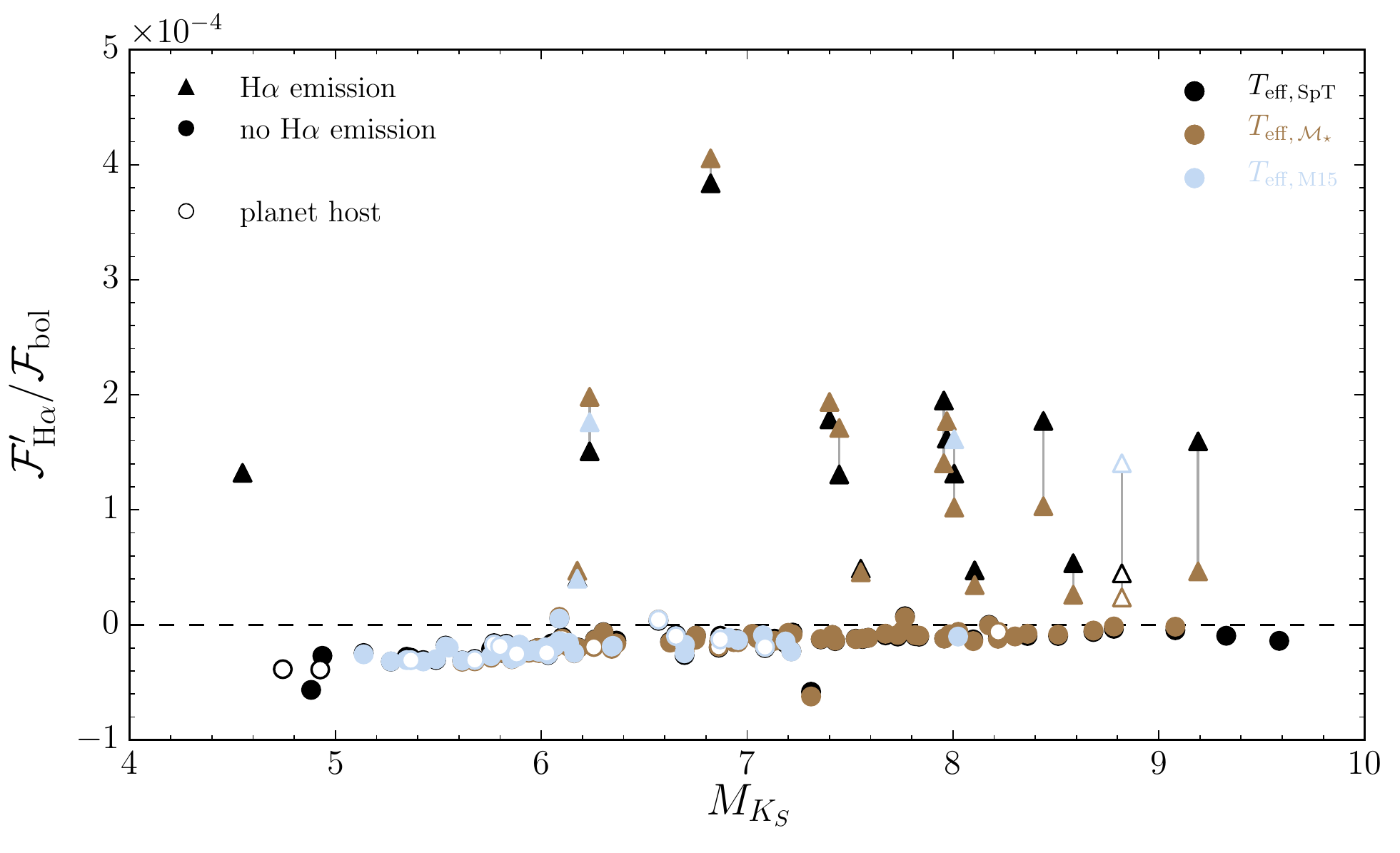}}
  \caption{
    Fractional chromospheric H$\alpha$ flux,
    ${\mathcal{F}'_\mathrm{H\alpha} / \mathcal{F}_\mathrm{bol}}$,
    as a function of absolute magnitude $M_{K_S}$.
    Vertical gray lines connect different measurements of the same star.
    Triangles indicate stars exhibiting H$\alpha$ emission.
    Filled symbols indicate stars without known planetary systems,
    and open symbols indicate stars with known planetary systems.
    Values > 0 indicate that H$\alpha$ is in emission,
    and values < 0 indicate that H$\alpha$ is in absorption.
    Values near 0 indicate filling-in of the line to the continuum.
  \label{FigFpHAlphaMk}
  }
\end{figure}
% -----------------------------------------------------------------------------
\subsection{Chromospheric H-alpha flux}
In Fig.~\ref{FigFpHAlphaMk},
we correct for the photospheric contribution and plot
the chromospheric flux ratio of H$\alpha$, \fpha.
This is 
not
on a logarithmic scale, unlike how
we plot $\log \rphk$.
This is because the sign of $\fpha$ indicates whether the line is
in emission or absorption.
Values of $\fpha~\sim~0$ indicate that H$\alpha$ is filled to the continuum.
Values $\fpha~>~0$ indicate that H$\alpha$ is in emission,
as in Fig.~\ref{FigHAlphaLineEmm},
while values $\fpha~<~0$ indicate that H$\alpha$ is in absorption
as in Fig.~\ref{FigHAlphaLineAbs}.
The plotting colors and symbols are the same as used in Fig.~\ref{FigRpHKMk}.

For stars with H$\alpha$ in absorption,
\fpha increases toward a filling-in of the 
continuum toward larger $M_{K_S}$.
The stars in which H$\alpha$ is in emission tend to have significantly higher \fpha
than stars for which H$\alpha$ is near the continuum level.
We note that the absorption of H$\alpha$ stays in a relatively narrow range.
Values of $\fpha$ are $-0.6 \cdot 10^{-4} < \fpha \lesssim 0$.
However, for stars in which H$\alpha$ is in emission,
the range of $\fpha$ is on the order of several $10^{-4}$.
larger $\Delta \fpha$ due to the different \teff calibrations.
Stars for which H$\alpha$ is in emission also have significantly higher
values of $\rphk$ than those in which H$\alpha$ is not in emission.

\subsection{H-alpha versus calcium II H and K flux}

\begin{figure}
\resizebox{\hsize}{!}{\includegraphics{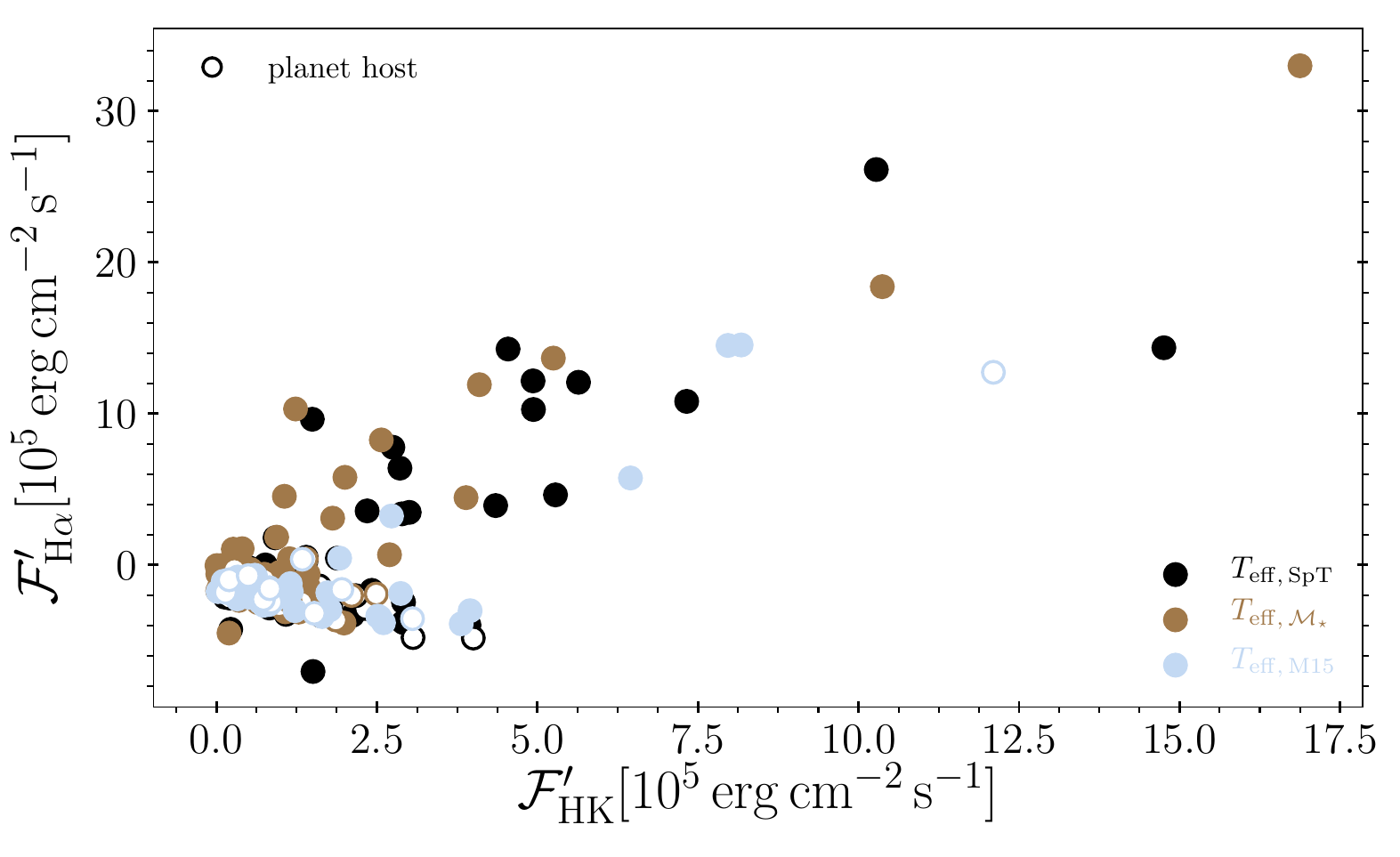}}
  \caption{
    Fractional chromospheric H$\alpha$ flux,
    ${\mathcal{F}'_\mathrm{H\alpha} }$,
    as a function of fractional chromospheric $\ion{Ca}{II}$ H and K flux, ${\mathcal{F}'_\mathrm{HK}}$ flux.
    Filled symbols indicate stars without known planetary systems,
    and open symbols indicate stars with known planetary systems.
    Values > 0 indicate that H$\alpha$ is in emission,
    and values < 0 indicate that H$\alpha$ is in absorption.
    Values near 0 indicate filling-in of the line to the continuum.
  \label{FigFpHAlphaFpHK}
  }
\end{figure}

\begin{figure}
\resizebox{\hsize}{!}{\includegraphics{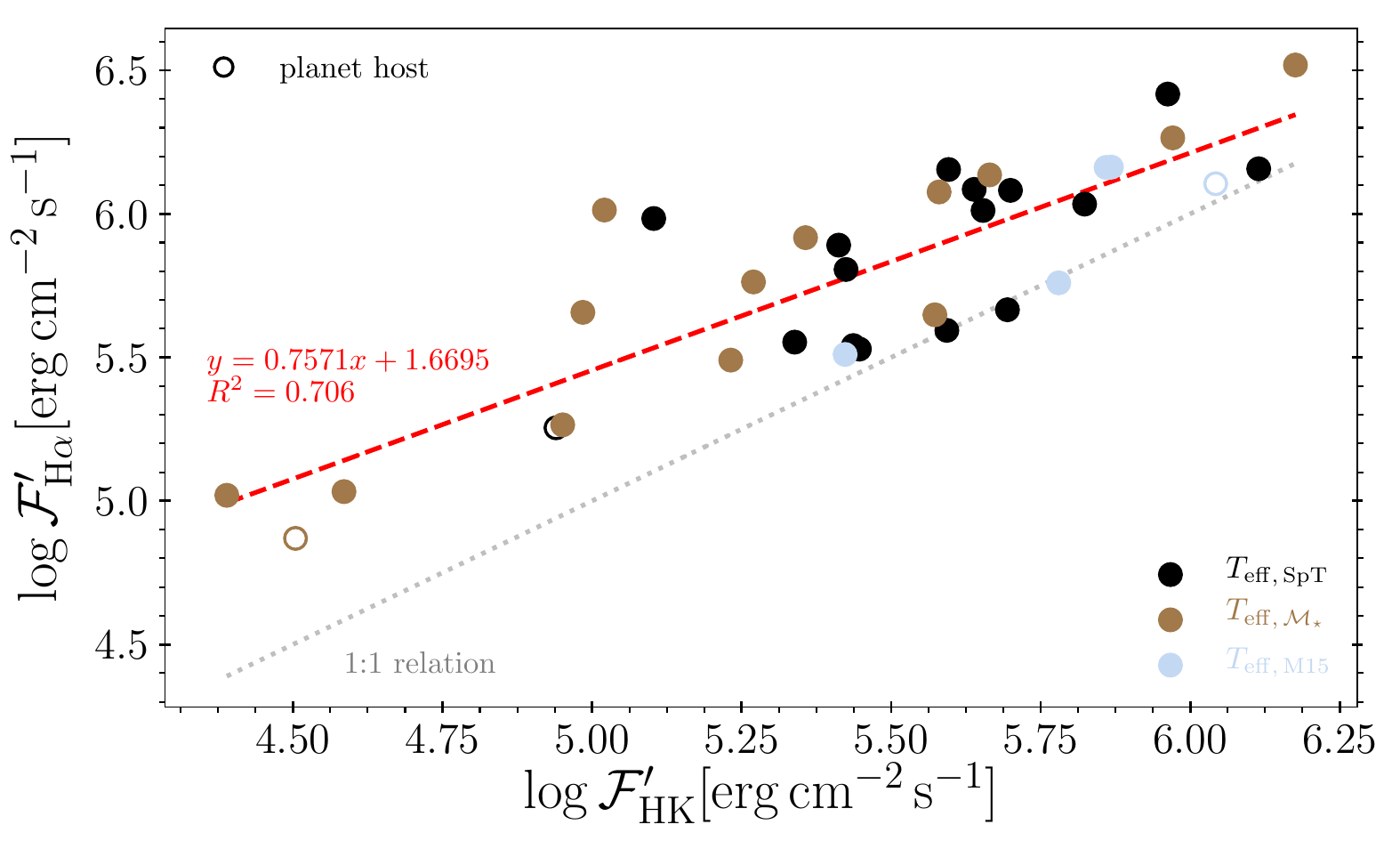}}
  \caption{
    Fractional chromospheric H$\alpha$ flux,
    ${\mathcal{F}'_\mathrm{H\alpha} }$,
    as a function of fractional chromospheric $\ion{Ca}{II}$ H and K flux, ${\mathcal{F}'_\mathrm{HK}}$ flux,
    both on a logarithmic scale.
    Filled symbols indicate stars without known planetary systems,
    and open symbols indicate stars with known planetary systems.
    In this plot, only stars with H$\alpha$ 
    in emission
     (> 0) are shown.
    A linear fit is shown with a dashed red line,
    and a 1:1 relation is plotted as a dotted gray line.
  \label{FiglogFpHAlphalogFpHK}
  }
\end{figure}

We plot chromospheric flux values of H$\alpha$
against \ion{Ca}{II} H and K
in Figs.~\ref{FigFpHAlphaFpHK} and~\ref{FiglogFpHAlphalogFpHK}.
Figure~\ref{FigFpHAlphaFpHK} contains flux in absolute flux units,
and therefore contains negative values of H$\alpha$ that correspond to absorption.
Figure~\ref{FiglogFpHAlphalogFpHK} shows flux on a log-scale
and excludes inactive stars with H$\alpha$ in absorption.
Symbols and markers are the same as in Figures~\ref{FigRpHKMk}~and~\ref{FigFpHAlphaMk}.

Fig.~\ref{FigFpHAlphaFpHK} shows H$\alpha$ only in absoprtion 
($\mathcal{F}^{'}_\mathrm{H\alpha} \lesssim 0$), and a
 decreasing trend is apparent where it
goes deeper into absorption with increasing $\mathcal{F}^{'}_\mathrm{HK}$.
The trend appears to be linear from 
$0.0 \le \mathcal{F}^{'}_\mathrm{HK} \le 4.0 \cdot 10^{-5}\ \mathrm{erg\,cm^{-2}\,s^{-1}}$.
When H$\alpha$ is in emission, the trend 
of $\mathcal{F}^{'}_\mathrm{H\alpha}$ increases with increasing $\mathcal{F}^{'}_\mathrm{HK}$.
This deepening of the H$\alpha$ line before
filling-in and going into emission was reported by
\citet{WalkowiczHawley2009} and \citet{Scandariato2017}.
\citet{Scandariato2017} only observed a decreasing trend from $0.0 \le \mathcal{F}^{'}_\mathrm{HK} \le 1.0$.
In Fig.~\ref{FiglogFpHAlphalogFpHK}, we overplot as a dashed red line
a linear fit that we find to be
\begin{equation}
  \log{\mathcal{F}^{'}_\mathrm{H\alpha}}  = 0.7571 \log{\mathcal{F}^{'}_\mathrm{HK}} + 1.6695,
\end{equation}
with $R^2 = 0.706$.

\subsection{Ca II H and K surface flux calibrations}
% ============================================================================
\label{SecCcf}
\begin{figure}
\resizebox{\hsize}{!}{\includegraphics{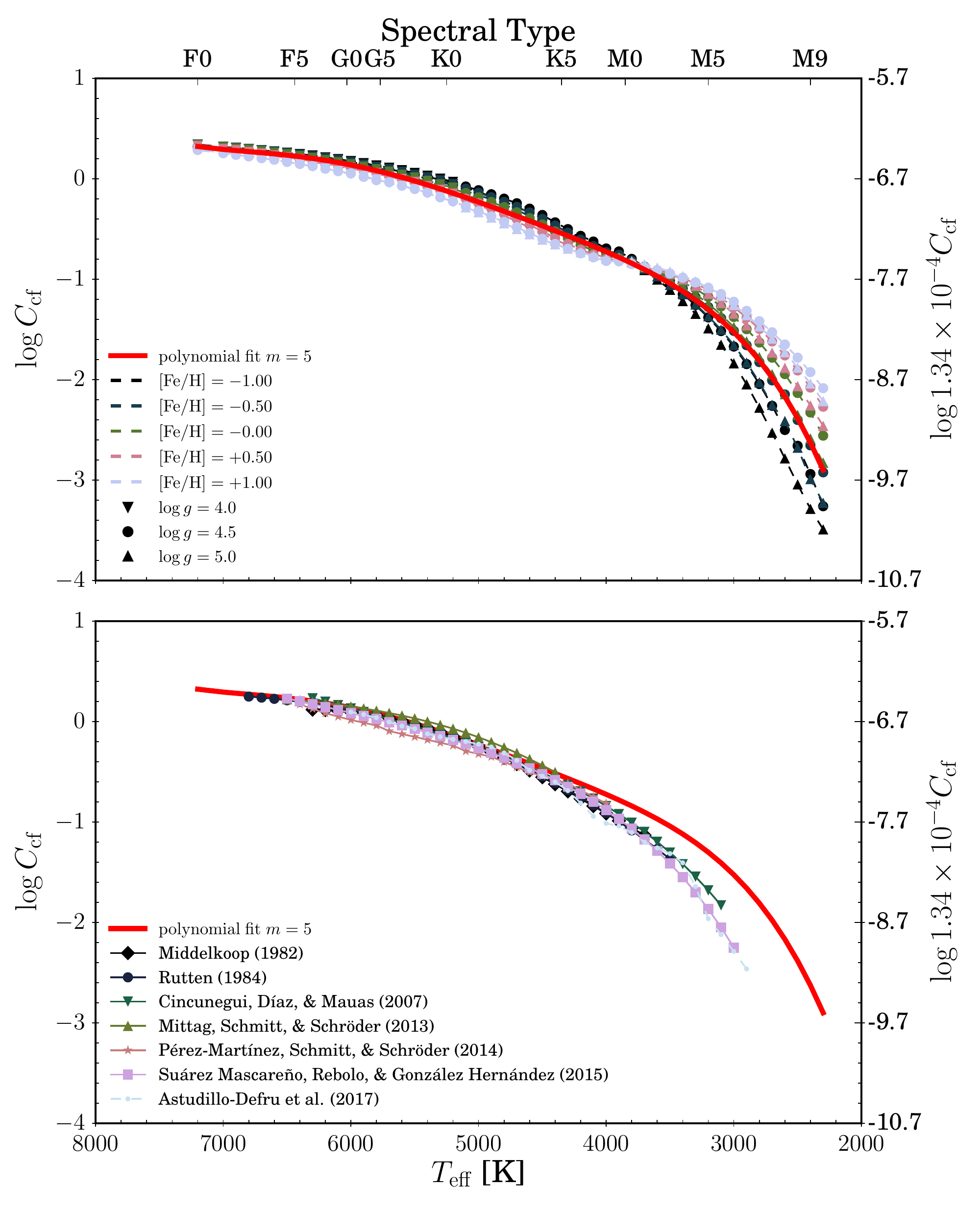}}
\caption{
  Different $\log{C_\mathrm{cf}}$ calibrations as a function of \teff.
  The approximate spectral type is shown on the top axis.
  \emph{Top}: PHOENIX stellar atmosphere models with different $\mbox{[Fe/H]}$ and \logg,
  where $\mbox{[Fe/H]}$ is indicated by color and \logg indicated by symbol.
  The solid red line indicates a fifth-order polynomial fit.
  \emph{Bottom}: Calibrations of $\log{C_\mathrm{cf}}$ from other works
  where only the valid calibration region is plotted.
  Overplotted in red is the same fifth-order polynomial fit from the top panel.
\label{FigCcf}
}
\end{figure}

% ============================================================================
We compared calibrations of the continuum conversion factor $C_\mathrm{cf}$
(used for measuring $F_\mathrm{HK}$)
of other studies 
with $C_\mathrm{cf}$ calibrations using the PHOENIX model grid.
To derive our \teff dependent conversion factor $C_\mathrm{cf}$,
we combined Eq.~\ref{EqFHKR84}, Eq.~\ref{EqFHKAbs}, and 
$S = 8 \alpha \left(\mathcal{F}_\mathrm{HK} / \mathcal{F}_{RV}\right)$,
to arrive at
\begin{equation}
  C_\mathrm{cf} = \frac{10^{14}}{8 \alpha K}
  \frac{\mathcal{F}_\mathrm{RV}}{T_\mathrm{eff}^4}.
\end{equation}
Using values of $\alpha = 2.4$ and $K = 1.07 \cdot 10^6$ erg cm$^{-2}$ s$^{-1}$
\citep{Hall2007},
we then obtained
\begin{align}
  \label{EqCcfOurs}
  C_\mathrm{cf} =  4.8676 \cdot 10^6 \cdot \frac{\mathcal{F}_\mathrm{RV}}{T_\mathrm{eff}^4}.
\end{align}

The top panel of Fig.~\ref{FigCcf} shows our computed $\log C_\mathrm{cf}$
as a function of \teff.
Different values of \feh are shown as a function of color,
and different values of \logg are plotted with different symbols.
For $\teff > 5000$~K, we used \logg values of 4.0 and 4.5,
and for $\teff \le 5000$~K, we used \logg values of 4.5 and 5.0.
A fifth-order polynomial,
\begin{equation}
  \label{EqCcfPoly}
  \log{C_\mathrm{cf}} = a + b\teff +c\teff^2 +d\teff^3 +e\teff^4 +f\teff^5,
\end{equation}
was fit to all of the points and overplotted as a solid red line.
The coefficients of this polynomial are listed in Table~\ref{TableCcfCoeffs}.
We provide the computed PHOENIX stellar atmosphere $\log C_\mathrm{cf}$ values
in Table~\ref{TableRpHKCalib} for the entire model grid.

Even for the wide range of metallicity values,
the spread of $\log C_\mathrm{cf}$ is not too wide for 
$\teff > 3400$~K.
For these higher temperatures, $\log C_\mathrm{cf}$
varies by 0.25~dex at most.
At $\teff < 3100$~K, metallicity begins to contribute
to a larger spread of $\log C_\mathrm{cf}$ values,
around 0.5~dex.
This spread of $\log C_\mathrm{cf}$ increases to 1.4~dex
as $\teff$ decreases to 2300~K.

%%%%%%%%%%%%%%%%%%%%%%%%%%%%%%%%%%%%%%%%%%%%%%%%%%%%%%%%%%%%%%%%%%%%%%%%%%%%%%%
\begin{table}
  \caption{\label{TableCcfCoeffs} $\log{C_\mathrm{cf}}$ vs. \teff fifth-order polynomial fit coefficients}
  \centering
    \begin{tabular}{c r}
      \hline\hline
      Parameter & Value \\
      % \hline\\
      \hline
      a & -2.9679E+01 \\
      b & 2.6864E-02 \\
      c & -1.0268E-05 \\
      d & 1.9866E-09 \\
      e & -1.9017E-13 \\
      f & 7.1548E-18 \\
      \hline %inserts single line
    \end{tabular}
    \tablefoot{Coefficients for calculating $\log{C_\mathrm{cf}}$ as a function of \teff using the equation $ \log{C_\mathrm{cf}} = a + b \teff +c{\teff}^2 +d{\teff}^3 +e{\teff}^4 +f{\teff}^5 $.}
\end{table}
%%%%%%%%%%%%%%%%%%%%%%%%%%%%%%%%%%%%%%%%%%%%%%%%%%%%%%%%%%%%%%%%%%%%%%%%%%%%%%%

The bottom panel of Fig.~\ref{FigCcf} compares the polynomial fit
in the upper panel with previously published 
$\log C_\mathrm{cf}$-\teff relations
(see Sec.~\ref{app-ccf} for details about the relations).
For the relations of the other authors, we plot 
the range for which the relations are calibrated.
Our $\log C_\mathrm{cf}$ values agree well with those from other studies
for $\teff \gtrsim 4100$~K to $\sim 0.1$~dex.
At temperatures cooler than 4100~K,
our $\log C_\mathrm{cf}$ values start to deviate from those of
\citet{Middelkoop1982},
\citet{Rutten1984},
\citet{Cincunegui2007},
\citet{SuarezMascareno2015},
and \citet{Astudillo2017};
our values are higher for cooler temperatures.
For temperatures cooler than $\teff = 4100$~K,
$\log C_\mathrm{cf}$ diverges by more than 0.2~dex.
The difference between our relation and \citet{Cincunegui2007}
increases to 0.4~dex at $\teff = 3100$~K,
while the difference between our relation and 
both \citet{SuarezMascareno2015} and \citet{Astudillo2017}
increases to 0.8~dex at $\teff = 3000$~K.
This discrepency in $\log C_\mathrm{cf}$ might be attributed
to the use of empirical calibrations with these studies as opposed to
this work using stellar models (see Sec.~\ref{SecDiscussion}).

As an example, when we take for the Sun $B-V = 0.642$
and $S_\mathrm{MWO} = 0.164$,
the \citet{Noyes1984} calibration of \rhk{} will give us $\log \rhk{} = -4.62$.
Our calibration using Eq.~\ref{EqCcfPoly} results in
$\log \rhk = -4.60$.
This means that for a Sun-like star, $\Delta \log \rhk = 0.02$.

\subsection{Ca II H and K photospheric flux calibrations}
\label{SecRphot}

\begin{figure}
\resizebox{\hsize}{!}{\includegraphics{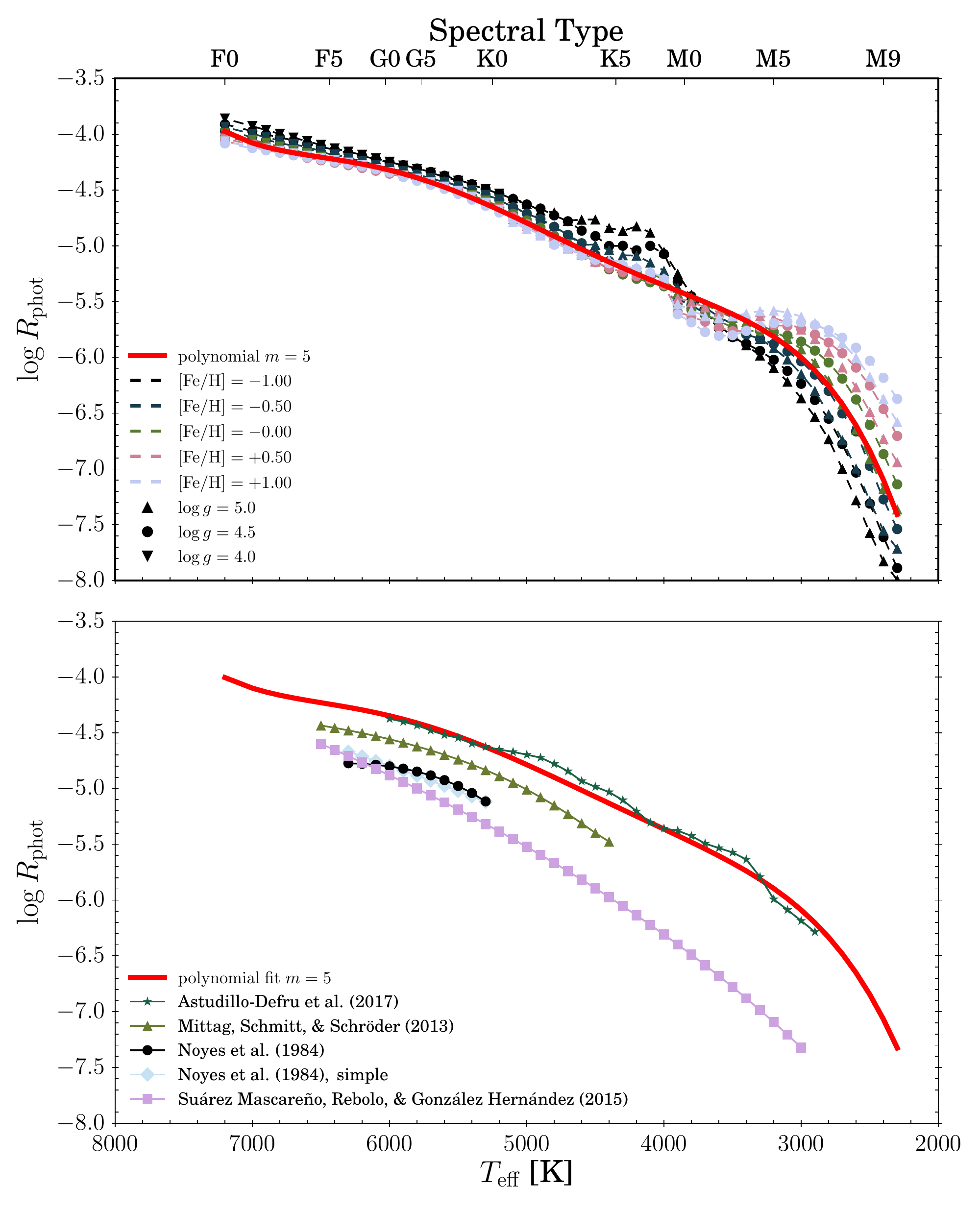}}
\caption{
  Different $\log \rphot{}$ as a function of \teff{}.
  The approximate spectral type is shown on the top axis.
  \emph{Top}: PHOENIX stellar atmosphere models with different $\mbox{[Fe/H]}$ and \logg,
  where $\mbox{[Fe/H]}$ is indicated by color and \logg indicated by symbol.
  The solid red line indicates a fifth-order polynomial fit.
  \emph{Bottom}: Calibrations of $\log \rphot{}$ from other works
  where only the valid calibration region is plotted.
  Overplotted in red is the same fifth-order polynomial fit from the top panel.
\label{FigRphot}
}
\end{figure}
% ============================================================================

In the top panel of Fig.~\ref{FigRphot}, we plot $\log \rphot$ as a function of \teff,
computed as described in Sec.~\ref{SecPhotFlux}.
The colors and symbols are assigned in the same manner as in Fig.~\ref{FigCcf}.
A fifth-order polynomial fit
\begin{equation}
  \label{EqRphotPoly}
  \log \rphot = a + b\teff +c\teff^2 +d\teff^3 +e\teff^4 +f\teff^5,
\end{equation}
is overplotted as a solid red line,
and coefficients of this polynomial are listed in Table~\ref{TableRphotCoeffs}.
Similarly as with $\log C_\mathrm{cf}$,
metallicity has an effect on the spread of $\log \rphot$ values.
The spread increases at $4000 \le \teff \le 5100$~K
and $\teff \le 3800$~K.
At $\teff > 5100$~K, the spread of $\log \rphot$ values is
lower than 0.2~dex.
The spread increases from $\teff = 5100$~K to $\teff = 4200$~K,
where it is about 0.4~dex.
At $\teff = 3100$~K, the spread of $\log \rphot$ is 0.6~dex.
Here, a change in metallicity of $\pm 0.5$~dex can result in
a $\Delta \log \rphot \sim 0.1$~dex.
At $\teff < 3300$~K, the spread continually increases with
lower temperatures to almost 1.6~dex at $\teff = 2400$~K.
At $\teff = 2400$~K., a change in metallicity of $\pm 0.5$ can result in
a $\Delta \log \rphot \sim 0.5$~dex.
We provide the computed PHOENIX stellar atmosphere $\log \rphot$ values
in Table~\ref{TableRpHKCalib} for the entire model grid.

In the bottom panel of Fig.~\ref{FigRphot}, we compare our $\log \rphot$-\teff
polynomial fit with previous literature relations
(see Sec.~\ref{app-rphot} for details about the relations).
The higher $\log \rphot$ values of both
\citet{Mittag2013} and our work in comparison with \citet{Noyes1984}
and \cite{SuarezMascareno2015} are clear.
The difference between \citet{Mittag2013} and \citet{Noyes1984}
is $\sim 0.3$~dex,
and the difference between our work and \citet{Noyes1984}
ranges from 0.4 to 0.5~dex.
The difference between our work and \citet{SuarezMascareno2015}
can be as large as 1.3~dex at $\teff{} = 3000~\mbox{K}$.
Our values of $\log \rphot$ agree very well with those of
\citet{Astudillo2017}, with the largest difference being 0.12~dex at $\teff{} = 4800~\mbox{K}$, and other differences on the order of 0.1~dex at
$\teff{} = 3500~\mbox{K}, 3400~\mbox{K}, \mathrm{and}\ 3100~\mbox{K}$.

Our $\log \rphot$ calibration has much higher values than the
calibration used by \citet{Noyes1984}.
This systematic offset was also observed by~\citet{Astudillo2017},
who also used synthetic spectra to obtain an \rphot calibration.
While the exact reason for this offset is not known (see the discussion in~\citet{Astudillo2017}),
an offset correction can be applied to scale our $\log \rphot$
calibration to \citet{Noyes1984},
\begin{equation}
  \log R_\mathrm{phot,\,N84} = \log R_\mathrm{phot,\,ours} - 0.4612,
  \label{EqRphotCorrect}
\end{equation}
where 0.4612 is the offset correction.
This simple offset correction scales our $\log \rphot$ values
to \citet{Noyes1984} values in their valid calibration region,
so that our calibration can be used to obtain 
comparable results to
\citet{Noyes1984},
and also to extend the calibration to later-type stars.
We note that although they are widely used, the \citet{Noyes1984} calibrations
are only calibrated in the range $5300~\mbox{K} \lesssim \teff{} \lesssim 6300~\mbox{K}$.

If we take the same $B-V$ and $S_\mathrm{MWO}$ values as in Sec.~\ref{SecCcf},
the \citet{Noyes1984} calibration of \rphot will give us $\log \rphot = -4.92$.
This will give us an activity measurement of $\log \rphk = -4.92$.
Our calibration using Eq.~\ref{EqRphotCorrect} results in
$\log \rphot = -4.89$,
which gives us $\log \rphk = -4.91$.
Then, for a Sun-like star, $\Delta \log \rphot = 0.03$
and $\Delta \log \rphk = 0.01$.

% ============================================================================
%%%%%%%%%%%%%%%%%%%%%%%%%%%%%%%%%%%%%%%%%%%%%%%%%%%%%%%%%%%%%%%%%%%%%%%%%%%%%%

\begin{table}
  \caption{\label{TableRphotCoeffs} $\log{R_\mathrm{HK,\,phot}}$ vs. \teff fifth-order polynomial fit coefficients}
  \centering
    \begin{tabular}{c r} % centered columns (4 columns)
      \hline\hline % inserts double horizontal lines
      Parameter & Value \\ % table heading
      \hline % inserts single horizontal line
      a & -3.7550E+01 \\
      b & 3.2131E-02 \\
      c & -1.3177E-05 \\
      d & 2.7133E-09 \\
      e & -2.7466E-13 \\
      f & 1.0887E-17 \\
      \hline %inserts single line
    \end{tabular}
    \tablefoot{Coefficients for calculating $\log{R_\mathrm{HK,\,phot}}$ as a function of \teff using the equation $\log{R_\mathrm{HK,\,phot}} = a + b\teff +c\teff^2 +d\teff^3 +e\teff^4 +f\teff^5$.}
\end{table}
%%%%%%%%%%%%%%%%%%%%%%%%%%%%%%%%%%%%%%%%%%%%%%%%%%%%%%%%%%%%%%%%%%%%%%%%%%%%%%

\subsection{$\chi$-factor}

% -----------------------------------------------------------------------------
\begin{figure}
    \resizebox{\hsize}{!}{\includegraphics{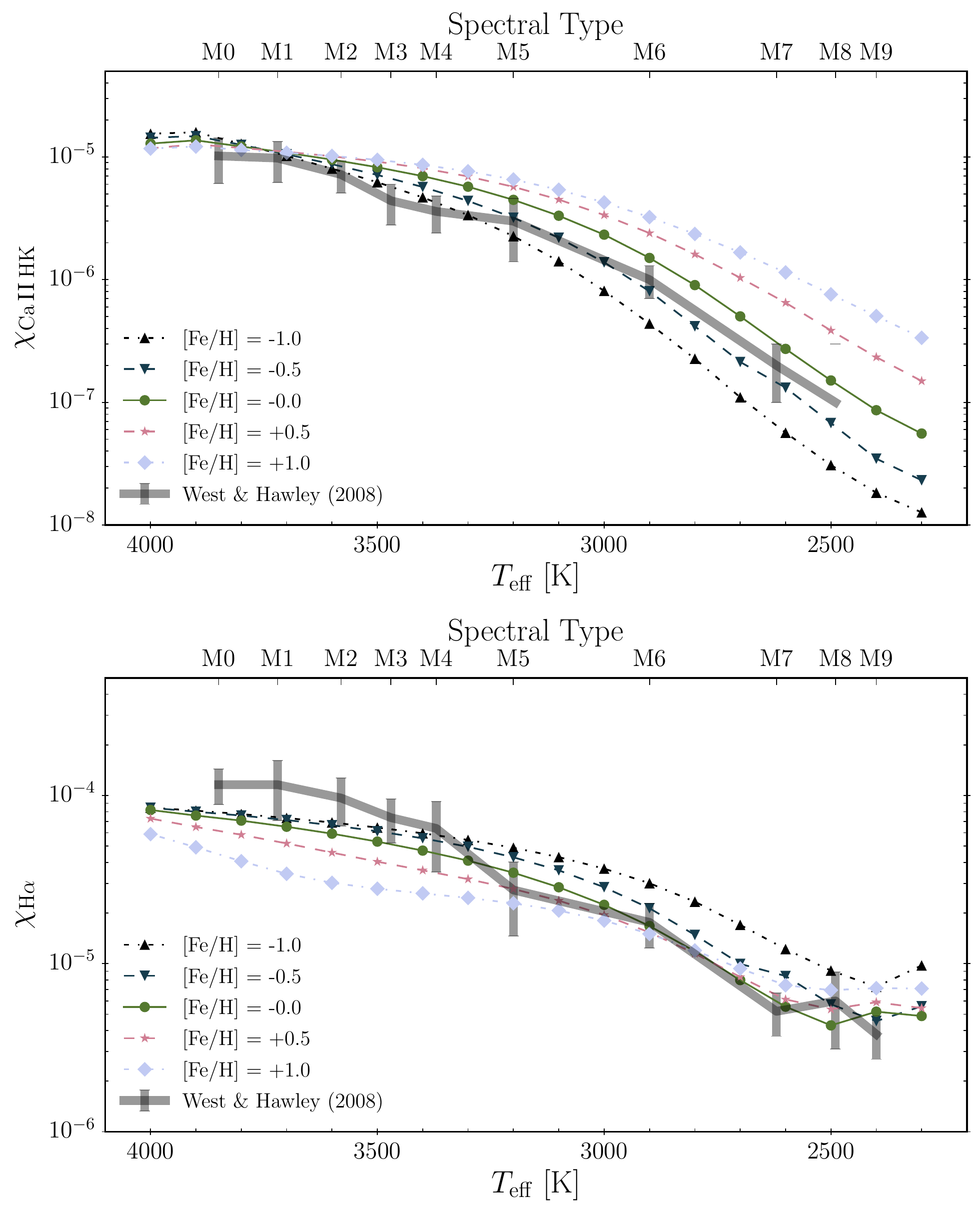}}
    \caption{
        \emph{Top}:
        Continuum flux normalized to bolometric flux of \ion{Ca}{II} H and K,
        $\chi_\mathrm{CaIIHK}$, as a function of \teff{} using PHOENIX stellar
        atmospheres with \logg{} = 5.0 and different metallicities.
        Values from \citet{WestHawley2008} are plotted in gray.
        \emph{Bottom}:
        Continuum flux normalized to bolometric flux of H$\alpha$,
        $\chi_\mathrm{H\alpha}$, as a function of \teff{} using PHOENIX stellar
        atmospheres with \logg{} = 5.0 and different metallicities.
        Values from \citet{WestHawley2008} are plotted in gray.
    \label{FigChi}
    }
\end{figure}
% -----------------------------------------------------------------------------

The $\chi$-factor provides a method to convert the equivalent width
of \ion{Ca}{II} H and K into \rhk and H$\alpha$ indto $\fha$ in M dwarfs.
\citet{WalkowiczHawleyWest2004} define the $\chi$ factor as the ratio of the 
$H\alpha$ line continuum luminosity to bolometric luminosity, namely
\begin{equation}
  \chi = L_\mathrm{H\alpha,\,cont} / L_\mathrm{bol},
\end{equation}
where $L_\mathrm{H\alpha,\,cont}$ is the luminosity of a selected
continuum region near H$\alpha$.
\citet{WestHawley2008} extend the selection of $\chi$ values to higher-order
Balmer lines and \ion{Ca}{II} H and K.
When multiplied by the equivalent width of the respective line, we can obtain
the ratio of the line luminosity to the bolometric luminosity,
\begin{equation}
  L_\mathrm{line} / L_\mathrm{bol} = \chi_\mathrm{line} \cdot \mathrm{EW_{line}}.
\end{equation}

For both \ion{Ca}{II} H and K and H$\alpha$,
we calculated the $\chi$ values of \citet{WestHawley2008} of the PHOENIX
model grid for M dwarf \teff values and different metallicities,
constraining $\logg = 5.0$.
For \ion{Ca}{II} H and K, we used the continuum regions given by
\citet{WalkowiczHawley2009},
and for H$\alpha$, we used the continuum regions given by \citet{WestHawley2008}.
We plot these with the values listed in \citet{WestHawley2008}
in Fig.~\ref{FigChi},
with \ion{Ca}{II} H and K in the top panel and H$\alpha$ in the lower panel.
We provide the computed $\chi_\mathrm{line}$ values in Table~\ref{TableChi}.

Similar to Fig.~\ref{FigCcf}, the continuum values diverge as \teff decreases
for different metallicities.
For $\chi_\mathrm{CaIIHK}$,
a metallicity of $\feh = -0.5$ agrees best with 
\citet{WestHawley2008}.
We note that all of our $\chi_\mathrm{CaIIHK}$ values are higher
for M3 and M4 spectral types.
However, this difference is very small
(on the order of $10^{-7}$), and the error bars of \citet{WestHawley2008}
are much smaller than the spread of $\chi_\mathrm{CaIIHK}$ by metallicity.
Our $\chi_\mathrm{H\alpha}$ values also agree well with \citet{WestHawley2008},
especially for $\feh = 0.0$.
Our $\chi_\mathrm{H\alpha}$ values deviate from \citet{WestHawley2008}
at spectral types M4 and earlier.
However, like with $\chi_\mathrm{CaIIHK}$, the difference is small,
on the order of $10^{-6}$.

\subsection{Proxima Centauri}
Proxima Centauri, or GJ\,551, is the only star in this study that exhibits H$\alpha$ emission
and is a known planet host~\citep{AngladaEscude2016}.
This indicator of high activity is consistent with the findings of \citet{Ribas2016},
who reported that Proxima b receives ten times more far-UV flux than the current Earth.

Except for the \teffmal calibration of Proxima Centauri ($\log \rphk = -3.92$, $\teffmal = 3555$ K),
we did not measure $\log \rphk > -4.5$ for any planet hosts.
However, this particular value may be overestimated.
For example, for Proxima Centauri, \citet{Boyajian2012} reported $\teff{} \sim 3050$ K,
and the calibration of \teffspt{} = 2900 K, 
which has $\log \rphk = -4.59$.
The similarities between these two temperatures mean that
$\log \rphk$ of Proxima Centauri probably sits closer to $\sim -4.5$ than $\sim -3.9$.
%%%%%%%%%%%%%%%%%%%%%%%%%%%%%%%%%%%%%%%%%%%%%%%%%%%%%%%%%%%%%%%%%%%%%%%%%%%%%%%

%%%%%%%%%%%%%%%%%%%%%%%%%%%%%%%%%%%%%%%%%%%%%%%%%%%%%%%%%%%%%%%%%%%%%%%%%%%%%%%
\section{Discussion}
\label{SecDiscussion}

With the exception of Proxima Centauri, which exhibits H$\alpha$ in emission,
the remaining known planet hosts exhibit low activity.
This is expected as activity can mimic planetary signals
and cause incorrect planet detections,
so that stars with lower activity stars are preferred for planet searches.

There is a divergence of $\log C_\mathrm{cf}$ values 
between our work and 
other works
that begins at the start of the M dwarf sequence near 4000 K.
\citet{Middelkoop1982}, \citet{Rutten1984}, \citet{Cincunegui2007}, \citet{SuarezMascareno2015},
and \citet{Astudillo2017}
 used observational data to calibrate
$\log C_\mathrm{cf}$ as a function of $B-V$,
while we used stellar models as a function of $\teff$.
The discrepency might be due to an insufficient
($B-V) - \teff{}$ relation in this region.
However, our $\chi_\mathrm{CaIIHK}$ values agree with those of \citet{WestHawley2008},
whose relations were also derived from observational data. This gives us confidence that the PHOENIX models
are accurate in the continuum near the 
\ion{Ca}{II} H and K lines,
although the continuum region of $\chi_\mathrm{CaIIHK}$
differs from the region used in $\log C_\mathrm{cf}$.

Our $\log \rphot{}$ values remain higher than 
$\log \rphot{}$ values obtained using observational data
\citep{Noyes1984, SuarezMascareno2016}.
As noted by \citet{Mittag2013}, this difference arises from the integration
region of the calcium line.
Our work and \citet{Mittag2013} both integrated the entire photospheric line.
The $\log \rphot{}$ relation used by \citet{Noyes1984} is that of 
\citet{Hartmann1984}, who only estimated the flux outside of the line core
exterior to the H1 and K1 points: they assumed the flux of the line core to be zero.
The $\log \rphot{}$ relation derived by \citet{SuarezMascareno2015}
also masks the line core: $0.7~\AA{}$ for FGK stars, and $0.4~\AA{}$ for M stars.
For this reason,
we provide the correction factor in Eq.~\ref{EqRphotCorrect} 
to keep the calculated \rphk{} values consistent with historical
published measurements based on \citet{Noyes1984}.
Moreover, although both use PHOENIX models,
our $\log \rphot$ values are still higher than those of \citet{Mittag2013}.
One reason for this difference might be the models that were used;
\citet{Mittag2013} used models computed in 
non-local thermodynamic equilibrium,
while we used models computed in 
local thermodynamic equilibrium.

\citet{Astudillo2017} measured \rphk of 403 stars of the HARPS M dwarf sample.
They extended the technique of \citet{Noyes1984} by calibrating their own
$\log C_\mathrm{cf}$ using 14 G or K and M dwarf pairs
and use BT-Settl models to arrive at \rphot through an S-index conversion.
Although we used different methods to arrive at the measurement of \rphk in M dwarfs,
we find our results to be consistent with \citet{Astudillo2017}.
Namely, our Fig.~\ref{FigRpHKMk} exhibits a very similar lower envelope of
\rphk values to their Fig. 10.
For brighter $M_{K_S}$ values (earlier-M dwarf types), we find a relatively constant lower envelope
of \rphk values (see Fig.~\ref{FigRpHKMk}),
while \citet{Astudillo2017} also reported a constant lower envelope of \rphk
for the higher M dwarf masses (see their lower Fig. 10).
As $M_{K_S}$ increases and M dwarf types move to later types,
the lower envelope of \rphk values begins to decrease.
Finally, this lower envelope flattens out again for lower masses, or later spectral types.
\citet{Mittag2013} similarly reported that the initially constant lower envelope
was followed by a decreasing lower envelope.
We note a key difference in our findings
in that the technique used to derive $T_\mathrm{eff}$
influences the extent of the decreasing lower envelope and then
the level of the constant envelope for the lowest masses.
However, each method individually still exhibits this behavior.

Surveys that focus on determining stellar parameters
(e.g., \citet{Maldonado2015}) would be the more reliable source
of stellar parameters
if the given star were included in that survey.
Moreover, when the S-index of a given star is the sole measurement and no access to any spectra is possible,
then we recommend the use of
Eq.~\ref{EqCcfPoly} and Eq.~\ref{EqRphotPoly} to calculate \rphk{}.

% ===========================================================================
% ===========================================================================
% ===========================================================================

\section{Summary and conclusions}
  \label{SecSummary}

In this work, 
%we have performed the following:
%
    we have measured \ion{Ca}{II} H and K and H$\alpha$ activity in a large
    sample of HARPS M dwarf spectra using high S/N template spectra
    and PHOENIX model atmospheres.
    We compared three different \teff calibrations and find
    $\Delta \teff \sim\,\mathrm{several}\,100$~K for mid- to late-M dwarfs.
    This uncertainty in \teff contributes up to
    $\Delta \log \rphk = 1.31$~dex
    and $\Delta \log \fpha = 2.93$~dex.
    We have extended \rphk calibrations to the M dwarf regime
    using PHOENIX model spectra.
    We compared these new \rphk calibrations with previous calibrations.
    Our $\log C_\mathrm{cf}$ calibration agrees very well with
    previous calibrations within 0.2 dex, and extends the calibration
    from $3100\ \mbox{K}\ \le \teff \le 6800$~K
    to $2300\ \mbox{K}\ \le \teff \le 7200$~K.
    Our $\log \rphot$ calibration overstimates the \citet{Noyes1984} 
    calibration by 0.46 dex.
    However, our calibration extends $\log \rphot$
    to $2300\ \mbox{K}\ \le \teff \le 7200$~K,
    and a simple offset correction can be applied to scale our $\log \rphot$
    to that of \citet{Noyes1984}.
    We have provided a grid of
    $\log C_\mathrm{cf}$ and $\log \rphot$ values
    as functions of \teff, \feh, and \logg.
    This grid can be used to compute \rphk from S-index values
    using either polynomial fits to or an interpolation of the grid,
    and can be further beneficial when constraints on 
    the stellar parameters of the targets are established.
    We have calculated
    $\chi_\mathrm{CaIIHK}$ and $\chi_\mathrm{H\alpha}$
    for $-1.0 \le \feh \le +1.0$
    in steps of $\Delta \feh = 0.5$
    for the entire M dwarf \teff range.
    We find that our $\chi$ values from PHOENIX models agree well with
    the $\chi$ values of \citet{WestHawley2008}.

We find that the lower boundary of $\log \rphk$ either stays constant
or decreases with later-M dwarfs depending on the \teff calibration used.
Because of conflicting \teff measurements toward later-M dwarfs,
an accurate determination of \rphk cannot be made beyond $M_{K_S} > 8$.
For $\fpha$, the lower boundary of inactive stars begins with early-M dwarfs
in deeper absorption,
and fills in to the continuum towards later-M dwarfs.
Stars with known planetary systems 
do not exhibit unexpected or peculiar levels
of \ion{Ca}{II} H, K, and H$\alpha$ activity in relation to stars of
similar spectral type or absolute magnitude.

Our surface flux calibration of $\log C_\mathrm{cf}$
agrees very well with that of \citet{Middelkoop1982} for FGK stars,
and our surface flux calibrations of
$\chi_\mathrm{CaIIHK}$ and $\chi_\mathrm{H\alpha}$
also agree well with those of \citet{WestHawley2008} for M stars.
Our \rphk calibrations agree very well with those of \citet{Noyes1984}
to within $\Delta \log \rphk = 0.01$~dex for the Sun.
We conclude that our calibrations are a reliable extension to previous \rphk
calibrations,
provide a consistent way to measure \rphk across spectral types early F to late M,
and allow the comparison of activity of Sun-like stars to M dwarfs.
%%%%%%%%%%%%%%%%%%%%%%%%%%%%%%%%%%%%%%%%%%%%%%%%%%%%%%%%%%%%%%%%%%%%%%%%%%%%%%%

%%%%%%%%%%%%%%%%%%%%%%%%%%%%%%%%%%%%%%%%%%%%%%%%%%%%%%%%%%%%%%%%%%%%%%%%%%%%%%
\begin{acknowledgements}
  We thank Tim-Oliver Husser for fruitful discussions and providing us
  with recalculated PHOENIX models.
  The authors acknowledge research funding by the
  Deutsche Forschungsgemeinschaft (DFG) under the grant SFB 963, project A04.
  SVJ acknowledges the support of the DFG priority program SPP 1992 “Exploring the Diversity of Extrasolar Planets" (JE 701/5-1).
  SBS acknowledges the support of the Austrian Science Fund (FWF) Lise Meitner project M2829-N.
  This work is based on data products from observations made with ESO Telescopes at the La Silla Observatory (Chile) under the program IDs
60.A-9036, 072.C-0488, 074.C-0364, 075.C-0202, 075.D-0614, 076.C-0155, 077.C-0364, 078.C-0044, 082.C-0718, 085.C-0019, 086.C-0284, 087.C-0831, 089.C-0050, 089.C-0732, 090.C-0395, 090.C-0421, 091.C-0034, 180.C-0886, 183.C-0437, 183.C-0972, 185.D-0056, 191.C-0505, 192.C-0224, and 283.C-5022.
We acknowledge the effort of all the observers of the aforementioned ESO projects whose data we have used.
\end{acknowledgements}

\bibliographystyle{aa}  % style aa.bst
\bibliography{references}

%%%%%%%%%%%%%%%%%%%%%%%%%%%%%%[ APPENDIX ]%%%%%%%%%%%%%%%%%%%%%%%%%%%%%%%%%%%%
\begin{appendix}
  \label{SecAppendix}

\section{Ca II H and K surface flux calibrations}
\label{app-ccf}

% MIDDELKOOP CCF
Measuring $\log C_\mathrm{cf}$ of 85 main-sequence stars
from the \citet{VaughanPreston1980} survey of the solar neighborhood,
\citet{Middelkoop1982} fit a polynomial to $\log C_\mathrm{cf}$ as a function of $B-V$,
\begin{equation}
\label{EqCcfM82}
    \log{C_\mathrm{cf,\,M82}} = 1.13 (B-V)^3 - 3.91 (B-V)^2 + 2.84 (B-V) - 0.47
\end{equation}
for $0.45 \le (B-V) \le 1.50$.
  
% RUTTEN CCF
\citet{Rutten1984} extended the relation further by
measuring 30 additional main-sequence stars and also 27 subgiants and giants,
\begin{equation}
    \label{EqCcfR84}
    \log{C_\mathrm{cf,\,R84}} = 0.25 (B-V)^3 - 1.33 (B-V)^2 + 0.43 (B-V) + 0.24
,\end{equation}
which is valid for main-sequence stars with $0.3 \le (B-V) \le 1.6$.

% CINCUNEGUI 2007
\citet{Cincunegui2007} extended the calibration using 109 stars ranging from
spectral type F6 to M5,
\begin{equation}
    \label{EqCcfC07}
    \log{C_\mathrm{cf,\,CDM07}} = -0.33 (B-V)^3 - 0.55 (B-V)^2 - 1.41 (B-V) + 0.8
\end{equation}
for $0.45 \le (B-V) \le 1.81$.
\citet{Middelkoop1982}, \citet{Rutten1984}, and \citet{Cincunegui2007}
all made use of the relation of \citet{Noyes1984} to convert
$B-V$ into \teff.
  
% MITTAG 2013
Instead of using observational spectra,
\citet{Mittag2013} used PHOENIX stellar atmosphere models to
measure $\mathcal{F}_{RV}$ directly in absolute flux units,
\begin{equation}
  \log{ \left( \mathcal{F}_{RV}/19.2 \right)}_\mathrm{MSS13} = 8.33-1.79(B-V),
\end{equation}
where $0.44 \le (B-V) \le 1.60$.
They transformed this to $\log C_\mathrm{cf}$ using
\begin{equation}
    \label{EqCcfM13}
    \log C_\mathrm{cf} = 8.33-1.79(B-V) - \log \alpha - \log \teff^4 -14,
\end{equation}
where $\alpha=19.2$,
and used the $B-V$ to \teff relation of \citet{Gray2005}.

% PEREZ MARTINEZ 2013
Similarly, \citet{PerezMartinez2014} also used PHOENIX models to derive
a conversion factor
\begin{equation}
    \log{C_\mathrm{cf,\,PMSH14}} = 0.66 - 1.11 (B-V)
\end{equation}
for $0.44 \le (B-V) \le 1.33$.
\citet{PerezMartinez2014} used the $B-V$ to \teff tables of Schmidt-Kaler (1982)
(which they also included in their publication).

% SUAREZ MASCARENO 2015
Using HARPS spectra of FGKM stars,
\citet{SuarezMascareno2015} derived a continuum correction factor
\begin{equation}
    \log{C_\mathrm{cf,\,SM15}} = 0.668 -1.270(B-V) +0.645(B-V)^2 -0.443(B-V)^3
\end{equation}
for $0.4 \lesssim (B-V) \lesssim 1.9$.

  % Astudillo 2017
\citet{Astudillo2017} empirically determine a continuum correction factor,
  \begin{equation}
    \log{C_\mathrm{cf,\,AD17}} = -0.203(B-V)^3 + 0.109(B-V)^2 -0.972(B-V) +0.669(B-V)
\end{equation}
for $0.54 \le (B-V) \le 1.9$.

\section{Ca II H and K photospheric flux calibrations}
\label{app-rphot}

In their publication, \citet{Noyes1984} also provided an alternative,
simpler form of Eq.~\ref{EqRphotN84} and obtained similar results within
the valid calibration range,
\begin{equation}
  \label{EqRphotN84Simple}
  \log R_\mathrm{phot,\,N84,simple} = -4.02 -1.40(B-V).
\end{equation}

\citet{Mittag2013}, using PHOENIX stellar atmosphere models,
reported two linear dependences of $\log \rphot$ on $B-V$ with for two $B-V$ ranges: 
\begin{equation}
  \label{EqRphotM13A}
  \log R_\mathrm{phot,\,MSS13} = 7.49 - 2.06(B-V) - \log \mathcal{F}_\mathrm{bol}
\end{equation}
for $0.44 \le B-V < 1.28$, and
\begin{equation}
  \label{EqRphotM13B}
  \log R_\mathrm{phot,\,MSS13} = 6.19 - 1.04(B-V) - \log \mathcal{F}_\mathrm{bol}
\end{equation}
for $1.28 \le B-V < 1.60$.

Using HARPS spectra
and a 0.7 \AA{} rectangular mask of the line core for the most inactive FGK stars
and a 0.4 \AA{} rectangular mask for the most inactive M stars,
\citet{SuarezMascareno2015} fit a photospheric flux relation,
\begin{equation}
    \log R_\mathrm{phot,\,SM15} = \log{1.48 \cdot 10^{-4} \exp{-4.3658(B-V)}}
,\end{equation}
for $0.4 \lesssim (B-V) \lesssim 1.9$.

\citet{Astudillo2017} determined a photospheric flux relation using a synthetic spectrum,
  resulting in,
  \begin{equation}
    \log{R_\mathrm{phot,\,AD17}} = - 0.045(B-V)^3 - 0.026(B-V)^2 - 1.036(B-V) - 3.749(B-V)
\end{equation}
for $0.54 \le (B-V) \le 1.9$.

  \section{Effective temperature comparison with other calibrations}
\label{teff-mann}

  In Section~\ref{SecParams} we presented three  different \teff calibrations that we used as
  stellar parameters in this study (see Fig.~\ref{FigTeffCompare}).
  Here in this appendix we present the \teff measurements of common targets by \citet{Mann2015}
  to show agreement and the degree of agreement with the different \teff calibration techniques.
  \begin{figure}
\resizebox{\hsize}{!}{\includegraphics{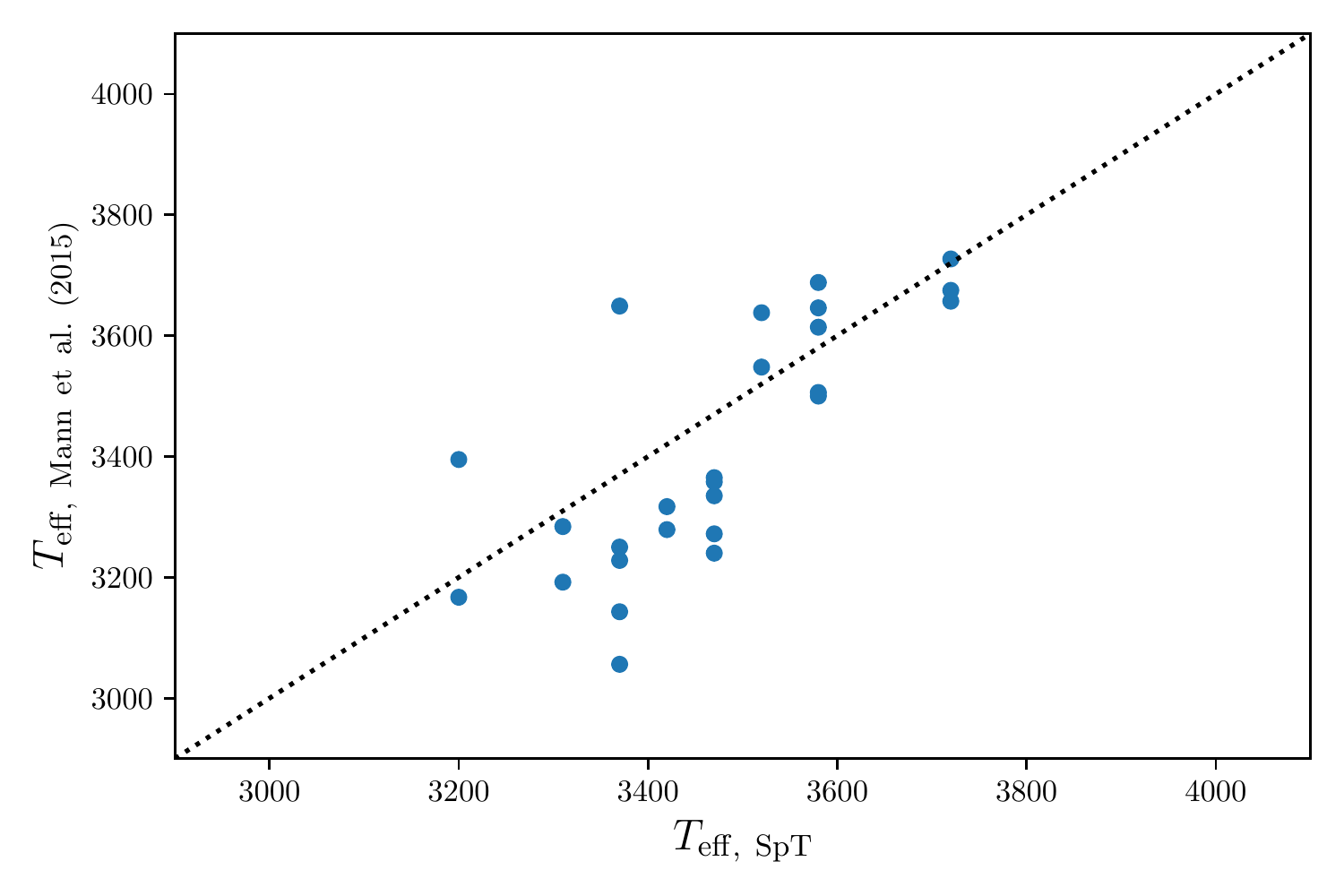}}
\caption{
\teff of \citet{Mann2015} vs \teffspt.
The dotted line shows the 1:1 relation.
\label{FigTeffSptVsMann}
}
\end{figure}

\begin{figure}
\resizebox{\hsize}{!}{\includegraphics{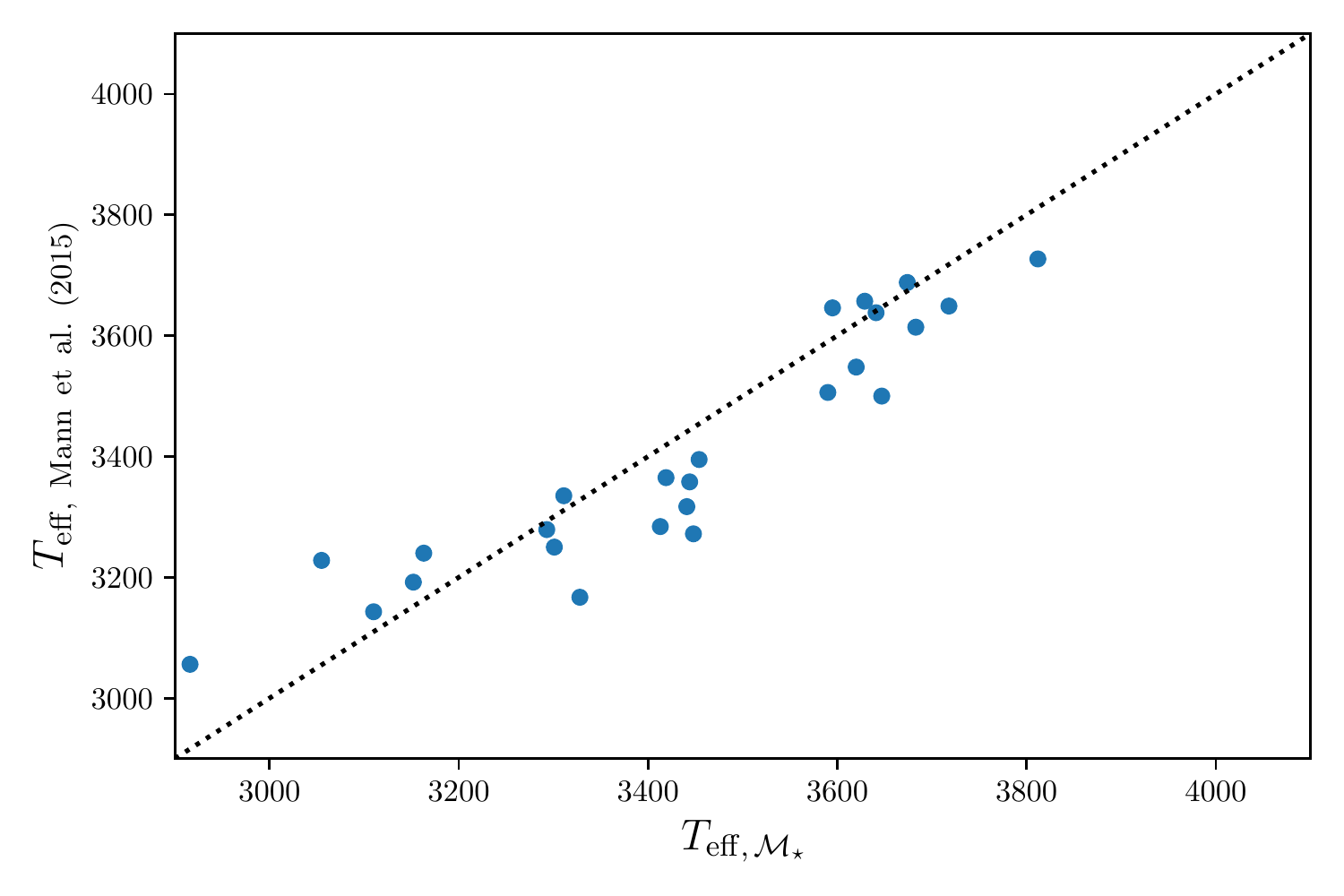}}
\caption{
\teff of \citet{Mann2015} vs \teffmass.
The dotted line shows the 1:1 relation.
\label{FigTeffMassVsMann}
}
\end{figure}

\begin{figure}
\resizebox{\hsize}{!}{\includegraphics{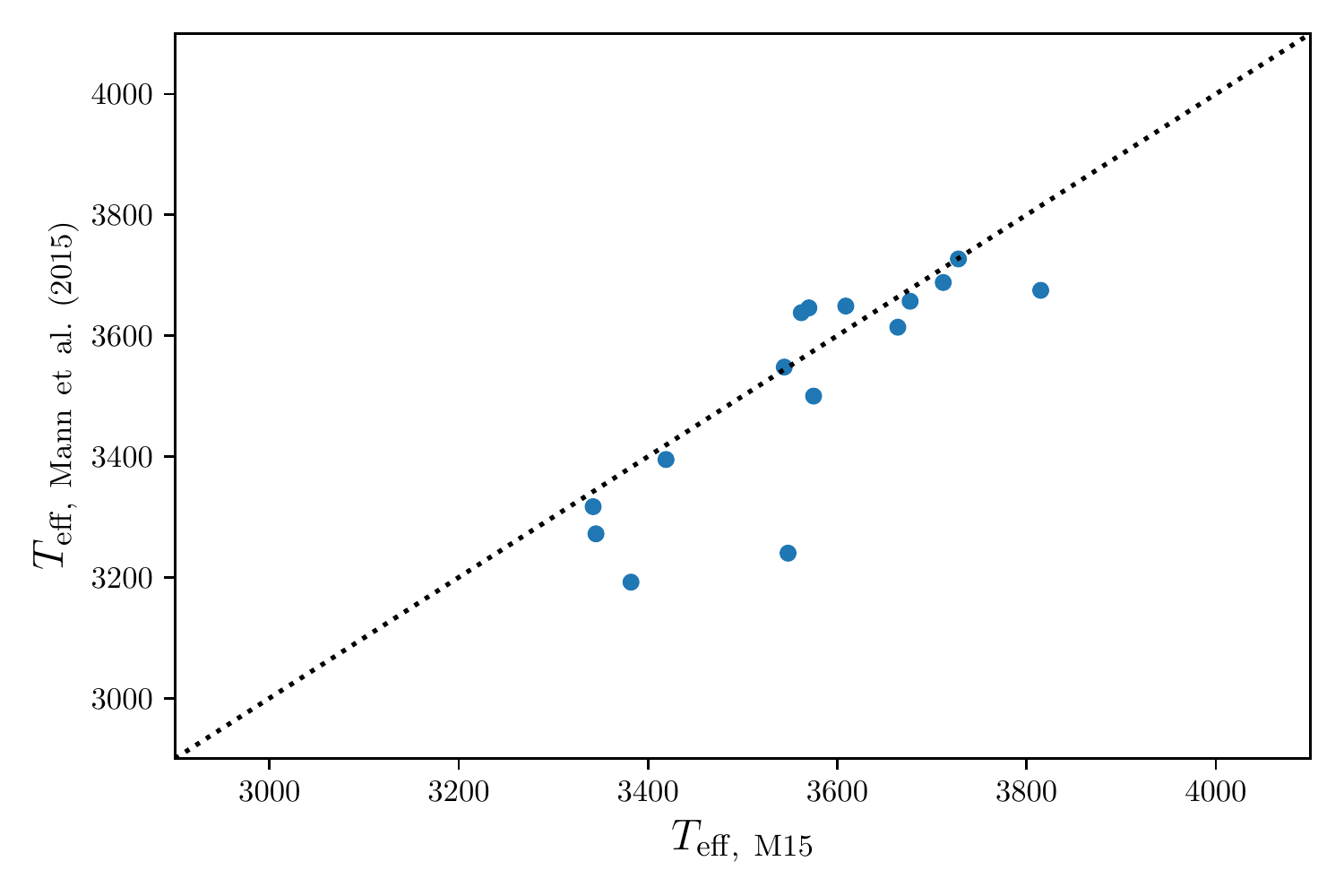}}
\caption{
\teff of \citet{Mann2015} vs \teffmal.
The dotted line shows the 1:1 relation.
\label{FigTeffM15VsMann}
}
\end{figure}

\section{Tables}

%%%%%%%%%%%%%%%%%%%%%%%%%%%%%%%%%%%%%%%%%%%%%%%%%%%%%%%%%%%%%%%%%%%%%%%%%%%%%%%
\longtab[1]{
\begin{landscape}
\begin{longtable}{lllrrrlrrrrr}
\caption{\label{TableSimbad}Astronomical measurements of the sample of stars.}\\
\hline\hline
Name
& $\alpha$ (2000)
& $\delta$ (2000)
& $\mu_{\alpha}$
& $\mu_{\delta}$
& $\pi$
& Spec. type
& $B$
& $V$
& $K_S$
& $M_{K_S}$
& $N_\mathrm{spectra}$
\\
{}
& {}
& {}
& [mas yr$^{-1}$]
& [mas yr$^{-1}$]
& [mas]
& {}
& [mag]
& [mag]
& [mag]
& [mag]
& {}
\\
\hline
\endfirsthead
\caption{continued.}\\
\hline\hline
Name
& $\alpha$ (2000)
& $\delta$ (2000)
& $\mu_{\alpha}$
& $\mu_{\delta}$
& $\pi$
& Spec. type
& $B$
& $V$
& $K_S$
& $M_{K_S}$
& $N_\mathrm{spectra}$
\\
{}
& {}
& {}
& [mas yr$^{-1}$]
& [mas yr$^{-1}$]
& [mas]
& {}
& [mag]
& [mag]
& [mag]
& [mag]
& {}
\\
\hline
\endhead
\hline
\endfoot
GJ 1        &    00:05:24.42828  &    -37:21:26.5010  &   5634.68 &  -2337.71 &  230.42 &                       M1.5  & 10.0 &  8.6 & 4.5 &  6.3 &     47 \\
GJ 1002     &      00:06:43.255  &      -07:32:14.71  &   -749.00 &  -1922.00 &  213.00 &                      M5.5V  & 15.7 & 13.8 & 7.4 &  9.1 &      7 \\
GJ 12       &       00:15:49.20  &       +13:33:21.9  &    621.00 &    333.00 &   84.00 &                         M3  & 14.0 & 12.5 & 7.8 &  7.4 &      7 \\
GJ 3049     &      00:43:26.031  &      -41:17:33.71  &   -775.00 &   -780.00 &  101.00 &                         M3  & 14.0 &  $\cdots$ & 7.7 &  7.7 &      6 \\
GJ 54.1     &    01:12:30.63918  &    -16:59:56.3353  &   1208.53 &    640.73 &  271.01 &                       M4Ve  & 13.9 & 12.1 & 6.4 &  8.6 &     21 \\
L 707-74    &      01:23:18.029  &      -12:56:23.00  &    -12.30 &    335.20 &   97.80 &                          M  & 13.8 & 12.8 & 8.3 &  8.3 &      5 \\
GJ 87       &    02:12:20.98849  &    +03:34:32.2312  &  -1760.70 &  -1852.91 &   96.02 &                      M2.5V  & 11.5 & 10.0 & 6.1 &  6.0 &     28 \\
GJ 105B     &      02:36:15.357  &      +06:52:19.14  &   1813.00 &   1447.00 &  129.40 &                      M4.5V  & 13.3 & 11.7 & 6.6 &  7.1 &     22 \\
CD-44-836A  &      02:45:10.726  &      -43:44:31.64  &     40.00 &   -370.00 &  113.90 &                         M4  & 13.9 & 12.3 & 7.3 &  7.6 &      8 \\
CD-44-836B  &       02:45:14.31  &       -43:44:10.2  &     40.00 &   -350.00 &  142.86 &                         M5  & 14.3 & 12.7 & 7.2 &  8.0 &      4 \\
GJ 3189     &       02:58:10.21  &       -12:53:06.7  &    275.00 &    551.00 &   95.50 &                     sdM3.0  & 14.4 & 12.7 & 8.2 &  8.1 &     10 \\
GJ 3193B    &    03:01:51.38955  &    -16:35:36.1103  &   -356.78 &   -302.31 &  106.16 &                       M3.0  & 12.2 & 10.5 & 6.5 &  6.6 &      9 \\
GJ 3207     &       03:11:35.20  &       -38:47:23.4  &    838.00 &   -195.00 &  130.00 &                       M3.5  & 13.0 & 11.5 & 9.0 &  9.6 &      6 \\
GJ 1057     &      03:13:22.995  &      +04:46:29.37  &   1749.00 &     84.00 &  117.10 &                      M5.0V  & 15.6 & 13.8 & 7.8 &  8.2 &      9 \\
GJ 145      &    03:32:55.85736  &    -44:42:07.0185  &   -311.53 &    131.98 &   93.11 &                      M2.5V  & 13.1 & 11.5 & 6.9 &  6.8 &      7 \\
GJ 1061     &       03:35:59.69  &       -44:30:45.3  &    730.00 &   -330.00 &  271.92 &                      M5.0V  & 15.0 & 13.1 & 6.6 &  8.8 &      7 \\
GJ 1065     &       03:50:44.32  &       -06:05:40.0  &   -450.00 &  -1373.00 &  104.00 &                        M3V  & 14.5 & 12.8 & 7.8 &  7.8 &      8 \\
GJ 1068     &      04:10:28.156  &      -53:36:07.81  &  -2585.00 &  -2744.00 &  143.40 &                       M4.5  & 15.5 & 13.6 & 7.9 &  8.7 &     18 \\
GJ 166C     &      04:15:21.733  &      -07:39:17.36  &  -2239.00 &  -3419.00 &  198.24 &                       M5Ve  & 12.8 & 11.2 & 6.0 &  7.4 &      4 \\
GJ 176      &    04:42:55.77497  &    +18:57:29.4043  &    656.85 &  -1116.20 &  107.83 &                      M2.5V  & 11.5 & 10.0 & 5.6 &  5.8 &     71 \\
GJ 3323     &      05:01:57.469  &      -06:56:45.92  &   -550.00 &   -533.00 &  187.92 &                      M4.0V  & 13.9 & 12.2 & 6.7 &  8.1 &      7 \\
GJ 3325     &    05:03:20.08559  &    -17:22:24.7421  &   -226.51 &   -447.19 &  108.61 &                        M3V  & 13.4 & 11.7 & 6.9 &  7.1 &      9 \\
GJ 191      &    05:11:40.58112  &    -45:01:06.2899  &   6505.08 &  -5730.84 &  255.66 &                     sdM1.0  & 10.4 &  8.9 & 5.0 &  7.1 &     97 \\
GJ 203      &    05:28:00.15251  &    +09:38:38.1399  &   -190.30 &   -759.45 &  113.50 &                       M3.5  & 14.1 & 12.4 & 7.5 &  7.8 &     11 \\
GJ 205      &    05:31:27.39595  &    -03:40:38.0311  &    761.86 &  -2093.60 &  176.77 &                      M1.5V  &  9.4 &  8.0 & 3.9 &  5.1 &     75 \\
GJ 213      &    05:42:09.26772  &    +12:29:21.6225  &   2000.53 &  -1569.63 &  171.55 &                      M4.0V  & 13.1 & 11.5 & 6.4 &  7.6 &     20 \\
GJ 229      &    06:10:34.61533  &    -21:51:52.7117  &   -137.09 &   -713.66 &  173.81 &                     M1/M2V  &  9.6 &  8.1 & 4.2 &  5.4 &     25 \\
HIP 31293   &    06:33:43.29814  &    -75:37:47.9759  &   -290.15 &    277.00 &  110.88 &                        M2V  & 11.9 & 10.3 & 5.9 &  6.1 &     10 \\
HIP 31292   &    06:33:46.84863  &    -75:37:30.2972  &   -291.38 &    221.24 &  115.19 &                        M3V  & 12.9 & 11.4 & 6.6 &  6.9 &     18 \\
GJ 3404A    &      06:42:11.193  &      +03:34:52.63  &     38.50 &   -259.30 &   64.00 &                       M3.5  & 13.3 & 11.7 & 7.3 &  6.4 &     15 \\
GJ 3405B    &       06:42:13.34  &       +03:35:31.1  &     35.00 &   -285.00 &   64.00 &                         M4  & 11.6 & 13.3 & 8.3 &  7.3 &      4 \\
GJ 250B     &       06:52:18.05  &       -05:11:24.2  &   -541.00 &      0.00 &  114.80 &                         M2  & 11.5 & 10.1 & 5.7 &  6.0 &     13 \\
GJ 273      &    07:27:24.49975  &    +05:13:32.8332  &    572.51 &  -3693.51 &  262.98 &                      M3.5V  & 11.4 &  9.9 & 4.9 &  7.0 &    169 \\
GJ 3459     &    07:38:40.95640  &    -21:13:28.5046  &    450.28 &   -477.71 &   94.31 &                         M3  & 13.3 & 11.7 & 7.1 &  6.9 &      8 \\
GJ 285      &    07:44:40.17401  &    +03:33:08.8350  &   -345.25 &   -450.70 &  167.88 &                      M4.5V  & 12.8 & 11.2 & 5.7 &  6.8 &      7 \\
GJ 299      &      08:11:57.575  &      +08:46:22.05  &   1099.00 &  -5123.00 &  148.00 &                      M4.5V  & 14.6 & 12.8 & 7.7 &  8.5 &     22 \\
GJ 300      &      08:12:40.881  &      -21:33:05.68  &      9.00 &   -694.00 &  125.60 &                         M4  & 13.7 & 12.1 & 6.7 &  7.2 &     39 \\
GJ 2066     &    08:16:07.98235  &    +01:18:09.2631  &   -374.45 &     60.10 &  109.62 &                      M2.0V  & 11.6 & 10.1 & 5.8 &  6.0 &      8 \\
GJ 1123     &      09:17:05.328  &      -77:49:23.37  &    695.00 &   -866.00 &  110.90 &                         M4  & 14.7 & 13.1 & 7.4 &  7.7 &      7 \\
GJ 341      &    09:21:37.60463  &    -60:16:55.0407  &   -837.79 &    180.67 &   95.58 &                       M0.0  & 11.0 &  9.5 & 5.6 &  5.5 &     57 \\
GJ 1125     &    09:30:44.58334  &    +00:19:21.5651  &   -571.90 &   -553.76 &  103.46 &                       M3.5  & 13.3 & 11.7 & 6.9 &  6.9 &      9 \\
GJ 357      &    09:36:01.63592  &    -21:39:38.8679  &    136.67 &   -989.13 &  110.82 &                      M2.5V  & 12.5 & 10.9 & 6.5 &  6.7 &     53 \\
GJ 358      &    09:39:46.36930  &    -41:04:03.2100  &   -526.56 &    356.39 &  105.63 &                         M3  & 12.2 & 10.7 & 6.1 &  6.2 &     38 \\
GJ 367      &    09:44:29.83739  &    -45:46:35.4273  &   -461.73 &   -584.28 &  101.31 &                       M1.0  & 11.4 & 10.0 & 5.8 &  5.8 &     24 \\
GJ 1129     &      09:44:47.315  &      -18:12:48.93  &  -1606.00 &   -172.00 &   90.90 &                         M4  & 14.2 & 12.6 & 7.3 &  7.0 &      8 \\
GJ 382      &    10:12:17.66904  &    -03:44:44.3966  &   -151.09 &   -244.31 &  127.08 &                      M2.0V  & 10.8 &  9.3 & 5.0 &  5.5 &     33 \\
GJ 388      &      10:19:36.277  &      +19:52:12.06  &   -501.80 &    -42.80 &  213.00 &                     M4.5Ve  &  9.6 &  8.1 & 4.6 &  6.2 &     43 \\
GJ 393      &    10:28:55.55087  &    +00:50:27.6218  &   -603.75 &   -728.94 &  141.50 &                      M2.5V  & 11.2 &  9.7 & 5.3 &  6.1 &     29 \\
GJ 3618     &      10:44:21.320  &      -61:12:38.44  &   -334.00 &   1626.00 &  208.95 &                         M4  & 15.7 & 13.9 & 7.7 &  9.3 &      7 \\
GJ 402      &    10:50:52.03129  &    +06:48:29.2278  &   -855.92 &   -822.34 &  147.92 &                      M5.0V  & 13.3 & 11.7 & 6.4 &  7.2 &      6 \\
GJ 406      &      10:56:28.865  &      +07:00:52.77  &  -3842.00 &  -2725.00 &  418.30 &                      M6.0V  & 15.5 & 13.5 & 6.1 &  9.2 &     29 \\
GJ 413.1    &    11:09:31.34559  &    -24:35:55.1194  &   -796.18 &   -446.82 &   93.00 &                         M2  & 12.0 & 10.4 & 6.1 &  5.9 &     18 \\
GJ 433      &    11:35:26.94663  &    -32:32:23.8952  &    -72.51 &   -851.92 &  112.58 &                       M1.5  & 11.3 &  9.8 & 5.6 &  5.9 &     86 \\
GJ 436      &      11:42:11.094  &      +26:42:23.65  &    896.07 &   -813.54 &   99.00 &                      M3.5V  & 12.1 & 10.6 & 6.1 &  6.1 &    169 \\
GJ 438      &      11:43:19.811  &      -51:50:25.98  &    654.10 &   -538.50 &  118.90 &                         M0  & 11.9 & 10.3 & 6.3 &  6.7 &     19 \\
GJ 447      &      11:47:44.405  &      +00:48:16.44  &    605.26 &  -1219.28 &  298.04 &                      M4.5V  & 12.9 & 11.2 & 5.7 &  8.0 &     41 \\
GJ 465      &    12:24:52.50278  &    -18:14:32.2435  &   1095.66 &  -2308.56 &  112.98 &                         M2  & 12.9 & 11.3 & 7.0 &  7.2 &     24 \\
GJ 479      &    12:37:52.21838  &    -52:00:05.3142  &  -1032.41 &     30.39 &  103.18 &                       M3Ve  & 12.2 & 10.7 & 6.0 &  6.1 &     58 \\
GJ 3737     &      12:38:49.141  &      -38:22:52.80  &   -567.00 &  -1399.00 &  156.78 &                      M4.5V  & 14.4 & 12.7 & 7.4 &  8.4 &     10 \\
GJ 480.1    &    12:40:46.28898  &    -43:33:58.9535  &   -781.58 &    693.43 &  128.52 &                        M3V  & 14.0 & 12.2 & 7.4 &  8.0 &     10 \\
GJ 486      &    12:47:56.62550  &    +09:45:05.0355  &  -1006.76 &   -460.19 &  119.47 &                      M4.0V  & 13.0 & 11.4 & 6.4 &  6.7 &     12 \\
GJ 514      &    13:29:59.78622  &    +10:22:37.7908  &   1128.32 &  -1073.47 &  130.62 &                      M1.0V  & 10.5 &  9.0 & 5.0 &  5.6 &     57 \\
GJ 526      &    13:45:43.77665  &    +14:53:29.4635  &   1778.45 &  -1456.44 &  185.49 &                      M4.0V  &  9.9 &  8.5 & 4.4 &  5.8 &     31 \\
GJ 536      &    14:01:03.18912  &    -02:39:17.5212  &   -823.47 &    598.19 &   99.72 &                      M1.5V  & 11.2 &  9.7 & 5.7 &  5.7 &     52 \\
GJ 551      &    14:29:42.94853  &    -62:40:46.1631  &  -3775.75 &    765.54 &  771.64 &                       M6Ve  & 12.9 & 11.1 & 4.4 &  8.8 &    261 \\
GJ 555      &    14:34:16.81183  &    -12:31:10.3965  &   -354.45 &    595.35 &  164.99 &                      M4.0V  & 12.9 & 11.3 & 5.9 &  7.0 &     14 \\
GJ 569A     &      14:54:29.241  &      +16:06:03.76  &    277.70 &   -132.70 &  101.90 &                      M2.5V  & 12.3 & 10.4 & $\cdots$ &  $\cdots$ &     25 \\
GJ 581      &      15:19:26.823  &      -07:43:20.21  &  -1227.67 &    -97.78 &  160.91 &                      M5.0V  & 11.8 & 10.6 & 5.8 &  6.9 &    243 \\
GJ 588      &    15:32:12.93186  &    -41:16:32.1081  &  -1177.12 &  -1028.54 &  168.66 &                      M2.5V  & 10.8 &  9.3 & 4.8 &  5.9 &    255 \\
GJ 618A     &      16:20:03.509  &      -37:31:44.80  &   -740.10 &    997.40 &  119.80 &                         M3  & 12.1 & 10.6 & 6.0 &  6.3 &     21 \\
GJ 628      &    16:30:18.05803  &    -12:39:45.3232  &    -94.81 &  -1183.43 &  232.98 &                        M3V  & 11.6 & 10.1 & 5.1 &  6.9 &     88 \\
GJ 643      &      16:55:25.272  &      -08:19:20.78  &   -813.83 &   -894.40 &  148.92 &                      M3.5V  & 13.4 & 11.8 & 6.7 &  7.6 &     10 \\
GJ 667C     &      17:18:58.838  &      -34:59:48.64  &   1155.00 &   -214.40 &  146.30 &                      M1.5V  & 11.8 & 10.2 & 6.0 &  6.9 &    182 \\
GJ 674      &    17:28:39.94461  &    -46:53:42.6930  &    571.26 &   -880.83 &  220.24 &                        M3V  & 11.0 &  9.4 & 4.9 &  6.6 &     93 \\
GJ 678.1A   &      17:30:22.727  &      +05:32:54.71  &     29.13 &   -248.15 &  100.20 &                        M1V  & 10.8 &  9.3 & 5.4 &  5.4 &     75 \\
GJ 680      &    17:35:13.61562  &    -48:40:51.1341  &     82.62 &    454.25 &  102.83 &                        M3V  & 11.7 & 10.1 & 5.8 &  5.9 &     39 \\
GJ 682      &    17:37:03.66242  &    -44:19:09.1697  &   -708.98 &   -937.40 &  196.90 &                       M3.5  & 12.6 & 10.9 & 5.6 &  7.1 &     21 \\
GJ 686      &    17:37:53.34690  &    +18:35:30.1613  &    926.78 &    984.20 &  123.67 &                      M1.0V  & 11.1 &  9.6 & 5.6 &  6.0 &     20 \\
GJ 693      &    17:46:34.22971  &    -57:19:08.5550  &  -1119.14 &  -1352.78 &  171.48 &                       M2.0  & 12.4 & 10.8 & 6.0 &  7.2 &     54 \\
GJ 699      &    17:57:48.49803  &    +04:41:36.2072  &   -798.58 &  10328.12 &  548.31 &                      M4.0V  & 11.2 &  9.5 & 4.5 &  8.2 &    234 \\
GJ 701      &    18:05:07.57896  &    -03:01:52.7575  &    570.26 &   -333.47 &  128.89 &                      M2.0V  & 10.9 &  9.4 & 5.3 &  5.9 &     63 \\
GJ 1224     &      18:07:32.927  &      -15:57:46.46  &   -617.00 &   -342.00 &  132.60 &                      M4.5V  & 15.4 & 13.6 & 7.8 &  8.4 &      9 \\
GJ 4071     &       18:42:44.99  &       +13:54:16.8  &    -25.00 &    347.00 &   93.30 &                       M4.5  & 14.8 & 12.8 & 7.6 &  7.4 &      5 \\
GJ 729      &    18:49:49.36216  &    -23:50:10.4291  &    637.02 &   -191.64 &  336.72 &                       M3Ve  & 12.2 & 10.5 & 5.4 &  8.0 &      8 \\
GJ 1232     &      19:09:50.980  &      +17:40:07.45  &   -637.00 &   -426.00 &   93.60 &                       M4.5  & 15.4 & 13.6 & 7.9 &  7.8 &      9 \\
GJ 752A     &    19:16:55.25687  &    +05:10:08.0510  &   -578.78 &  -1331.95 &  170.36 &                        M3V  & 10.6 &  9.1 & 4.7 &  5.8 &     57 \\
GJ 754      &      19:20:47.955  &      -45:33:28.33  &    792.00 &  -3008.00 &  169.00 &                       M4.5  & 13.9 & 12.2 & 6.8 &  8.0 &     57 \\
GJ 1236     &       19:22:02.07  &       +07:02:31.0  &   -747.00 &   -440.00 &   92.90 &                         M3  & 14.0 & 12.3 & 7.7 &  7.5 &     12 \\
GJ 1256     &       20:40:33.64  &       +15:29:57.2  &   1324.00 &    660.00 &  100.80 &                       M4.5  & 15.1 & 13.4 & 7.8 &  7.8 &     10 \\
GJ 803      &    20:45:09.53147  &    -31:20:27.2425  &    279.96 &   -360.61 &  100.91 &                       M1Ve  & 10.1 &  8.6 & 4.5 &  4.5 &      4 \\
GJ 4154     &       20:46:37.26  &       -81:43:14.1  &    637.00 &   -577.00 &   77.10 &                       M2.5  & 13.3 &  $\cdots$ & 6.8 &  6.3 &      8 \\
LP 816-60   &    20:52:33.01679  &    -16:58:29.0249  &   -307.74 &     34.79 &  175.03 &                        M4V  & 13.0 & 11.5 & 6.2 &  7.4 &     14 \\
GJ 832      &    21:33:33.97533  &    -49:00:32.4192  &    -46.05 &   -817.63 &  201.87 &                       M1.5  & 10.2 &  8.7 & 4.5 &  6.0 &     61 \\
GJ 846      &    22:02:10.27388  &    +01:24:00.8292  &   -454.53 &   -279.09 &   97.61 &                         M0  & 10.6 &  9.1 & 5.3 &  5.3 &     55 \\
GJ 4248     &      22:02:29.353  &      -37:04:51.22  &    910.00 &   -337.00 &  134.29 &                       M3.5  & 13.4 & 11.8 & 6.7 &  7.4 &      5 \\
GJ 849      &    22:09:40.34327  &    -04:38:26.6210  &   1130.27 &    -19.27 &  109.94 &                      M3.5V  & 11.9 & 10.4 & 5.6 &  5.8 &     48 \\
GJ 1265     &       22:13:42.78  &       -17:41:08.2  &    849.00 &   -302.00 &   96.00 &                         M4  & 15.3 & 13.6 & 8.1 &  8.0 &     13 \\
GJ 4274     &       22:23:06.97  &       -17:36:25.0  &    248.00 &   -895.00 &  134.10 &                         M4  & 15.1 & 13.2 & 7.3 &  8.0 &      4 \\
GJ 876      &    22:53:16.73352  &    -14:15:49.3186  &    959.84 &   -675.33 &  213.28 &                      M5.0V  & 11.7 & 10.2 & 5.0 &  6.7 &    196 \\
GJ 877      &    22:55:45.50948  &    -75:27:31.2069  &  -1027.76 &  -1060.78 &  116.07 &                        M3V  & 11.9 & 10.4 & 5.8 &  6.1 &     46 \\
GJ 880      &    22:56:34.80475  &    +16:33:12.3541  &  -1034.34 &   -284.09 &  146.09 &                      M2.0V  & 10.1 &  8.6 & 4.5 &  5.3 &     36 \\
GJ 887      &    23:05:52.03604  &    -35:51:11.0475  &   6768.20 &   1327.52 &  305.26 &                        M2V  &  8.8 &  7.3 & 3.5 &  5.9 &     77 \\
GJ 4333     &    23:21:37.44994  &    +17:17:25.3922  &   -534.13 &  -1382.74 &   91.00 &                         M4  & 13.2 & 11.7 & 6.5 &  6.3 &      7 \\
GJ 908      &    23:49:12.52790  &    +02:24:04.4072  &    996.96 &   -967.88 &  167.29 &                        M2V  & 10.4 &  9.0 & 5.0 &  6.2 &     83 \\
LTT 9759    &    23:53:50.11043  &    -75:37:57.1048  &    243.04 &   -378.12 &  100.07 &                         Ma  & 11.5 & 10.0 & 5.5 &  5.6 &     11 \\
GJ 676A     &    17:30:11.20344  &    -51:38:13.1046  &   -260.02 &   -184.29 &   60.79 &                        M0V  & 11.0 &  9.6 & 5.8 &  4.7 &    123 \\
GJ 1214     &       17:15:18.94  &       +04:57:49.7  &    585.00 &   -752.00 &   77.20 &                      M4.5V  & 16.4 & 14.7 & 8.8 &  8.2 &     88 \\
HIP 12961   &    02:46:42.88581  &    -23:05:11.8047  &    292.62 &    140.88 &   43.45 &                         M0  & 11.7 & 10.2 & 6.7 &  4.9 &     48 \\
GJ 163      &    04:09:15.66350  &    -53:22:25.3050  &   1041.43 &    583.01 &   66.69 &                       M3.5  & 13.3 & 11.8 & 7.1 &  6.3 &    170 \\
GJ 317      &       08:40:59.24  &       -23:27:23.3  &   -438.00 &    794.00 &   65.30 &                      M3.5V  & 13.2 & 12.0 & 7.0 &  6.1 &     80 \\
GJ 570B     &    14:57:26.53330  &    -21:24:41.5778  &    961.78 &  -1677.83 &  168.77 &                      M1.5V  &  9.6 &  8.1 & 3.8 &  4.9 &    347 \\
GJ 160.2    &    04:06:34.84136  &    -20:51:11.2391  &     51.90 &   -780.04 &   43.25 &                         K7  & 10.9 &  9.7 & 6.7 &  4.9 &     37 \\
GJ 180      &    04:53:49.97976  &    -17:46:24.2944  &    408.07 &   -642.82 &   82.52 &                         M2  & 12.4 & 10.9 & 6.6 &  6.2 &     77 \\
GJ 3634     &      10:58:35.133  &      -31:08:38.29  &   -583.00 &    -92.00 &   50.55 &                       M2.5  & 14.4 & 11.9 & 7.5 &  6.0 &     77 \\
GJ 3470     &       07:59:05.87  &       +15:23:29.5  &   -175.00 &    -52.00 &   39.68 &                       M1.5  & 13.5 & 12.3 & 8.0 &  6.0 &     98 \\

\end{longtable}
\end{landscape}
}

%%%%%%%%%%%%%%%%%%%%%%%%%%%%%%%%%%%%%%%%%%%%%%%%%%%%%%%%%%%%%%%%%%%%%%%%%%%%%%%

%%%%%%%%%%%%%%%%%%%%%%%%%%%%%%%%%%%%%%%%%%%%%%%%%%%%%%%%%%%%%%%%%%%%%%%%%%%%%%%
\longtab[2]{
\begin{longtable}{lrrrrrrrrrrrrr}
\caption{ \label{TableStellarParams} Stellar parameters}\\
\hline\hline
Name
& $\teffspt$
& $\teffmass$
& $\teffmal$
& [Fe/H]
& [Fe/H]
\\
& [K]
& [K]
& [K]
&
& source
\\
\hline
\endfirsthead

\caption{continued.}\\
\hline\hline
Name
& $\teffspt$
& $\teffmass$
& $\teffmal$
& [Fe/H]
& [Fe/H]
\\
& [K]
& [K]
& [K]
&
& source
\\
\hline

\hline
\endhead
\hline
\endfoot
GJ 1       &     3650 &      3546 &     3482 & -0.45 &        1 \\
GJ 1002    &     3070 &      2536 &      ... & -0.27 &        1 \\
GJ 12      &     3470 &      3333 &      ... & -0.29 &        1 \\
GJ 3049    &     3470 &      3253 &      ... & -0.13 &        1 \\
GJ 54.1    &     3370 &      2916 &      ... & -0.38 &        1 \\
L 707-74    &      ... &      3021 &      ... & -0.38 &        1 \\
GJ 87      &     3520 &      3641 &     3562 & -0.32 &        1 \\
GJ 105B    &     3310 &      3413 &      ... & -0.02 &        1 \\
CD-44-836A  &     3370 &      3304 &      ... & -0.07 &        1 \\
CD-44-836B  &     3200 &      3183 &      ... &   ... &        ... \\
GJ 3189    &     3470 &      3113 &      ... & -0.76 &        1 \\
GJ 3193B   &     3470 &      3491 &      ... & -0.34 &        1 \\
GJ 3207    &     3420 &       ... &      ... &   ... &        ... \\
GJ 1057    &     3200 &      3074 &      ... & -0.10 &        1 \\
GJ 145     &     3520 &      3471 &      ... & -0.28 &        1 \\
GJ 1061    &     3070 &      2752 &      ... & -0.09 &        1 \\
GJ 1065    &     3470 &      3226 &      ... & -0.23 &        1 \\
GJ 1068    &     3310 &      2849 &      ... & -0.43 &        1 \\
GJ 166C    &     3200 &      3328 &      ... & -0.12 &        1 \\
GJ 176     &     3520 &      3713 &     3603 & -0.01 &        1 \\
GJ 3323    &     3370 &      3110 &      ... & -0.24 &        1 \\
GJ 3325    &     3470 &      3419 &      ... & -0.19 &        1 \\
GJ 191     &     3720 &      3423 &     3587 & -0.85 &        1 \\
GJ 203     &     3420 &      3231 &      ... & -0.22 &        1 \\
GJ 205     &     3650 &       ... &     3800 &  0.19 &        1 \\
GJ 213     &     3370 &      3301 &      ... & -0.11 &        1 \\
GJ 229     &     3650 &       ... &     3779 & -0.03 &        1 \\
HIP 31293  &     3580 &      3615 &     3526 & -0.05 &        1 \\
HIP 31292  &     3470 &      3454 &      ... & -0.06 &        1 \\
GJ 3404A   &     3420 &      3542 &      ... & -0.02 &        1 \\
GJ 3405B   &     3370 &      3360 &      ... &   ... &        ... \\
GJ 250B    &     3580 &      3632 &     3557 & -0.08 &        1 \\
GJ 273     &     3420 &      3441 &     3342 & -0.01 &        1 \\
GJ 3459    &     3470 &      3444 &      ... & -0.22 &        1 \\
GJ 285     &     3310 &      3461 &      ... &  0.27 &        1 \\
GJ 299     &     3310 &      2942 &      ... & -0.53 &        1 \\
GJ 300     &     3370 &      3392 &      ... &  0.13 &        1 \\
GJ 2066    &     3580 &      3647 &     3575 & -0.17 &        1 \\
GJ 1123    &     3310 &      3269 &      ... &  0.15 &        1 \\
GJ 341     &     3850 &       ... &     3783 & -0.14 &        1 \\
GJ 1125    &     3420 &      3443 &      ... & -0.09 &        1 \\
GJ 357     &     3520 &      3480 &     3477 & -0.30 &        1 \\
GJ 358     &     3470 &      3592 &     3450 & -0.01 &        1 \\
GJ 367     &     3720 &      3700 &     3559 & -0.07 &        1 \\
GJ 1129    &     3370 &      3429 &      ... &  0.05 &        1 \\
GJ 382     &     3580 &       ... &     3653 &  0.02 &        1 \\
GJ 388     &     3310 &      3576 &     3473 &  0.12 &        1 \\
GJ 393     &     3520 &      3620 &     3544 & -0.20 &        1 \\
GJ 3618    &     3370 &       ... &      ... & -0.55 &        1 \\
GJ 402     &     3200 &      3385 &      ... &  0.03 &        1 \\
GJ 406     &     2900 &      2509 &      ... &  0.19 &        1 \\
GJ 413.1   &     3580 &      3655 &     3570 & -0.10 &        1 \\
GJ 433     &     3650 &      3675 &     3618 & -0.17 &        1 \\
GJ 436     &     3420 &      3624 &      ... &   ... &        ... \\
GJ 438     &     3850 &      3480 &     3647 & -0.36 &        1 \\
GJ 447     &     3310 &      3152 &     3382 & -0.17 &        1 \\
GJ 465     &     3580 &      3387 &     3403 & -0.62 &        1 \\
GJ 479     &     3470 &      3614 &     3476 &  0.01 &        1 \\
GJ 3737    &     3310 &      2997 &      ... & -0.27 &        1 \\
GJ 480.1   &     3470 &      3190 &      ... & -0.48 &        1 \\
GJ 486     &     3370 &      3472 &      ... &  0.03 &        1 \\
GJ 514     &     3720 &      3812 &     3728 & -0.16 &        1 \\
GJ 526     &     3370 &      3718 &     3609 & -0.22 &        1 \\
GJ 536     &     3650 &      3773 &     3685 & -0.14 &        1 \\
GJ 551     &     2900 &      2718 &     3555 &  0.16 &        1 \\
GJ 555     &     3370 &      3432 &      ... &  0.14 &        1 \\
GJ 569A    &     3520 &       ... &     3608 & -0.06 &        1 \\
GJ 581     &     3200 &      3454 &     3419 & -0.20 &        1 \\
GJ 588     &     3520 &      3670 &     3525 &  0.06 &        1 \\
GJ 618A    &     3470 &      3548 &     3451 & -0.06 &        1 \\
GJ 628     &     3470 &      3448 &     3345 & -0.02 &        1 \\
GJ 643     &     3420 &      3293 &      ... & -0.26 &        1 \\
GJ 644A    &      ... &       ... &     3463 &   ... &        ... \\
GJ 667C    &     3650 &      3455 &      ... & -0.50 &        1 \\
GJ 674     &     3470 &      3500 &     3484 & -0.23 &        1 \\
GJ 678.1A  &     3720 &       ... &     3815 & -0.14 &        1 \\
GJ 680     &     3470 &      3672 &     3585 & -0.19 &        1 \\
GJ 682     &     3420 &      3425 &     3393 &  0.10 &        1 \\
GJ 686     &     3720 &      3629 &     3677 & -0.35 &        1 \\
GJ 693     &     3580 &      3396 &     3390 & -0.28 &        1 \\
GJ 699     &     3370 &      3055 &      ... & -0.51 &        1 \\
GJ 701     &     3580 &      3683 &     3664 & -0.27 &        1 \\
GJ 1224    &     3310 &      2967 &      ... & -0.25 &        1 \\
GJ 4071    &     3310 &      3339 &      ... & -0.08 &        1 \\
GJ 729     &     3470 &      3163 &     3548 & -0.40 &        1 \\
GJ 1232    &     3310 &      3246 &      ... &  0.03 &        1 \\
GJ 752A    &     3470 &      3692 &     3551 &  0.05 &        1 \\
GJ 754     &     3310 &      3175 &      ... & -0.14 &        1 \\
GJ 1236    &     3470 &      3311 &      ... & -0.47 &        1 \\
GJ 1256    &     3310 &      3244 &      ... &  0.06 &        1 \\
GJ 803     &     3720 &       ... &      ... &   ... &        ... \\
GJ 4154    &     3520 &      3569 &      ... & -0.22 &        1 \\
LP 816-60   &     3370 &      3336 &      ... & -0.07 &        1 \\
GJ 832     &     3650 &      3631 &     3580 & -0.17 &        1 \\
GJ 846     &     3850 &       ... &     3835 &  0.01 &        1 \\
GJ 4248    &     3420 &      3349 &      ... & -0.13 &        1 \\
GJ 849     &     3420 &      3703 &     3486 &  0.22 &        1 \\
GJ 1265    &     3370 &      3152 &      ... & -0.20 &        1 \\
GJ 4274    &     3370 &      3191 &      ... &  0.10 &        1 \\
GJ 876     &     3200 &      3486 &     3357 &  0.14 &        1 \\
GJ 877     &     3470 &      3602 &     3428 &  0.00 &        1 \\
GJ 880     &     3580 &       ... &     3736 &  0.03 &        1 \\
GJ 887     &     3580 &      3674 &     3712 & -0.24 &        1 \\
GJ 4333    &     3370 &      3558 &      ... &  0.23 &        1 \\
GJ 908     &     3580 &      3595 &     3570 & -0.44 &        1 \\
LTT 9759   &      ... &       ... &     3581 &  0.17 &        1 \\
GJ 676A    &     3850 &       ... &      ... &  0.26 &        1 \\
GJ 1214    &     3310 &      3055 &      ... &  0.05 &        1 \\
HIP 12961  &     3850 &       ... &      ... &  0.22 &        1 \\
GJ 163     &     3420 &      3571 &      ... &  0.07 &        1 \\
GJ 317     &     3420 &      3610 &      ... &  0.22 &        1 \\
GJ 570B    &     3650 &       ... &      ... &  0.29 &        2 \\
GJ 160.2   &     3850 &       ... &      ... &   ... &        ... \\
GJ 180     &     3580 &      3590 &      ... & -0.20 &        2 \\
GJ 3634    &     3520 &      3641 &      ... & -0.07 &        1 \\
GJ 3470    &     3650 &      3642 &      ... &   ... &        ... \\
\end{longtable}
\tablebib{(1)~\citet{Neves2014}; (2)~\citet{Neves2012}.}
}
%%%%%%%%%%%%%%%%%%%%%%%%%%%%%%%%%%%%%%%%%%%%%%%%%%%%%%%%%%%%%%%%%%%%%%%%%%%%%%%

%%%%%%%%%%%%%%%%%%%%%%%%%%%%%%%%%%%%%%%%%%%%%%%%%%%%%%%%%%%%%%%%%%%%%%%%%%%%%%%

\longtab[3]{
\begin{longtable}{lrrrrrr}
\caption{ \label{TableResults} Results }\\
\hline\hline
% -------------------------------------------------------
Name
& $R'_\mathrm{HK}$\tablefootmark{a}
& $R'_\mathrm{HK}$\tablefootmark{a}
& $R'_\mathrm{HK}$\tablefootmark{a}
& $\mathcal{F}'_\mathrm{H\alpha}/\mathcal{F}_\mathrm{bol}$
& $\mathcal{F}'_\mathrm{H\alpha}/\mathcal{F}_\mathrm{bol}$
& $\mathcal{F}'_\mathrm{H\alpha}/\mathcal{F}_\mathrm{bol}$
\\
% -------------------------------------------------------
% -------------------------------------------------------
{}
& $\left( \teffspt \right)$
& $\left( \teffmass \right)$
& $\left( \teffmal \right)$
& $\left( \teffspt \right)$
& $\left( \teffmass \right)$
& $\left( \teffmal \right)$
\\
% -------------------------------------------------------
\hline
\endfirsthead

\caption{continued.}\\
\hline
\hline
% -------------------------------------------------------
Name
& $R'_\mathrm{HK}$\tablefootmark{a}
& $R'_\mathrm{HK}$\tablefootmark{a}
& $R'_\mathrm{HK}$\tablefootmark{a}
& $\mathcal{F}'_\mathrm{H\alpha}/\mathcal{F}_\mathrm{bol}$
& $\mathcal{F}'_\mathrm{H\alpha}/\mathcal{F}_\mathrm{bol}$
& $\mathcal{F}'_\mathrm{H\alpha}/\mathcal{F}_\mathrm{bol}$
\\
% -------------------------------------------------------
% -------------------------------------------------------
{}
& $\left( \teffspt \right)$
& $\left( \teffmass \right)$
& $\left( \teffmal \right)$
& $\left( \teffspt \right)$
& $\left( \teffmass \right)$
& $\left( \teffmal \right)$
\\
% -------------------------------------------------------
\hline
\endhead
\hline
\endfoot

  GJ 1      & -5.46   &        -5.54       &  -5.60           &   -1.95e-05              &     -1.87e-05           &        -1.82e-05  \\
  GJ 1002   & -5.31   &        -6.62       &    $\dots$       &           -6.77e-06      &             -1.57e-06   &                $\dots$  \\
  GJ 12     & -5.22   &        -5.35       &    $\dots$       &           -1.49e-05      &             -1.36e-05   &                $\dots$  \\
  GJ 3049   & -5.22   &        -5.43       &    $\dots$       &           -1.03e-05      &             -8.71e-06   &                $\dots$  \\
  GJ 54.1   & -4.36   &        -5.05       &    $\dots$       &            6.23e-05      &              2.63e-05   &                $\dots$  \\
 L 707-74   & $\dots$ &          -5.86     &      $\dots$     &               $\dots$    &               -1.00e-05 &                  $\dots$  \\
  GJ 87     & -5.28   &        -5.19       &  -5.24           &   -2.30e-05              &     -2.41e-05           &               -2.34e-05  \\
  GJ 105B   & -5.41   &        -5.31       &    $\dots$       &           -1.21e-05      &             -1.38e-05   &                $\dots$  \\
 CD-44-836A & -4.57   &        -4.63       &    $\dots$       &            5.03e-05      &              4.58e-05   &                $\dots$  \\
 CD-44-836B & -4.76   &        -4.78       &    $\dots$       &            1.82e-04      &              1.77e-04   &                $\dots$  \\
  GJ 3189   & -5.54   &        -6.13       &    $\dots$       &           -1.52e-05      &             -1.41e-05   &                $\dots$  \\
  GJ 3193B  & -5.40   &        -5.38       &    $\dots$       &           -1.53e-05      &             -1.54e-05   &                $\dots$  \\
  GJ 3207   & $\dots$ &            $\dots$ &          $\dots$ &                  $\dots$ &               $\dots$   &                $\dots$  \\
  GJ 1057   & -5.04   &        -5.21       &    $\dots$       &           -1.06e-07      &             -3.98e-07   &                $\dots$  \\
  GJ 145    & -4.92   &        -4.96       &    $\dots$       &           -9.62e-06      &             -9.40e-06   &                $\dots$  \\
  GJ 1061   & -5.17   &        -5.81       &    $\dots$       &           -3.90e-06      &             -1.37e-06   &                $\dots$  \\
  GJ 1065   & -5.16   &        -5.42       &    $\dots$       &           -1.05e-05      &             -9.35e-06   &                $\dots$  \\
  GJ 1068   & -5.32   &        -6.16       &    $\dots$       &           -7.39e-06      &             -4.93e-06   &                $\dots$  \\
  GJ 166C   & -4.41   &        -4.28       &    $\dots$       &            1.40e-04      &              1.71e-04   &                $\dots$  \\
  GJ 176    & -4.78   &        -4.68       &  -4.74           &   -1.59e-05              &     -1.78e-05           &  -1.69e-05  \\
  GJ 3323   & -4.50   &        -4.80       &    $\dots$       &            5.19e-05      &              3.47e-05   &                $\dots$  \\
  GJ 3325   & -5.14   &        -5.18       &    $\dots$       &           -1.54e-05      &             -1.47e-05   &                $\dots$  \\
  GJ 191    & -5.61   &        -5.96       &  -5.74           &   -1.91e-05              &     -1.93e-05           &  -1.92e-05  \\
  GJ 203    & -5.34   &        -5.56       &    $\dots$       &           -1.05e-05      &             -8.89e-06   &                $\dots$  \\
  GJ 205    & -4.59   &          $\dots$   &      -4.53       &           -2.28e-05      &               $\dots$   &          -2.55e-05  \\
  GJ 213    & -5.54   &        -5.62       &    $\dots$       &           -1.30e-05      &             -1.19e-05   &                $\dots$  \\
  GJ 229    & -4.70   &          $\dots$   &      -4.63       &           -2.86e-05      &               $\dots$   &          -3.07e-05  \\
  HIP 31293 & -4.89   &        -4.87       &  -4.92           &   -1.46e-05              &     -1.49e-05           &  -1.41e-05  \\
  HIP 31292 & -5.11   &        -5.12       &    $\dots$       &           -1.44e-05      &             -1.42e-05   &                $\dots$  \\
  GJ 3404A  & -5.37   &        -5.28       &    $\dots$       &           -1.38e-05      &             -1.60e-05   &                $\dots$  \\
  GJ 3405B  & -5.64   &        -5.65       &    $\dots$       &           -6.29e-05      &             -6.22e-05   &                $\dots$  \\
  GJ 250B   & -4.91   &        -4.88       &  -4.92           &   -1.99e-05              &     -2.05e-05           &  -1.95e-05  \\
  GJ 273    & -5.28   &        -5.26       &  -5.34           &   -1.46e-05              &     -1.50e-05           &  -1.33e-05  \\
  GJ 3459   & -5.20   &        -5.22       &    $\dots$       &           -1.52e-05      &             -1.50e-05   &                $\dots$  \\
  GJ 285    & -3.83   &        -3.74       &    $\dots$       &            3.33e-04      &              4.06e-04   &                $\dots$  \\
  GJ 299    & -5.39   &        -6.08       &    $\dots$       &           -1.24e-05      &             -8.42e-06   &                $\dots$  \\
  GJ 300    & -5.01   &        -4.99       &    $\dots$       &           -6.76e-06      &             -6.93e-06   &                $\dots$  \\
  GJ 2066   & -5.09   &        -5.05       &  -5.09           &   -2.22e-05              &     -2.34e-05           &  -2.22e-05  \\
  GJ 1123   & -4.95   &        -4.98       &    $\dots$       &         -8.16e-06        &           -7.66e-06     &              $\dots$  \\
  GJ 341    & -4.67   &          $\dots$   &      -4.71       &        -3.03e-05         &               $\dots$   &          -2.98e-05  \\
  GJ 1125   & -5.38   &        -5.37       &    $\dots$       &         -1.25e-05        &           -1.29e-05     &              $\dots$  \\
  GJ 357    & -5.53   &        -5.56       &  -5.57           &   -1.75e-05              &     -1.72e-05           &  -1.72e-05  \\
  GJ 358    & -4.50   &        -4.43       &  -4.51           &    4.13e-05              &      4.71e-05           &   4.03e-05  \\
  GJ 367    & -4.91   &        -4.92       &  -4.99           &   -2.43e-05              &     -2.41e-05           &  -2.18e-05  \\
  GJ 1129   & -5.08   &        -5.03       &    $\dots$       &        -1.09e-05         &          -1.17e-05      &             $\dots$  \\
  GJ 382    & -4.63   &          $\dots$   &      -4.60       &     -1.79e-05            &             $\dots$     &        -1.87e-05  \\
  GJ 388    & -4.17   &        -4.01       &  -4.06           &    1.42e-04              &      1.98e-04           &   1.76e-04  \\
  GJ 393    & -4.99   &        -4.92       &  -4.97           &   -1.83e-05              &     -1.93e-05           &  -1.86e-05  \\
  GJ 3618   & -5.32   &          $\dots$   &        $\dots$   &           -1.25e-05      &               $\dots$   &                $\dots$  \\
  GJ 402    & -5.29   &        -5.10       &    $\dots$       &           -6.54e-06      &             -8.18e-06   &                $\dots$  \\
  GJ 406    & -4.10   &        -4.97       &    $\dots$       &            1.52e-04      &              4.66e-05   &                $\dots$  \\
  GJ 413.1  & -5.08   &        -5.04       &  -5.09           &   -2.30e-05              &     -2.42e-05           &  -2.29e-05  \\
  GJ 433    & -5.06   &        -5.05       &  -5.08           &   -2.56e-05              &     -2.60e-05           &  -2.52e-05  \\
  GJ 436    & -5.36   &        -5.23       &    $\dots$       &           -1.62e-05      &             -1.95e-05   &                $\dots$  \\
  GJ 438    & -5.06   &        -5.32       &  -5.19           &   -2.55e-05              &     -2.28e-05           &  -2.47e-05  \\
  GJ 447    & -5.26   &        -5.45       &  -5.19           &   -9.74e-06              &     -8.65e-06           &  -1.01e-05  \\
  GJ 465    & -5.80   &        -6.07       &  -6.04           &   -2.51e-05              &     -2.30e-05           &  -2.31e-05  \\
  GJ 479    & -4.68   &        -4.60       &  -4.68           &    5.42e-06              &      7.00e-06           &   5.47e-06  \\
  GJ 3737   & -5.40   &        -5.92       &    $\dots$       &        -1.07e-05         &          -7.74e-06      &             $\dots$  \\
  GJ 480.1  & -5.17   &        -5.52       &    $\dots$       &           -1.28e-05      &             -1.19e-05   &                $\dots$  \\
  GJ 486    & -5.47   &        -5.38       &    $\dots$       &           -1.16e-05      &             -1.29e-05   &                $\dots$  \\
  GJ 514    & -4.86   &        -4.82       &  -4.86           &   -3.10e-05              &     -3.20e-05           &  -3.11e-05  \\
  GJ 526    & -5.28   &        -5.03       &  -5.09           &   -2.27e-05              &     -2.86e-05           &  -2.72e-05  \\
  GJ 536    & -4.88   &        -4.82       &  -4.86           &   -3.00e-05              &     -3.17e-05           &  -3.05e-05  \\
  GJ 551    & -4.59   &        -5.01       &  -3.92           &    4.34e-05              &      2.39e-05           &   1.41e-04  \\
  GJ 555    & -5.23   &        -5.19       &    $\dots$       &        -7.31e-06         &          -7.81e-06      &             $\dots$  \\
  GJ 569A   & -4.27   &          $\dots$   &      -4.22       &       5.44e-05           &              $\dots$    &          5.98e-05  \\
  GJ 581    & -5.96   &        -5.61       &  -5.64           &   -1.10e-05              &     -1.29e-05           &  -1.26e-05  \\
  GJ 588    & -5.07   &        -5.00       &  -5.06           &   -1.71e-05              &     -1.92e-05           &  -1.72e-05  \\
  GJ 618A   & -5.23   &        -5.18       &  -5.25           &   -1.93e-05              &     -2.08e-05           &  -1.89e-05  \\
  GJ 628    & -5.27   &        -5.29       &  -5.37           &   -1.28e-05              &     -1.25e-05           &  -1.12e-05  \\
  GJ 643    & -5.43   &        -5.57       &    $\dots$       &        -1.20e-05         &          -1.13e-05      &             $\dots$  \\
  GJ 667C   & -5.33   &        -5.50       &    $\dots$       &        -2.00e-05         &          -1.89e-05      &             $\dots$  \\
  GJ 674    & -4.84   &        -4.82       &  -4.83           &    4.42e-06              &      4.69e-06           &   4.55e-06  \\
  GJ 678.1A & -4.76   &          $\dots$   &      -4.71       &      -3.07e-05           &              $\dots$    &         -3.18e-05  \\
  GJ 680    & -5.12   &        -5.00       &  -5.05           &   -2.30e-05              &     -2.63e-05           &  -2.51e-05  \\
  GJ 682    & -5.17   &        -5.16       &  -5.18           &   -9.35e-06              &     -9.41e-06           &  -9.01e-06  \\
  GJ 686    & -5.14   &        -5.19       &  -5.16           &   -2.60e-05              &     -2.53e-05           &  -2.57e-05  \\
  GJ 693    & -5.47   &        -5.63       &  -5.64           &   -1.61e-05              &     -1.45e-05           &  -1.44e-05  \\
  GJ 699    & -5.56   &        -6.12       &    $\dots$       &           -1.35e-05      &             -1.20e-05   &                $\dots$  \\
  GJ 701    & -5.00   &        -4.94       &  -4.95           &   -2.81e-05              &     -2.98e-05           &  -2.96e-05  \\
  GJ 1224   & -4.20   &        -4.67       &    $\dots$       &            1.99e-04      &              1.03e-04   &                $\dots$  \\
  GJ 4071   & -4.22   &        -4.19       &    $\dots$       &            1.86e-04      &              1.94e-04   &                $\dots$  \\
  GJ 729    & -4.17   &        -4.50       &  -4.10           &    1.51e-04              &      1.02e-04           &   1.61e-04  \\
  GJ 1232   & -5.15   &        -5.21       &    $\dots$       &           -5.02e-06      &             -4.75e-06   &                $\dots$  \\
  GJ 752A   & -5.01   &        -4.90       &  -4.96           &   -1.60e-05              &     -1.93e-05           &  -1.74e-05  \\
  GJ 754    & -5.26   &        -5.43       &    $\dots$       &           -9.27e-06      &             -8.22e-06   &                $\dots$  \\
  GJ 1236   & -5.24   &        -5.42       &    $\dots$       &           -1.30e-05      &             -1.24e-05   &                $\dots$  \\
  GJ 1256   & -4.82   &        -4.89       &    $\dots$       &            7.36e-06      &              6.56e-06   &                $\dots$  \\
  GJ 803    & $\dots$ &            $\dots$ &          $\dots$ &                 $\dots$  &                $\dots$  &                 $\dots$  \\
  GJ 4154   & -5.04   &        -5.01       &    $\dots$       &           -1.29e-05      &             -1.31e-05   &                $\dots$  \\
 LP 816-60  & -5.08   &        -5.11       &    $\dots$       &           -8.99e-06      &             -8.72e-06   &                $\dots$  \\
  GJ 832    & -5.10   &        -5.11       &  -5.14           &   -2.55e-05              &     -2.52e-05           &  -2.43e-05  \\
  GJ 846    & -4.56   &          $\dots$   &      -4.56       &       -3.18e-05          &               $\dots$   &          -3.17e-05  \\
  GJ 4248   & -5.26   &        -5.32       &    $\dots$       &           -1.30e-05      &             -1.22e-05   &                $\dots$  \\
  GJ 849    & -5.10   &        -4.98       &  -5.06           &   -1.73e-05              &     -2.32e-05           &       -1.87e-05  \\
  GJ 1265   & -5.19   &        -5.45       &    $\dots$       &           -6.82e-06      &             -6.03e-06   &                $\dots$  \\
  GJ 4274   & -4.27   &        -4.43       &    $\dots$       &            1.85e-04      &              1.41e-04   &                $\dots$  \\
  GJ 876    & -5.40   &        -5.15       &  -5.23           &   -7.74e-06              &     -1.14e-05           &       -9.78e-06  \\
  GJ 877    & -5.21   &        -5.14       &  -5.24           &   -1.65e-05              &     -1.92e-05           &       -1.57e-05  \\
  GJ 880    & -4.77   &          $\dots$   &      -4.69       &       -2.74e-05          &               $\dots$   &          -3.06e-05  \\
  GJ 887    & -4.91   &        -4.85       &  -4.83           &   -2.56e-05              &     -2.70e-05           &       -2.75e-05  \\
  GJ 4333   & -5.01   &        -4.91       &    $\dots$       &           -5.48e-06      &             -6.58e-06   &                $\dots$  \\
  GJ 908    & -5.39   &        -5.37       &  -5.39           &   -2.45e-05              &     -2.46e-05           &       -2.44e-05  \\
  LTT 9759  & $\dots$ &            $\dots$ &        -4.77     &              $\dots$     &                $\dots$  &           -1.97e-05  \\
  GJ 676A   & -4.67   &          $\dots$   &        $\dots$   &            -3.66e-05     &                $\dots$  &                 $\dots$  \\
  GJ 1214   & -5.07   &        -5.39       &    $\dots$       &           -9.69e-06      &             -6.01e-06   &                $\dots$  \\
  HIP 12961 & -4.56   &          $\dots$   &        $\dots$   &            -3.69e-05     &                 $\dots$ &                  $\dots$  \\
  GJ 163    & -5.29   &        -5.20       &    $\dots$       &           -1.59e-05      &             -1.92e-05   &                $\dots$  \\
  GJ 317    & -5.05   &        -4.96       &    $\dots$       &           -9.09e-06      &             -1.19e-05   &                $\dots$  \\
  GJ 570B   & -4.79   &          $\dots$   &        $\dots$   &            -2.39e-05     &                $\dots$  &                 $\dots$  \\
  GJ 160.2  & -4.93   &          $\dots$   &        $\dots$   &             -5.73e-05    &                 $\dots$ &                  $\dots$  \\
  GJ 180    & -5.11   &        -5.11       &    $\dots$       &           -2.00e-05      &             -2.01e-05   &                $\dots$  \\
  GJ 3634   & -5.06   &        -4.99       &    $\dots$       &           -2.23e-05      &             -2.44e-05   &                $\dots$  \\
  GJ 3470   & -4.72   &        -4.73       &    $\dots$       &           -2.03e-05      &             -2.02e-05   &                $\dots$  \\

\end{longtable}
\tablefoot{ \tablefoottext{a}{Values are on a log scale.} }
}
%%%%%%%%%%%%%%%%%%%%%%%%%%%%%%%%%%%%%%%%%%%%%%%%%%%%%%%%%%%%%%%%%%%%%%%%%%%%%%%

%%%%%%%%%%%%%%%%%%%%%%%%%%%%%%%%%%%%%%%%%%%%%%%%%%%%%%%%%%%%%%%%%%%%%%%%%%%%%%%
\longtab[4]{
\begin{longtable}{lrr|rr|rr|rr|rr|rr}
\caption{ \label{TableRpHKCalib} PHOENIX-ACES grid calibrations }\\
\hline
\hline
{[Fe/H]}
& {}
& {}
& \multicolumn{2}{|c}{-1.00}
& \multicolumn{2}{|c}{-0.50}
& \multicolumn{2}{|c}{0.00}
& \multicolumn{2}{|c}{+0.50}
& \multicolumn{2}{|c}{+1.00}
\\
\hline
{}
& $T_\mathrm{eff}$
& $\log{g}$
& $\log{}$
& $\log{}$
& $\log{}$
& $\log{}$
& $\log{}$
& $\log{}$
& $\log{}$
& $\log{}$
& $\log{}$
& $\log{}$
\\
{}
& {[K]}
& {}
& $C_\mathrm{cf}$
& $R_\mathrm{HK,\,phot}$
& $C_\mathrm{cf}$
& $R_\mathrm{HK,\,phot}$
& $C_\mathrm{cf}$
& $R_\mathrm{HK,\,phot}$
& $C_\mathrm{cf}$
& $R_\mathrm{HK,\,phot}$
& $C_\mathrm{cf}$
& $R_\mathrm{HK,\,phot}$
\\
\hline
\endfirsthead
\caption{continued.}\\
\hline
\hline
{[Fe/H]}
& {}
& {}
& \multicolumn{2}{|c}{-1.00}
& \multicolumn{2}{|c}{-0.50}
& \multicolumn{2}{|c}{0.00}
& \multicolumn{2}{|c}{+0.50}
& \multicolumn{2}{|c}{+1.00}
\\
\hline
{}
& $T_\mathrm{eff}$
& $\log{g}$
& $\log{}$
& $\log{}$
& $\log{}$
& $\log{}$
& $\log{}$
& $\log{}$
& $\log{}$
& $\log{}$
& $\log{}$
& $\log{}$
\\
{}
& {[K]}
& {}
& $C_\mathrm{cf}$
& $R_\mathrm{HK,\,phot}$
& $C_\mathrm{cf}$
& $R_\mathrm{HK,\,phot}$
& $C_\mathrm{cf}$
& $R_\mathrm{HK,\,phot}$
& $C_\mathrm{cf}$
& $R_\mathrm{HK,\,phot}$
& $C_\mathrm{cf}$
& $R_\mathrm{HK,\,phot}$
\\
\hline
\endhead
\hline
\endfoot
{} & 2300 & 4.0 & -2.95 & -7.63  & -2.65 & -7.27  & -2.34 & -6.86  & -2.13 & -6.50  & -2.00 & -6.26 \\
{} & 2400 & 4.0 & -2.68 & -7.34  & -2.40 & -6.97  & -2.14 & -6.61  & -1.96 & -6.31  & -1.86 & -6.13 \\
{} & 2500 & 4.0 & -2.42 & -7.04  & -2.17 & -6.69  & -1.96 & -6.40  & -1.81 & -6.16  & -1.73 & -6.03 \\
{} & 2600 & 4.0 & -2.26 & -6.82  & -2.01 & -6.52  & -1.80 & -6.24  & -1.67 & -6.06  & -1.61 & -5.96 \\
{} & 2700 & 4.0 & -2.04 & -6.61  & -1.83 & -6.34  & -1.66 & -6.12  & -1.55 & -5.98  & -1.50 & -5.91 \\
{} & 2800 & 4.0 & -1.86 & -6.43  & -1.68 & -6.21  & -1.53 & -6.02  & -1.44 & -5.92  & -1.40 & -5.88 \\
{} & 2900 & 4.0 & -1.69 & -6.30  & -1.54 & -6.10  & -1.42 & -5.96  & -1.34 & -5.89  & -1.31 & -5.87 \\
{} & 3000 & 4.0 & -1.55 & -6.19  & -1.42 & -6.03  & -1.31 & -5.92  & -1.24 & -5.88  & -1.24 & -5.88 \\
{} & 3100 & 4.0 & -1.42 & -6.11  & -1.31 & -5.99  & -1.22 & -5.91  & -1.17 & -5.89  & -1.17 & -5.90 \\
{} & 3200 & 4.0 & -1.30 & -6.05  & -1.21 & -5.96  & -1.14 & -5.92  & -1.10 & -5.91  & -1.12 & -5.93 \\
{} & 3300 & 4.0 & -1.20 & -6.01  & -1.13 & -5.96  & -1.07 & -5.94  & -1.05 & -5.95  & -1.06 & -5.98 \\
{} & 3400 & 4.0 & -1.10 & -5.97  & -1.06 & -5.93  & -1.02 & -5.95  & -1.00 & -5.99  & -1.01 & -6.02 \\
{} & 3500 & 4.0 & -1.01 & -5.92  & -0.99 & -5.88  & -0.97 & -5.91  & -0.97 & -5.99  & -0.99 & -6.04 \\
{} & 3600 & 4.0 & -0.93 & -5.85  & -0.93 & -5.82  & -0.94 & -5.86  & -0.94 & -5.95  & -0.96 & -6.01 \\
{} & 3700 & 4.0 & -0.86 & -5.74  & -0.88 & -5.77  & -0.90 & -5.81  & -0.91 & -5.88  & -0.92 & -5.96 \\
{} & 3800 & 4.0 & -0.79 & -5.62  & -0.82 & -5.73  & -0.85 & -5.78  & -0.87 & -5.82  & -0.89 & -5.85 \\
{} & 3900 & 4.0 & -0.72 & -5.54  & -0.76 & -5.70  & -0.80 & -5.74  & -0.83 & -5.77  & -0.85 & -5.79 \\
{} & 4000 & 4.0 & -0.65 & -5.46  & -0.70 & -5.66  & -0.74 & -5.72  & -0.78 & -5.75  & -0.82 & -5.75 \\
{} & 4100 & 4.0 & -0.58 & -5.45  & -0.63 & -5.63  & -0.68 & -5.69  & -0.73 & -5.72  & -0.78 & -5.73 \\
{} & 4200 & 4.0 & -0.51 & -5.40  & -0.56 & -5.60  & -0.61 & -5.67  & -0.67 & -5.70  & -0.73 & -5.72 \\
{} & 4300 & 4.0 & -0.44 & -5.38  & -0.48 & -5.54  & -0.55 & -5.63  & -0.61 & -5.69  & -0.68 & -5.71 \\
{} & 4400 & 4.0 & -0.37 & -5.30  & -0.42 & -5.46  & -0.48 & -5.58  & -0.55 & -5.66  & -0.63 & -5.70 \\
{} & 4500 & 4.0 & -0.31 & -5.21  & -0.35 & -5.38  & -0.42 & -5.51  & -0.49 & -5.62  & -0.58 & -5.68 \\
{} & 4600 & 4.0 & -0.26 & -5.13  & -0.30 & -5.29  & -0.36 & -5.44  & -0.43 & -5.55  & -0.52 & -5.65 \\
{} & 4700 & 4.0 & -0.21 & -5.05  & -0.25 & -5.22  & -0.30 & -5.36  & -0.38 & -5.48  & -0.46 & -5.61 \\
{} & 4800 & 4.0 & -0.17 & -4.98  & -0.20 & -5.14  & -0.25 & -5.28  & -0.32 & -5.40  & -0.41 & -5.55 \\
{} & 4900 & 4.0 & -0.12 & -4.91  & -0.16 & -5.06  & -0.21 & -5.19  & -0.27 & -5.31  & -0.37 & -5.47 \\
{} & 5000 & 4.0 & -0.08 & -4.82  & -0.11 & -4.99  & -0.16 & -5.11  & -0.23 & -5.23  & -0.31 & -5.41 \\
{} & 5100 & 4.0 & -0.06 & -4.62  & -0.09 & -4.71  & -0.13 & -4.75  & -0.19 & -4.76  & -0.27 & -4.76 \\
{} & 5200 & 4.0 & -0.03 & -4.59  & -0.05 & -4.63  & -0.09 & -4.70  & -0.15 & -4.70  & -0.23 & -4.67 \\
{} & 5300 & 4.0 &  0.00 & -4.53  & -0.02 & -4.60  & -0.05 & -4.64  & -0.11 & -4.65  & -0.18 & -4.62 \\
{} & 5400 & 4.0 &  0.03 & -4.49  &  0.02 & -4.55  & -0.02 & -4.60  & -0.07 & -4.60  & -0.14 & -4.57 \\
{} & 5500 & 4.0 &  0.06 & -4.45  &  0.04 & -4.51  &  0.01 & -4.55  & -0.03 & -4.55  & -0.11 & -4.52 \\
{} & 5600 & 4.0 &  0.08 & -4.42  &  0.06 & -4.48  &  0.05 & -4.51  & -0.02 & -4.52  & -0.09 & -4.50 \\
{} & 5700 & 4.0 &  0.11 & -4.38  &  0.10 & -4.44  &  0.07 & -4.47  &  0.01 & -4.47  & -0.05 & -4.46 \\
{} & 5800 & 4.0 &  0.13 & -4.35  &  0.11 & -4.41  &  0.10 & -4.44  &  0.05 & -4.44  & -0.02 & -4.42 \\
{} & 5900 & 4.0 &  0.15 & -4.32  &  0.14 & -4.38  &  0.12 & -4.41  &  0.07 & -4.41  &  0.01 & -4.40 \\
{} & 6000 & 4.0 &  0.17 & -4.28  &  0.15 & -4.34  &  0.15 & -4.38  &  0.10 & -4.38  &  0.04 & -4.35 \\
{} & 6100 & 4.0 &  0.19 & -4.25  &  0.17 & -4.32  &  0.16 & -4.36  &  0.13 & -4.35  &  0.06 & -4.33 \\
{} & 6200 & 4.0 &  0.21 & -4.22  &  0.21 & -4.29  &  0.19 & -4.33  &  0.15 & -4.33  &  0.10 & -4.30 \\
{} & 6300 & 4.0 &  0.22 & -4.18  &  0.23 & -4.26  &  0.21 & -4.30  &  0.18 & -4.30  &  0.12 & -4.28 \\
{} & 6400 & 4.0 &  0.25 & -4.16  &  0.24 & -4.23  &  0.23 & -4.27  &  0.19 & -4.28  &  0.15 & -4.25 \\
{} & 6500 & 4.0 &  0.26 & -4.13  &  0.26 & -4.19  &  0.25 & -4.25  &  0.22 & -4.26  &  0.17 & -4.23 \\
{} & 6600 & 4.0 &  0.28 & -4.09  &  0.27 & -4.16  &  0.26 & -4.21  &  0.24 & -4.23  &  0.19 & -4.21 \\
{} & 6700 & 4.0 &  0.29 & -4.06  &  0.29 & -4.13  &  0.28 & -4.19  &  0.25 & -4.20  &  0.21 & -4.19 \\
{} & 6800 & 4.0 &  0.30 & -4.03  &  0.30 & -4.10  &  0.29 & -4.16  &  0.27 & -4.18  &  0.22 & -4.17 \\
{} & 6900 & 4.0 &  0.31 & -3.99  &  0.31 & -4.06  &  0.30 & -4.13  &  0.29 & -4.15  &  0.23 & -4.15 \\
{} & 7000 & 4.0 &  0.32 & -3.96  &  0.32 & -4.03  &  0.32 & -4.10  &  0.30 & -4.12  &  0.25 & -4.13 \\
{} & 7200 & 4.0 &  0.34 & -3.89  &  0.34 & -3.97  &  0.34 & -4.03  &  0.33 & -4.06  &  0.29 & -4.07 \\
{} & 2300 & 4.5 & -3.26 & -7.86 & -2.92 & -7.53 & -2.56 & -7.15 & -2.27 & -6.73 & -2.08 & -6.41 \\
{} & 2400 & 4.5 & -2.94 & -7.60 & -2.65 & -7.28 & -2.33 & -6.89 & -2.08 & -6.50 & -1.92 & -6.22 \\
{} & 2500 & 4.5 & -2.66 & -7.32 & -2.40 & -6.99 & -2.13 & -6.64 & -1.91 & -6.29 & -1.78 & -6.07 \\
{} & 2600 & 4.5 & -2.50 & -7.04 & -2.15 & -6.70 & -1.94 & -6.42 & -1.76 & -6.13 & -1.65 & -5.96 \\
{} & 2700 & 4.5 & -2.26 & -6.80 & -2.01 & -6.54 & -1.78 & -6.23 & -1.62 & -6.01 & -1.53 & -5.87 \\
{} & 2800 & 4.5 & -2.04 & -6.58 & -1.82 & -6.34 & -1.63 & -6.09 & -1.49 & -5.92 & -1.42 & -5.81 \\
{} & 2900 & 4.5 & -1.84 & -6.42 & -1.66 & -6.20 & -1.50 & -5.99 & -1.38 & -5.85 & -1.31 & -5.77 \\
{} & 3000 & 4.5 & -1.67 & -6.28 & -1.51 & -6.08 & -1.37 & -5.91 & -1.28 & -5.80 & -1.22 & -5.75 \\
{} & 3100 & 4.5 & -1.52 & -6.16 & -1.39 & -5.99 & -1.27 & -5.86 & -1.19 & -5.78 & -1.15 & -5.75 \\
{} & 3200 & 4.5 & -1.38 & -6.06 & -1.28 & -5.93 & -1.18 & -5.83 & -1.11 & -5.77 & -1.08 & -5.76 \\
{} & 3300 & 4.5 & -1.26 & -5.99 & -1.18 & -5.88 & -1.10 & -5.82 & -1.04 & -5.79 & -1.03 & -5.79 \\
{} & 3400 & 4.5 & -1.15 & -5.93 & -1.09 & -5.84 & -1.03 & -5.81 & -0.99 & -5.82 & -0.98 & -5.83 \\
{} & 3500 & 4.5 & -1.05 & -5.87 & -1.01 & -5.79 & -0.97 & -5.78 & -0.95 & -5.82 & -0.94 & -5.86 \\
{} & 3600 & 4.5 & -0.96 & -5.79 & -0.94 & -5.73 & -0.92 & -5.73 & -0.91 & -5.79 & -0.91 & -5.86 \\
{} & 3700 & 4.5 & -0.87 & -5.67 & -0.88 & -5.66 & -0.88 & -5.67 & -0.88 & -5.74 & -0.88 & -5.83 \\
{} & 3800 & 4.5 & -0.80 & -5.52 & -0.82 & -5.59 & -0.83 & -5.63 & -0.84 & -5.69 & -0.85 & -5.74 \\
{} & 3900 & 4.5 & -0.72 & -5.40 & -0.75 & -5.54 & -0.79 & -5.59 & -0.81 & -5.65 & -0.82 & -5.67 \\
{} & 4000 & 4.5 & -0.65 & -5.28 & -0.70 & -5.49 & -0.74 & -5.57 & -0.77 & -5.61 & -0.79 & -5.62 \\
{} & 4100 & 4.5 & -0.59 & -5.21 & -0.64 & -5.45 & -0.68 & -5.56 & -0.72 & -5.58 & -0.75 & -5.59 \\
{} & 4200 & 4.5 & -0.52 & -5.22 & -0.57 & -5.42 & -0.63 & -5.53 & -0.67 & -5.57 & -0.71 & -5.57 \\
{} & 4300 & 4.5 & -0.46 & -5.18 & -0.51 & -5.40 & -0.56 & -5.51 & -0.62 & -5.55 & -0.67 & -5.56 \\
{} & 4400 & 4.5 & -0.40 & -5.17 & -0.44 & -5.35 & -0.50 & -5.46 & -0.56 & -5.52 & -0.62 & -5.55 \\
{} & 4500 & 4.5 & -0.33 & -5.11 & -0.38 & -5.28 & -0.43 & -5.40 & -0.50 & -5.48 & -0.57 & -5.53 \\
{} & 4600 & 4.5 & -0.28 & -5.06 & -0.32 & -5.20 & -0.37 & -5.34 & -0.44 & -5.43 & -0.51 & -5.51 \\
{} & 4700 & 4.5 & -0.22 & -4.95 & -0.26 & -5.12 & -0.31 & -5.26 & -0.38 & -5.37 & -0.46 & -5.46 \\
{} & 4800 & 4.5 & -0.18 & -4.92 & -0.21 & -5.06 & -0.27 & -5.19 & -0.33 & -5.30 & -0.41 & -5.42 \\
{} & 4900 & 4.5 & -0.13 & -4.85 & -0.17 & -4.99 & -0.22 & -5.12 & -0.28 & -5.22 & -0.35 & -5.34 \\
{} & 5000 & 4.5 & -0.09 & -4.78 & -0.12 & -4.93 & -0.17 & -5.04 & -0.23 & -5.15 & -0.30 & -5.27 \\
{} & 5100 & 4.5 & -0.08 & -4.62 & -0.10 & -4.71 & -0.14 & -4.76 & -0.20 & -4.77 & -0.27 & -4.76 \\
{} & 5200 & 4.5 & -0.04 & -4.57 & -0.07 & -4.64 & -0.10 & -4.70 & -0.16 & -4.72 & -0.23 & -4.70 \\
{} & 5300 & 4.5 & -0.01 & -4.55 & -0.03 & -4.59 & -0.06 & -4.65 & -0.11 & -4.66 & -0.18 & -4.64 \\
{} & 5400 & 4.5 &  0.02 & -4.48 & -0.00 & -4.55 & -0.03 & -4.60 & -0.10 & -4.63 & -0.14 & -4.59 \\
{} & 5500 & 4.5 &  0.04 & -4.44 &  0.03 & -4.50 &  0.00 & -4.55 & -0.04 & -4.55 & -0.10 & -4.54 \\
{} & 5600 & 4.5 &  0.06 & -4.40 &  0.05 & -4.47 &  0.03 & -4.51 & -0.01 & -4.51 & -0.10 & -4.52 \\
{} & 5700 & 4.5 &  0.09 & -4.37 &  0.08 & -4.43 &  0.06 & -4.47 &  0.01 & -4.49 & -0.06 & -4.48 \\
{} & 5800 & 4.5 &  0.11 & -4.34 &  0.10 & -4.40 &  0.09 & -4.44 &  0.05 & -4.45 & -0.02 & -4.45 \\
{} & 5900 & 4.5 &  0.14 & -4.31 &  0.12 & -4.38 &  0.10 & -4.42 &  0.07 & -4.42 &  0.01 & -4.40 \\
{} & 6000 & 4.5 &  0.16 & -4.28 &  0.15 & -4.35 &  0.13 & -4.38 &  0.09 & -4.39 &  0.04 & -4.37 \\
{} & 6100 & 4.5 &  0.18 & -4.25 &  0.17 & -4.32 &  0.16 & -4.36 &  0.12 & -4.37 &  0.07 & -4.34 \\
{} & 6200 & 4.5 &  0.20 & -4.22 &  0.19 & -4.29 &  0.18 & -4.33 &  0.14 & -4.34 &  0.09 & -4.31 \\
{} & 6300 & 4.5 &  0.22 & -4.19 &  0.21 & -4.26 &  0.19 & -4.31 &  0.17 & -4.31 &  0.12 & -4.29 \\
{} & 6400 & 4.5 &  0.23 & -4.17 &  0.23 & -4.23 &  0.22 & -4.27 &  0.18 & -4.29 &  0.14 & -4.27 \\
{} & 6500 & 4.5 &  0.24 & -4.14 &  0.24 & -4.20 &  0.23 & -4.25 &  0.21 & -4.26 &  0.16 & -4.24 \\
{} & 6600 & 4.5 &  0.26 & -4.11 &  0.25 & -4.17 &  0.25 & -4.22 &  0.22 & -4.24 &  0.18 & -4.22 \\
{} & 6700 & 4.5 &  0.27 & -4.08 &  0.27 & -4.14 &  0.26 & -4.20 &  0.24 & -4.22 &  0.20 & -4.20 \\
{} & 6800 & 4.5 &  0.28 & -4.05 &  0.28 & -4.11 &  0.28 & -4.17 &  0.26 & -4.19 &  0.22 & -4.18 \\
{} & 6900 & 4.5 &  0.29 & -4.02 &  0.29 & -4.08 &  0.29 & -4.14 &  0.27 & -4.17 &  0.24 & -4.16 \\
{} & 7000 & 4.5 &  0.30 & -3.99 &  0.30 & -4.06 &  0.30 & -4.11 &  0.29 & -4.14 &  0.26 & -4.14 \\
{} & 7200 & 4.5 &  0.31 & -3.93 &  0.32 & -4.00 &  0.32 & -4.06 &  0.31 & -4.09 &  0.28 & -4.10 \\
{} & 2300 & 5.0 & -3.49 & -7.96 & -3.23 & -7.69 & -2.83 & -7.36 & -2.46 & -6.96 & -2.21 & -6.61 \\
{} & 2400 & 5.0 & -3.29 & -7.80 & -2.99 & -7.54 & -2.58 & -7.19 & -2.26 & -6.76 & -2.04 & -6.41 \\
{} & 2500 & 5.0 & -3.04 & -7.55 & -2.68 & -7.29 & -2.35 & -6.93 & -2.06 & -6.52 & -1.88 & -6.22 \\
{} & 2600 & 5.0 & -2.78 & -7.27 & -2.41 & -7.02 & -2.14 & -6.67 & -1.88 & -6.31 & -1.73 & -6.05 \\
{} & 2700 & 5.0 & -2.53 & -7.01 & -2.25 & -6.76 & -1.94 & -6.46 & -1.73 & -6.13 & -1.59 & -5.92 \\
{} & 2800 & 5.0 & -2.28 & -6.75 & -2.03 & -6.54 & -1.77 & -6.25 & -1.59 & -5.99 & -1.47 & -5.82 \\
{} & 2900 & 5.0 & -2.05 & -6.56 & -1.83 & -6.34 & -1.62 & -6.09 & -1.46 & -5.88 & -1.36 & -5.75 \\
{} & 3000 & 5.0 & -1.84 & -6.40 & -1.65 & -6.19 & -1.48 & -5.97 & -1.34 & -5.80 & -1.25 & -5.69 \\
{} & 3100 & 5.0 & -1.65 & -6.25 & -1.50 & -6.06 & -1.35 & -5.88 & -1.24 & -5.74 & -1.16 & -5.66 \\
{} & 3200 & 5.0 & -1.49 & -6.13 & -1.37 & -5.96 & -1.24 & -5.81 & -1.15 & -5.70 & -1.09 & -5.64 \\
{} & 3300 & 5.0 & -1.34 & -6.02 & -1.25 & -5.88 & -1.15 & -5.76 & -1.07 & -5.69 & -1.02 & -5.65 \\
{} & 3400 & 5.0 & -1.22 & -5.93 & -1.15 & -5.81 & -1.07 & -5.73 & -1.00 & -5.69 & -0.97 & -5.67 \\
{} & 3500 & 5.0 & -1.10 & -5.85 & -1.05 & -5.75 & -1.00 & -5.69 & -0.95 & -5.68 & -0.92 & -5.69 \\
{} & 3600 & 5.0 & -1.00 & -5.77 & -0.97 & -5.68 & -0.93 & -5.63 & -0.90 & -5.65 & -0.88 & -5.70 \\
{} & 3700 & 5.0 & -0.91 & -5.65 & -0.90 & -5.60 & -0.88 & -5.57 & -0.86 & -5.60 & -0.85 & -5.68 \\
{} & 3800 & 5.0 & -0.82 & -5.50 & -0.82 & -5.53 & -0.83 & -5.52 & -0.82 & -5.56 & -0.82 & -5.63 \\
{} & 3900 & 5.0 & -0.74 & -5.32 & -0.76 & -5.43 & -0.78 & -5.48 & -0.79 & -5.53 & -0.79 & -5.58 \\
{} & 4000 & 5.0 & -0.67 & -5.18 & -0.69 & -5.35 & -0.73 & -5.44 & -0.75 & -5.50 & -0.76 & -5.52 \\
{} & 4100 & 5.0 & -0.60 & -5.07 & -0.64 & -5.32 & -0.68 & -5.41 & -0.71 & -5.47 & -0.73 & -5.48 \\
{} & 4200 & 5.0 & -0.54 & -5.01 & -0.58 & -5.27 & -0.63 & -5.39 & -0.67 & -5.46 & -0.70 & -5.45 \\
{} & 4300 & 5.0 & -0.47 & -5.00 & -0.52 & -5.24 & -0.58 & -5.38 & -0.63 & -5.41 & -0.66 & -5.42 \\
{} & 4400 & 5.0 & -0.42 & -4.98 & -0.46 & -5.23 & -0.52 & -5.33 & -0.58 & -5.38 & -0.62 & -5.41 \\
{} & 4500 & 5.0 & -0.36 & -4.95 & -0.40 & -5.13 & -0.46 & -5.32 & -0.52 & -5.38 & -0.58 & -5.40 \\
{} & 4600 & 5.0 & -0.30 & -4.95 & -0.35 & -5.14 & -0.40 & -5.26 & -0.46 & -5.33 & -0.52 & -5.38 \\
{} & 4700 & 5.0 & -0.25 & -4.89 & -0.29 & -5.07 & -0.34 & -5.19 & -0.40 & -5.27 & -0.47 & -5.34 \\
{} & 4800 & 5.0 & -0.20 & -4.85 & -0.24 & -5.00 & -0.28 & -5.12 & -0.35 & -5.21 & -0.42 & -5.29 \\
{} & 4900 & 5.0 & -0.15 & -4.81 & -0.19 & -4.94 & -0.24 & -5.05 & -0.29 & -5.15 & -0.36 & -5.24 \\
{} & 5000 & 5.0 & -0.11 & -4.74 & -0.14 & -4.88 & -0.19 & -4.99 & -0.25 & -5.08 & -0.31 & -5.18 \\
{} & 5100 & 5.0 & -0.09 & -4.61 & -0.12 & -4.70 & -0.16 & -4.78 & -0.22 & -4.80 & -0.28 & -4.79 \\
{} & 5200 & 5.0 & -0.06 & -4.55 & -0.08 & -4.64 & -0.12 & -4.72 & -0.17 & -4.73 & -0.24 & -4.73 \\
{} & 5300 & 5.0 & -0.03 & -4.51 & -0.05 & -4.59 & -0.08 & -4.66 & -0.13 & -4.67 & -0.20 & -4.68 \\
{} & 5400 & 5.0 &  0.00 & -4.46 & -0.02 & -4.54 & -0.04 & -4.60 & -0.09 & -4.62 & -0.15 & -4.62 \\
{} & 5500 & 5.0 &  0.03 & -4.43 &  0.02 & -4.50 & -0.01 & -4.56 & -0.06 & -4.58 & -0.12 & -4.57 \\
{} & 5600 & 5.0 &  0.05 & -4.39 &  0.03 & -4.46 &  0.02 & -4.52 & -0.04 & -4.55 & -0.10 & -4.55 \\
{} & 5700 & 5.0 &  0.07 & -4.35 &  0.06 & -4.43 &  0.05 & -4.48 & -0.01 & -4.50 & -0.07 & -4.49 \\
{} & 5800 & 5.0 &  0.10 & -4.32 &  0.09 & -4.39 &  0.07 & -4.44 &  0.02 & -4.47 & -0.04 & -4.45 \\
{} & 5900 & 5.0 &  0.12 & -4.29 &  0.11 & -4.36 &  0.10 & -4.41 &  0.05 & -4.43 & -0.00 & -4.42 \\
{} & 6000 & 5.0 &  0.14 & -4.26 &  0.13 & -4.33 &  0.12 & -4.38 &  0.08 & -4.40 &  0.03 & -4.38 \\
{} & 6100 & 5.0 &  0.16 & -4.24 &  0.15 & -4.30 &  0.14 & -4.35 &  0.10 & -4.36 &  0.06 & -4.35 \\
{} & 6200 & 5.0 &  0.18 & -4.21 &  0.18 & -4.27 &  0.16 & -4.33 &  0.13 & -4.34 &  0.08 & -4.33 \\
{} & 6300 & 5.0 &  0.20 & -4.18 &  0.19 & -4.25 &  0.18 & -4.29 &  0.15 & -4.31 &  0.10 & -4.30 \\
{} & 6400 & 5.0 &  0.21 & -4.16 &  0.21 & -4.23 &  0.20 & -4.27 &  0.17 & -4.29 &  0.13 & -4.27 \\
{} & 6500 & 5.0 &  0.23 & -4.13 &  0.23 & -4.20 &  0.22 & -4.25 &  0.19 & -4.27 &  0.15 & -4.25 \\
{} & 6600 & 5.0 &  0.24 & -4.12 &  0.24 & -4.17 &  0.23 & -4.22 &  0.21 & -4.24 &  0.17 & -4.23 \\
{} & 6700 & 5.0 &  0.25 & -4.09 &  0.25 & -4.15 &  0.25 & -4.20 &  0.23 & -4.22 &  0.19 & -4.21 \\
{} & 6800 & 5.0 &  0.26 & -4.07 &  0.26 & -4.12 &  0.26 & -4.17 &  0.24 & -4.20 &  0.21 & -4.19 \\
{} & 6900 & 5.0 &  0.27 & -4.04 &  0.28 & -4.10 &  0.27 & -4.15 &  0.26 & -4.18 &  0.23 & -4.17 \\
{} & 7000 & 5.0 &  0.28 & -4.02 &  0.28 & -4.07 &  0.29 & -4.12 &  0.27 & -4.16 &  0.24 & -4.15 \\
{} & 7200 & 5.0 &  0.30 & -3.97 &  0.30 & -4.02 &  0.30 & -4.07 &  0.30 & -4.11 &  0.27 & -4.11 \\
\end{longtable}
}
%%%%%%%%%%%%%%%%%%%%%%%%%%%%%%%%%%%%%%%%%%%%%%%%%%%%%%%%%%%%%%%%%%%%%%%%%%%%%%%

%%%%%%%%%%%%%%%%%%%%%%%%%%%%%%%%%%%%%%%%%%%%%%%%%%%%%%%%%%%%%%%%%%%%%%%%%%%%%%%
\longtab[5]{
\begin{longtable}{llrr}
\caption{ \label{TableChi} Chi values of PHOENIX model atmospheres }\\
\hline
\hline

$\teff$
& [Fe/H]
& $\chi_\mathrm{CaIIHK}$
& $\chi_\mathrm{H\alpha}$ \\

{[K]} 
&       
& $\times 10^{-4}$      
& $\times 10^{-4}$ \\

\hline
\endfirsthead

\caption{continued.}\\
\hline
\hline

$\teff$
& [Fe/H]
& $\chi_\mathrm{CaIIHK}$
& $\chi_\mathrm{H\alpha}$ \\

{[K]} 
&       
& $\times 10^{-4}$      
& $\times 10^{-4}$ \\

\hline

\hline
\endhead
\hline
\endfoot

2300 & -1.0 & 0.000551 & 0.202107 \\
2400 & -1.0 & 0.000788 & 0.192654 \\
2500 & -1.0 & 0.001321 & 0.257018 \\
2600 & -1.0 & 0.002428 & 0.374069 \\
2700 & -1.0 & 0.004713 & 0.573781 \\
2800 & -1.0 & 0.009718 & 0.857325 \\
2900 & -1.0 & 0.018792 & 1.180935 \\
3000 & -1.0 & 0.034808 & 1.526253 \\
3100 & -1.0 & 0.060607 & 1.872651 \\
3200 & -1.0 & 0.096861 & 2.212708 \\
3300 & -1.0 & 0.143898 & 2.551446 \\
3400 & -1.0 & 0.200437 & 2.881837 \\
3500 & -1.0 & 0.266487 & 3.204793 \\
3600 & -1.0 & 0.344311 & 3.524350 \\
3700 & -1.0 & 0.437417 & 3.839320 \\
3800 & -1.0 & 0.549425 & 4.150202 \\
3900 & -1.0 & 0.683523 & 4.462624 \\
4000 & -1.0 & 0.827883 & 4.754775 \\
4100 & -1.0 & 0.987884 & 5.042933 \\
4200 & -1.0 & 1.146199 & 5.301651 \\
4300 & -1.0 & 1.322262 & 5.590931 \\
2300 & -0.5 & 0.000998 & 0.171112 \\
2400 & -0.5 & 0.001491 & 0.157325 \\
2500 & -0.5 & 0.002967 & 0.216378 \\
2600 & -0.5 & 0.005836 & 0.330543 \\
2700 & -0.5 & 0.009153 & 0.403176 \\
2800 & -0.5 & 0.017915 & 0.628417 \\
2900 & -0.5 & 0.034500 & 0.940191 \\
3000 & -0.5 & 0.059440 & 1.295086 \\
3100 & -0.5 & 0.094095 & 1.678432 \\
3200 & -0.5 & 0.137106 & 2.056224 \\
3300 & -0.5 & 0.187938 & 2.429038 \\
3400 & -0.5 & 0.244146 & 2.784152 \\
3500 & -0.5 & 0.306810 & 3.129856 \\
3600 & -0.5 & 0.375939 & 3.460377 \\
3700 & -0.5 & 0.452755 & 3.778883 \\
3800 & -0.5 & 0.542023 & 4.091369 \\
3900 & -0.5 & 0.638553 & 4.381635 \\
4000 & -0.5 & 0.750407 & 4.663454 \\
4100 & -0.5 & 0.869224 & 4.931749 \\
4200 & -0.5 & 0.993874 & 5.173299 \\
4300 & -0.0 & 1.141496 & 5.448736 \\
2300 & 0.0 & 0.002397 & 0.188068 \\
2400 & 0.0 & 0.003729 & 0.203213 \\
2500 & 0.0 & 0.006476 & 0.186438 \\
2600 & 0.0 & 0.011695 & 0.244792 \\
2700 & 0.0 & 0.021565 & 0.362157 \\
2800 & 0.0 & 0.038689 & 0.555819 \\
2900 & 0.0 & 0.064316 & 0.808300 \\
3000 & 0.0 & 0.099613 & 1.106536 \\
3100 & 0.0 & 0.141954 & 1.432723 \\
3200 & 0.0 & 0.191728 & 1.779128 \\
3300 & 0.0 & 0.245980 & 2.132738 \\
3400 & 0.0 & 0.300718 & 2.478126 \\
3500 & 0.0 & 0.356153 & 2.833188 \\
3600 & 0.0 & 0.411375 & 3.182271 \\
3700 & 0.0 & 0.467986 & 3.522503 \\
3800 & 0.0 & 0.526894 & 3.854023 \\
3900 & 0.0 & 0.591553 & 4.162256 \\
4000 & 0.0 & 0.661128 & 4.451381 \\
4100 & 0.0 & 0.739789 & 4.728403 \\
4200 & 0.0 & 0.826032 & 4.978128 \\
4300 & 0.0 & 0.933288 & 5.260589 \\
2300 & +0.5 & 0.006413 & 0.238301 \\
2400 & +0.5 & 0.010061 & 0.260112 \\
2500 & +0.5 & 0.016492 & 0.247500 \\
2600 & +0.5 & 0.027800 & 0.292038 \\
2700 & +0.5 & 0.044263 & 0.407142 \\
2800 & +0.5 & 0.068607 & 0.575496 \\
2900 & +0.5 & 0.102022 & 0.786998 \\
3000 & +0.5 & 0.143249 & 1.020113 \\
3100 & +0.5 & 0.191720 & 1.265885 \\
3200 & +0.5 & 0.244465 & 1.513235 \\
3300 & +0.5 & 0.298218 & 1.760222 \\
3400 & +0.5 & 0.349878 & 2.011364 \\
3500 & +0.5 & 0.397360 & 2.282790 \\
3600 & +0.5 & 0.439667 & 2.586420 \\
3700 & +0.5 & 0.478450 & 2.923161 \\
3800 & +0.5 & 0.514756 & 3.277294 \\
3900 & +0.5 & 0.550537 & 3.628710 \\
4000 & +0.5 & 0.588930 & 3.979705 \\
4100 & +0.5 & 0.639626 & 4.358765 \\
4200 & +0.5 & 0.700250 & 4.708474 \\
4300 & +0.5 & 0.782585 & 5.064212 \\
2300 & +1.0 & 0.014411 & 0.333220 \\
2400 & +1.0 & 0.021711 & 0.338989 \\
2500 & +1.0 & 0.032464 & 0.340200 \\
2600 & +1.0 & 0.048820 & 0.374320 \\
2700 & +1.0 & 0.070960 & 0.481655 \\
2800 & +1.0 & 0.100049 & 0.628507 \\
2900 & +1.0 & 0.136855 & 0.803741 \\
3000 & +1.0 & 0.180487 & 0.983418 \\
3100 & +1.0 & 0.229947 & 1.156046 \\
3200 & +1.0 & 0.280350 & 1.308436 \\
3300 & +1.0 & 0.329080 & 1.442195 \\
3400 & +1.0 & 0.372528 & 1.561571 \\
3500 & +1.0 & 0.410402 & 1.683466 \\
3600 & +1.0 & 0.443302 & 1.836109 \\
3700 & +1.0 & 0.470929 & 2.065178 \\
3800 & +1.0 & 0.497274 & 2.413894 \\
3900 & +1.0 & 0.529802 & 2.873790 \\
4000 & +1.0 & 0.564174 & 3.366484 \\
4100 & +1.0 & 0.601777 & 3.868406 \\
4200 & +1.0 & 0.655195 & 4.370432 \\
4300 & +1.0 & 0.718588 & 4.824441 \\
\end{longtable}
}
%%%%%%%%%%%%%%%%%%%%%%%%%%%%%%%%%%%%%%%%%%%%%%%%%%%%%%%%%%%%%%%%%%%%%%%%%%%%%%%

\end{appendix}
%%%%%%%%%%%%%%%%%%%%%%%%%%%%%%%%%%%%%%%%%%%%%%%%%%%%%%%%%%%%%%%%%%%%%%%%%%%%%%%

\end{document}